\documentclass{aa}
\usepackage[varg]{txfonts}
\usepackage{graphicx}
\usepackage{natbib}
\usepackage{xcolor}
\usepackage{pdflscape}
\usepackage{caption}
\usepackage{subcaption}
\usepackage[linkcolor=black,citecolor=cyan,filecolor=black,urlcolor=magenta]{hyperref}
\usepackage[flushleft]{threeparttable}

\title{HST grism spectroscopy of $z\sim3$ massive quiescent galaxies} 
\subtitle{Approaching the metamorphosis}

\author{C.~D'Eugenio\inst{1}
\and E.~Daddi\inst{1} 
\and R.~Gobat \inst{2}
\and V.~Strazzullo \inst{3,}\inst{4,}\inst{5,}\inst{6}
\and P.~Lustig \inst{4}
\and I.~Delvecchio \inst{1,}\inst{5}
\and S.~Jin \inst{7,} \inst{8}
\and A.~Cimatti \inst{9,}\inst{10}
\and M.~Onodera \inst{11,}\inst{12}}

\institute{CEA, Irfu, DAp, AIM, Université Paris-Saclay, Université de Paris, CNRS, F-91191 Gif-sur-Yvette, France
\and 
Instituto de F\'isica, Pontificia Universidad Cat\'olica de Valpara\'iso, Casilla 4059, Valpara\'iso, Chile
\and 
Faculty of Physics, Ludwig-Maximilians-Universit\"at, Scheinerstr. 1, 81679 Munich, Germany
\and 
University of Trieste, Piazzale Europa, 1, 34127 Trieste TS, Italy
\and 
INAF - Osservatorio Astronomico di Brera, via Brera 28, I-20121, Milano, Italy \& via Bianchi 46, I-23807, Merate, Italy
\and 
INAF - Osservatorio Astronomico di Trieste, via Tiepolo 11, I-34131, Trieste, Italy
\and 
Instituto de Astrofísica de Canarias (IAC), E-38205 La Laguna, Tenerife, Spain
\and 
Universidad de La Laguna, Dpto. Astrofísica, E-38206 La Laguna, Tenerife, Spain
\and 
Università di Bologna, Dipartimento di Fisica e Astronomia,
Via Gobetti 93/2, I-40129, Bologna, Italy
\and
INAF - Osservatorio Astrofisico di Arcetri,
Largo E. Fermi 5, I-50125, Firenze, Italy
\and 
Subaru Telescope, National Astronomical Observatory of Japan, National Institutes of Natural Sciences, 650 North A’ohoku Place, Hilo, HI 96720, USA
\and 
Department of Astronomical Science, SOKENDAI (The Graduate University for Advanced Studies), 650 North A'ohoku Place, Hilo, HI, 96720, USA\\}

\offprints{C. D'Eugenio, \email{chiara.deugenio@cea.fr}}

\begin{document}

\date{Received December 7, 2020; accepted May 26, 2021}

\abstract{Tracing the emergence of the massive quiescent galaxy (QG) population requires the build-up of reliable quenched samples, disentangling these systems from red, dusty star-forming sources. We present \textit{Hubble Space Telescope} WFC3/G141 grism spectra of 10 quiescent galaxy candidates selected at $2.5<z<3.5$ in the COSMOS field. Spectroscopic confirmation for the whole sample is obtained within 1-3 orbits based on the presence of strong spectral breaks and Balmer absorption lines. Combining their spectra with optical to near-infrared (near-IR) photometry, star-forming solutions are formally rejected for the entire sample. Broad spectral indices are consistent with the presence of young A-type stars, which implies that the last major episode of star formation has taken place no earlier than $\sim$300-800 Myr prior to observation, confirming clues from their post-starburst UVJ colors. Marginalising over three different slopes of the dust attenuation curve, we obtain short mass-weighted ages and an average peak star formation rate of SFR$\sim10^3$ M$_{\odot}$ yr$^{-1}$ at $z_{formation}\sim3.5$. Despite mid- and far-IR data are too shallow to determine the obscured SFR on a galaxy-by-galaxy basis, the mean stack emission from 3GHz data constrains the level of residual obscured SFR to be globally below 50 M$_{\odot}$ yr$^{-1}$, hence three times below the scatter of the coeval main sequence. Alternatively, the very same radio detection suggests a widespread radio-mode feedback by active galactic nuclei (AGN) four times stronger than in z$\sim$1.8 massive QGs. This is accompanied by a 30\% fraction of X-ray luminous AGN with a black hole accretion rate per unit SFR enhanced by a factor of $\sim30$ with respect to similarly massive QGs at lower redshift. The average compact, high Sérsic index morphologies of the galaxies in this sample, coupled with their young mass-weighted ages, suggest that the mechanisms responsible for the development of a spheroidal component might be concomitant with (or preceding) those causing their quenching.}

\keywords{Quiescent galaxies -- Galaxy evolution -- Quenching}
\maketitle

\section{Introduction}
The formation channels that lead to the build-up of the red sequence of massive, quiescent galaxies (QGs) are still to be fully understood. In the present-day Universe, 75\% of the total stellar mass budget is locked into spheroids, either elliptical galaxies or bulges in spirals. These are several Gyr old, pressure-supported stellar systems that are unable to host a significant amount of star formation \citep{Renzini06}. The majority of massive galaxies (log(M$_{\star}$/M$_{\sun}$)>10.5) in the local Universe appears to be quenched, with QGs outnumbering star-forming galaxies by a factor of 10 at log(M$_{\star}$/M$_{\sun}$)>11.5 \citep{Baldry04}.
Fossil-record studies have established that the most massive galaxies are those where the oldest stellar populations are found, implying an anti-correlation between their stellar mass and the duration of their main star formation episode \citep{Gallazzi05, Thomas05, Citro16}. These results have been reinforced by the first identifications of QGs at increasing redshifts \citep{Dunlop96, Franx03, Cimatti04, Daddi05} allowed by near-infrared (NIR) sensitive detectors that sample the rest frame optical at $z\geq1.5$. These advances were essential to show that the assembly and quenching of massive systems took place at z>1-1.5, with little evolution thereafter of the high-mass end of their stellar mass function, especially compared to the progressive rise of the low-mass end \citep{Fontana04, Cimatti06, Arnouts07, Drory09, Pozzetti10, vd14, Gargiulo16, Davidzon17, Kawinwanichakij20}. The growth of the lower-mass end of the red sequence can be thought of as the result of the generalised progressive decline in global star formation rate density through gas consumption, cluster-related processes, and cosmic starvation affecting the star-forming population mostly at z<1.5 \citep[e.g.][]{emsellem11, Saracco2011, carollo13, MD14, Sargent14, Schreiber2015, Wild16, maltby18, Matharu19, Kawinwanichakij20}. On the other hand, understanding how massive galaxies quenched in an epoch in which galaxies were generally gas rich and prodigiously star-forming \citep[2<z<4,][]{Daddi10,tacconi10, Genzel10, MD14} is in itself non-trivial.\\
Historically, the progressive discovery of populations of massive QGs already in place at high-z \citep{Dunlop96, Franx03, Cimatti04, Daddi05,Kriek09, Gobat12,G17} posed challenges to hierarchical models of structure formation \citep[e.g.][]{Cimatti06, Steinhardt16}. Hydrodynamical simulations and semi-analytical models have struggled to reproduce the comoving number densities of massive, passively evolving galaxies at z >2--3, falling short by roughly an order of magnitude \citep{Wellons15, Nelson15, Steinhardt16, dave16, Cecchi19, Schreiber18}. In recent years, however, confirmations of quiescent galaxies extended up to z$\sim$4 \citep{G17, Schreiber18, Valentino20, Forrest20a, Forrest2020b} while cosmological simulations progressively increased in volume, spatial and mass resolution, as well as in improvements of feedback schemes and subgrid physics regulating star formation (\citet[Illustris:][]{Vogelsberger14, Vogelsberger14b,  Genel2014, Nelson15}; \citet[Illustris TNG:][]{Pillepich18,Nelson2019}; \citet[EAGLE:][]{Schaye15};  \citet[SIMBA:][]{dave19}). The most recent estimates of the number density of QGs, either from observational samples or from state-of-the-art cosmological simulations, appear to broadly agree on the number density of quiescent galaxies up to z$\sim$3, but it is unclear whether or not this holds at z$\sim$4 and if a population of passive objects exists as early as z$\sim$5 in significant numbers. Additionally, the degree to which this agreement is robust against systematics, such as the exact details of sample selection, is not firmly established \citep{Merlin19, Valentino20}.\\
Despite the major progress accomplished so far in this field, it remains unclear whether there is a dominant process that causes quenching \citep{ManBelli18}. The extreme stellar densities of compact quiescent systems at $z\sim2$ \citep{Franx03, Daddi05, vDokkum08, Newman12} suggest that they are remnants of an intense burst of star formation triggered by the rapid collapse of a large amount of gas that occurred at z>4. This could be resulting, for instance, from starbursts plunging down into quiescence after dissipative gas-rich mergers \citep{Cimatti08, Elbaz18, GGuijarro18, GGuijarro19, Puglisi19} or from a more “secular-like” evolution of high-z dusty star-forming galaxies quickly accreting and consuming gas through disk instabilities, leaving compact passive remnants (e.g. \citet{Dekel09, Barro13, Toft14, Zolotov15}).
Increased consensus among semi-analytic models and hydrodynamical simulations has been reached on ascribing the shut-down of star formation in massive galaxies to AGN feedback \citep{DLB07,Henriques17,Girelli2020}. Major mergers or violent disk instabilities can compress the gas into massive compact cores and trigger starburst events that rapidly consume the gas of the system \citep{Dekel86, Murray05}. Quasar activity is ignited by these processes, launching powerful outflows into the CGM and depleting the host galaxy from its reservoirs \citep{Sanders88, DiMatteo05, Hopkins06}. However, AGN feedback is not the only process that can halt or reduce star formation in high-z galaxies. Cosmological starvation \citep{Feldmann15} and the development of a stable virial shock in sufficiently massive haloes \citep{DB06, Cattaneo06} could play a role, followed by maintenance processes such as radio-mode feedback from radiatively inefficient accretion onto a supermassive black hole (SMBH) \citep{Best05, Croton06}, gravitational heating of the diffuse medium from infalling satellites \citep{DB06, Khochfar08, Johansson09, Johansson12} or morphological quenching \citep{Martig09}. The detailed study of large, statistically relevant samples of massive quiescent galaxies is crucial to distinguish among these mechanisms. Robust samples of quiescent galaxies are challenging to build up, however: the spectra necessary to reject low-redshift interlopers or star-forming contaminants become increasingly more difficult to obtain at $z>1.4$ because it requires either long integrations in 8-10m class ground-based telescopes or space-based observations. Moreover, the rapid drop in number density of massive quenched objects at $z>1.5$ requires large areas covered by deep observations. On a positive note, on the other hand, the boundedness of the age of the Universe can be exploited to investigate the demographics of the quiescent population across cosmic time as it keeps emerging. In fact, as the population becomes younger and younger at increasing redshift, the discerning power of rest frame optical spectra at mapping the early star formation of massive QGs surpasses the one at low redshift. This is because around z$\gtrsim$2, the Universe starts to be young enough to make stellar age differences of $\sim$1 Gyr, down to few hundreds Myr, visible through the rapid appearance of Balmer absorption lines in stellar populations of ages $<1$ Gyr, when A-type stars enter the turn-off, increasingly dominating the integrated stellar spectra. Relatively large samples of QGs have been assembled up to $z\sim2.5$ in COSMOS and the CANDELS fields exploiting large telescopes (e.g. \citet{Bezanson13, Belli15, Onodera15, Kriek16, Belli19, Stockmann20}) or the \textit{Hubble Space Telescope} (HST) in combination with strong lensing \citep{Whitaker12, Whitaker13, Newman18}. These studies have been tracing the progressive decrease in stellar age of log(M$_{\star}$/M$_{\odot}$)>11 QGs with look-back time, revealing an increasing spread in stellar age and dust extinction with bulk values of about 1–2 Gyr and A$_{\rm{V}}$ = 0--1.0 mag at $z\sim2$, respectively. This is apparently happening while keeping high metallicities and with velocity dispersions up to a factor of two higher than local scaling relations \citep{Toft2012, Ono12, vdSande13, Belli15, Kriek16, Belli2017, EstradaCarp19, EstradaCarp20, Stockmann20, Tanaka20}. The progressive appearance of the quenched population can therefore be quantified through the relative fraction of young versus old systems once an age threshold is defined \citep[e.g. age>1 Gyr,][]{Whitaker13}.
At even higher redshifts, the colour selections generally applied to photometric samples already indicate a substantial migration of QGs towards bluer colours \citep{Whitaker11, Muzzin13}, accompanied by the drop by roughly one order of magnitude of their number densities \citep{Straatman14, Davidzon17}. Spectroscopic follow-ups have the advantage of refining this picture by testing the colours of selected galaxies against photometric errors, star-forming interlopers, or AGN interfering with their spectral energy distribution (SED). Moreover, detailed spectra allow us to potentially break or at least reduce the degeneracy between age, dust extinction, and metallicity, provided that enough signal-to-noise ratio and spectral coverage are reached. Here we present one of the largest samples of spectroscopically confirmed QGs at $2.4<z<3.3$ obtained with HST dedicated observations. In section~\ref{sec:selection} we describe the sample selection. In section~\ref{sec:obs} we give details on the observational strategy and data reduction. In section~\ref{sec:specconf} we present the spectral analysis and the spectroscopic confirmation of our targets. Section~\ref{sec:combo} presents their formal classification, the effect of adding COSMOS2015 photometry with or without calibration of zero-points, as well as the use of marginalising over multiple attenuation laws. In section~\ref{sec:age} we constrain their recent star formation history (SFH) by comparing the relative strength of the Balmer and 4000\AA\ breaks. In section~\ref{sec:agn} we investigate the incidence of AGN in our sample. In section~\ref{sec:discussion} we discuss our results in the context of the current literature. Finally, in section~\ref{sec:summary}, we summarise our results and conclusions. We assume a $\Lambda$CDM cosmology with H$_{0}=70$ km s$^{-1}$ Mpc$^{-1}$, $\Omega_{M}=0.27$, $\Omega_{\Lambda}=0.73,$ and a \citet{salpeter} initial mass function (IMF), unless otherwise specified. Magnitudes are given in the AB photometric system.

\section{Sample selection}
\label{sec:selection}
Given the expected low number density of high-z massive quiescent galaxies, large fields with deep photometric coverage are required to identify and reliably assess their SEDs. For this reason, we exploited the 2deg$^2$ COSMOS field. Sources with $K_{\rm{tot}}<22.5$ were extracted from the catalogue of \citet{mc10}, limiting the selection to those satisfying the observed frame BzK colour criterion for passive systems \citep{Daddi04}. Targets formally classified as star-forming BzK (sBzK) with a signal-to-noise ratio S/N$<$5 in the B and z bands were retained, as these photometric candidates are degenerate with quiescent galaxies becoming fainter in such bands with increasing redshift and decreasing mass. Photometric redshifts specifically calibrated for high-z QGs were derived with EAZY \citep{Brammer08} as in \citet{Ono12} and \citet{strazz15}. This calibration was based on the sample of 34 spectroscopically confirmed passive galaxies at 1.3 < z$_{\rm{spec}}$ < 2.1 observed with VLT/VIMOS that later appeared in \citet{Gobat17} and the sample of 18 passive galaxies at 1.4<z$_{\rm{spec}}$ <1.9 of \citet{Ono12} observed with Subaru/MOIRCS. The calibrated $z_{\rm{phot}}$ were used to select galaxies within $2.5\leqslant z_{\rm{phot}} \leqslant3.5$ and to remove objects with UVJ rest frame colours inconsistent with passive evolution \citep{Pozzetti2000, Labbe05, Williams09}. 
SED fitting was performed using FAST \citep{Kriek09}, allowing for constant and delayed exponentially declining SFHs. Optical dust attenuation was left free to vary up to A$_{\rm{V}}$=5 mag assuming a \citet{Calzetti2000} attenuation law. Fits were repeated by adopting purely quiescent templates only. All passive UVJ candidates whose SED fits to optical-NIR photometry could not reject dusty star-forming solutions at high confidence were further discarded. Contamination from dusty star-forming galaxies was further minimised by removing objects with Spitzer/MIPS 24$\mu$m S/N$\geq$4 detections in the catalogue of \citet{Lefloch09}, except for galaxies with high-confidence passive SEDs, which are indicative of mid-IR emission caused by a central dusty AGN torus. This selection provided a total of 174 passive candidates with UVJ-quiescent colours (47 of which were pBzK in the original selection, plus 67 and 60 uncertain sBzK with and without significant MIPS detections, respectively). The maximum required magnitude to obtain HST/G141 spectra with sufficient S/N in order to secure spectroscopic confirmation within one to two orbits was assessed by simulating their grism spectra based on their best-fitting SED templates. This yielded a magnitude cut of  H$_{AB}$<22, which narrowed the sample down to 23 objects. A total of 10 galaxies were targeted for HST WFC3/IR G141 near-IR observations: 9 randomly drawn galaxies with H$_{AB}<$22 ($M_{\star}>$ 1.1$\times$ 10$^{11}$ M$_{\odot}$) plus 1 robust candidate with H$_{AB}$=22.9 (M$_{\star}$=8$\times$ 10$^{10}$ M$_{\odot}$) selected to be the highest-z candidate based on its high-confidence passive SED. More details of the selection are available in \citet{lustig}.

\section{HST WFC3 F160W imaging and G141 grism spectroscopy}
\label{sec:obs}
Ground-based observations have already confirmed the existence of quiescent galaxies at $z\sim4$ \citep{Schreiber18, Valentino20, Forrest20a, Forrest2020b}. However, in the framework of high-z galaxy evolution, their statistical power is mostly modulated by the time-expensiveness of these campaigns. Moreover, OH sky emission lines and related background notoriously affect the quality of the spectra, sometimes effectively cutting them in correspondence to spectral regions that are crucial to estimate stellar ages. Space-based observations, on the other hand, ensure extended and continuous spectral coverage. The HST WFC3 G141 slitless spectrograph has a spectral coverage from 1.1 to 1.7 $\mu$m, reaching maximum transmission at 1.45 $\mu$m. This allows access to the rest frame near-UV/optical and specifically to the Balmer/4000 \AA\ break region up to z$\sim$3.2. The G141 dispersion in the first spectral order is 46.5 \AA\ pixel$^{-1}$ and $R\sim130$ for unresolved sources and compact objects such as those considered in this paper. For resolved sources, the spectral resolution is determined by their morphology, namely their size along the dispersion axis. For these reasons, HST WFC3 observations of the selected sources include F160W imaging exposures. As this paper focuses mostly on the spectroscopic analysis, we refer to \citet{lustig} for a thorough analysis of their morphology.

\subsection{Observing strategy}
Observations for program GO 15229 took place from January 11, 2018, to December 2, 2018. Each pointing was observed from one to three orbits according to each target H$_{AB}$ magnitude for a total of 17 orbits (see Table~\ref{tab:orbits}). For each target the first orbit was split into a direct F160W exposure (for a total of 984 s) and a grism G141 exposure (for a total of 1498 s) adopting the WFC3-IR-DITHER-LINE-3PT dither pattern. For targets with two orbits, the second orbit was also split in two, with a total exposure of 73 s in F160W and 2496 s in G141, adopting the WFC3-IR-DITHER-BLOB dither pattern. The third orbit for ID 4 was a repetition of the second orbit. The ORIENT was carefully chosen for each target in order to avoid any contamination from neighbouring sources.

\begin{table*}
\centering
\caption{Coordinates and orbit details for the targeted galaxies. The $z_{\rm{phot, cal}}$ column lists the calibrated $z_{\rm{phot}}$ used for the sample selection (see Sect. ~\ref{sec:selection}). H-band magnitudes are taken from \citet{lustig} and result from fitting the F160W images of our sources with PSF-convolved Sérsic profiles.}
\label{tab:orbits}
\renewcommand\arraystretch{1.2}
\begin{tabular}{cccc cccccc}
\hline
ID & ID$_{\rm{Laigle}}$ & RA & DEC &  H$_{tot}$&  z$_{\rm{phot, cal}}$ & N$_{\rm{orbits}}$ & \multicolumn{2}{c}{total int. time (s)}\\
     &                           &       &          &                   &                            &  &                        F160W        & G141  \\
\hline
\hline
1 &135730  & 10:01:39.9790 & +01:29:34.49 & 21.99$^{+0.10}_{-0.09}$ & 2.6$\pm0.2$ & 1 & 983.8       & 1497.7 \\ 
2 &137182  & 10:00:57.3452 & +01:29:39.46 & 21.32$^{+0.02}_{-0.03}$ & 2.7$\pm0.1$ & 1 & 983.8       & 1497.7 \\ 
3 &252568  & 09:57:48.5727 & +01:39:57.82 & 21.88$^{+0.10}_{-0.10}$ & 2.8$\pm0.2$& 2 & 1056.7     & 3993.9 \\ 
4 &361413  & 10:02:0.9700  & +01:50:24.30 & 23.37$^{+0.08}_{-0.11}$ & 3.2$\pm0.1$ & 3 & 1129.7       & 6490.1 \\
5 &447058 & 09:59:11.7700 & +01:58:32.96 & 22.20$^{+0.02}_{-0.02}$ & 2.5$\pm0.1$ & 2 & 1056.7     & 3993.9 \\
6 &478302 & 09:59:1.3123  & +02:01:34.15 & 22.23$^{+0.02}_{-0.02}$ & 2.6$\pm0.2$ & 2 & 1056.7      & 3993.9 \\
7 &503898 & 10:01:31.8594 & +02:03:58.79 & 21.60$^{+0.13}_{-0.11}$ & 2.6$\pm0.1$& 1 & 983.8      & 1497.7 \\
8 &575436 & 10:00:43.7668 & +02:10:28.71 & 22.30$^{+0.03}_{-0.04}$ & 2.8$\pm0.2$& 2 & 1056.7     & 3993.9 \\
9 &707962 & 09:59:32.5170 & +02:22:21.99 & 21.66$^{+0.15}_{-0.17}$ & 2.6$\pm0.1$& 1 & 983.8      & 1497.7 \\ 
10 &977680 & 10:00:12.6549 & +02:47:23.47 & 22.39$^{+0.03}_{-0.03}$& 2.5$\pm0.1$& 2 & 1056.7     & 3993.9 \\

\hline 
\end{tabular}
\end{table*}

\subsection{Data reduction}
\label{sec:reduction}
\begin{figure*}
\centering
   \includegraphics[width=\textwidth]{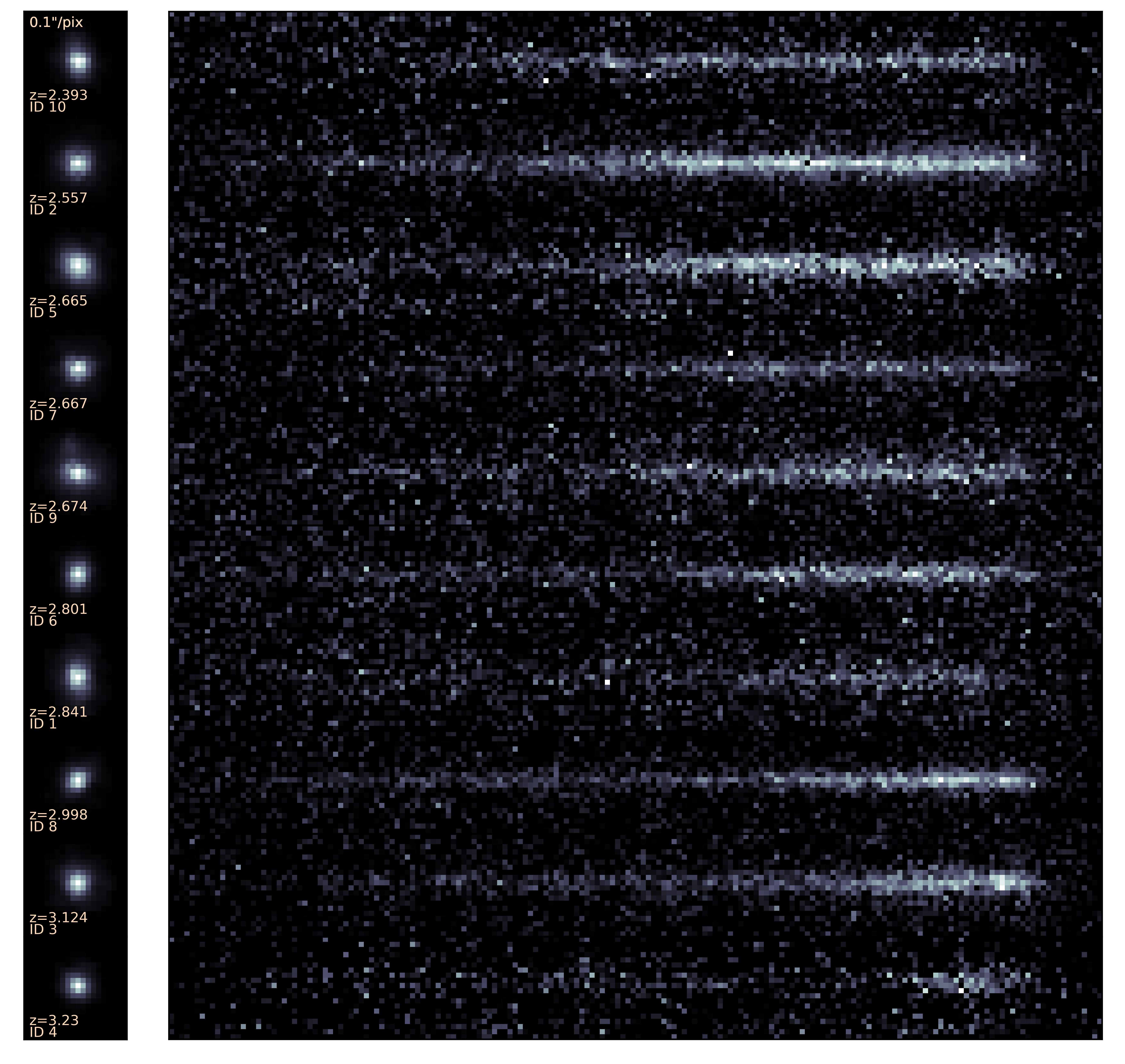}
     \caption{F160W imaging cutouts (left) and 2D G141 spectra (right) at the native WFC3 pixel scale. Redshift increases from top to bottom. The color scale is in linear scale.}
     \label{fig:2D}
\end{figure*}

The data reduction of direct F160W and grism G141 exposures was performed by adopting the pipeline \textit{grizli}, version number \textit{0.7.0-34-g91c9412} \footnote{https://github.com/gbrammer/grizli}. After relative alignment of each direct and grism exposures, absolute astrometric registration was performed, providing the pipeline with COSMOS ACS I-band reference catalogues of RA-DEC positions of sources brighter than I$_{\rm{AB}}$<27 mag within a radius of 5' from each target. Grism sky background subtraction was performed by \textit{grizli} by means of the Master sky images from \citet{brammer15}, applying a grey correction using the F160W flat-field for the G141 grism. The residuals of the background subtraction are generally of the order of 0.5-1\% of the overall background level and are further subtracted, removing a column-average of the sky pixels in the grism exposures. During this phase, the pipeline also runs Astrodrizzle to reject cosmic rays, persistence, and other artifacts. The final drizzling parameters were kept as default. A segmentation map is produced out of the drizzled and combined direct exposure. This is later used to identify and model the spectral trace of each object in each of the grism exposures and to generate model contaminants to be subtracted from the target cutout. The reduced and decontaminated 2D spectra of our targets are presented in Fig.~\ref{fig:2D} together with their relative F160W cutouts. At this stage, the detector's FOV is parsed and modelled assuming simple linear continua for all objects in the field (brighter than 25 mag).
Grizli then refines the modelling of the brightest objects ([16, 24] mag) with a second-order polynomial fit, fitting spectra directly after subtracting the model for contaminants. At this stage, the background was further fitted for in each of the target beams to account for further residuals. The coefficients of this fit were used to optimally extract any residual background level, which was subtracted from the optimally extracted 1D spectrum of the target. The final scale of the extracted 1D spectra is 0.8 in units of the native dispersion of the G141 spectrograph. Sources brighter than H$_{AB}$<22 provide a mean S/N $\sim$ 15 over 100 \AA\ at 1.6 $\mu$m within a relatively low number of one to two orbits per target.\\ 

In the following sections we give a detailed description of the method we used to measure the redshifts, as well as the criteria we adopted to establish the nature of our sources. In brief, we first fitted the spectra alone to extract $z_{spec}$ and check the quality of the $z_{phot}$ calibration (Sect.~\ref{sec:specconf}). Secondly, we added COSMOS2015 photometry from \citet{Laigle16} adopting the newly derived $z_{spec}$ for SED fitting. We tested the performance of the combined fits when the zero-point (ZP) corrections were dropped as proposed in \citet{Laigle16} (Sect.~\ref{sec:zps}) and when different dust attenuation laws were used (Sect.~\ref{sec:dustlaws}). After the best configuration was identified, we tested the quiescence of individual targets on the basis of their rest frame UV-to-NIR emission as well as judging from the constraints from the mid-infrared (MIR), far-infrared (FIR), and radio emission on dust-obscured star formation (Sect.~\ref{sec:quiescencetest}). Lastly, using the combined information from both spectroscopy and photometry, we characterised our targets in terms of mass-weighted age and dust extinction (Sect.~\ref{sec:age}). We also attempted a direct estimation of the relative light-weighted strength of the Balmer and 4000 \AA\, breaks to determine the post-starburst nature of individual sources suggested by their UVJ colours.


\section{Spectroscopic confirmation}
\label{sec:specconf}

In this section we describe the procedure we adopted to obtain our grism redshifts. We also compare our results with the calibrated $z_{phot}$ used for the sample selection as well as with $z_{phot}$ available in \citet{Muzzin13}.

\subsection{Fitting setup}
\label{sec:setup}
The optimally extracted spectra were fitted with a custom IDL routine that compares the 1D spectra with composite stellar population templates by $\chi^2$-minimization. These templates were generated on the fly by combining a grid of \citet{bc03} (hereafter BC03) simple stellar population (SSP) models with a set of parametric SFHs: a constant, an exponentially declining, a delayed exponentially declining, and a truncated SFH. In the latter, the initial SFR drops to zero after a cutoff time $\tau_{\rm{tr}}$ from the onset of star formation. In standard $\tau$-models however, $\tau$ is the e-folding timescale of the SFH. The ratios of t/$\tau$ and t/$\tau_{tr}$, when above unity, approximate the description of an SSP.
Due to an apparent systematic excessive broadening when the imaging cutouts are adopted to estimate the line spread function, the adopted templates were smoothed to the G141 resolution adopting a FWHM computed by fitting the stacked absorption lines of each galaxy with the IRAF task \textit{splot}, resulting in relatively good agreement after visual inspection.
The variety in SFHs was chosen in order to allow for both passive and dusty star-forming solutions. The routine fits the observed spectra simultaneously with a set of emission lines complexes using standard line ratios \citep{anders2003} and the SFR–H$\alpha$ calibration of \citet{Kennicutt98}, where the SFR is taken from the model grid (see \citet{Gobat12} for further details). In Table~\ref{tab:params} we list the grid of parameters used for the fit. When performing the redshift identification, the \citet{Calzetti2000} attenuation law was adopted and the stellar metallicity was left free to vary from 0.4 Z$_{\odot}$ to 2.5 Z$_{\odot}$.  The age of templates was constrained to be lower than the age of the Universe at a limiting redshift as described in the following: in order to reduce the computational cost, we fitted the observed spectra through a first pass from z=0.01 to z=5.0 with a low-resolution redshift grid (dz=0.01). The probability of a given redshift was computed by comparing the $\chi^2$ difference between the best-fitting solution and all other solutions following \citet{Avni76} to determine the 1$\sigma$ confidence range. After the most probable peaks were identified, we narrowed down the redshift grid around the 1$\sigma$ peaks, allowing for an interval of $\Delta z\sim0.2$ spanned at dz=0.001. This time we limited the template library to the age of the Universe corresponding to the lowest redshift among the identified 1$\sigma$ peaks. This approach led to the redshift probability distributions shown in the upper panels of Fig.~\ref{fig:test}.

\begin{table}
\centering
\caption{Grid of parameters for the spectral fit and the optical-to-NIR SED modeling. All steps are constant in linear scale.}
\label{tab:params}
\begin{tabular}{cccc}
\hline
parameter [units] & min & max & step\\
\hline
\hline
z   & 0.01 & 5 & 0.01 (first pass) \\
     &         &    & 0.001 (second pass) \\ 
age [Gyr] & 0.1  & age$_{H}$(z) & 0.1\\
$\tau$ [Gyr]  & 0.001 & 1 & 0.005-0.05 for $\tau<0.1$\\ 
                      &           &    & 0.1 for $\tau>0.1$\\
A$_{\rm{V}}$ [mag]   & 0 & 8 & 0.1\\
\hline
\end{tabular}
\begin{tabular}{cc}
Z/Z$\odot$   & 0.4, 1, 2.5 \\ 
$\delta$ attenuation curve & -0.7, -0.4, 0 (Calzetti)\\
(Noll et al. 2009)                 & \\
IMF & Salpeter 1955 \\
Stellar population templates & Bruzual \& Charlot 2003\\
\hline 
\end{tabular}
\end{table}

\begin{figure*}
\centering
\begin{subfigure}{.4\textwidth}
  \centering
  \includegraphics[width=0.9\linewidth]{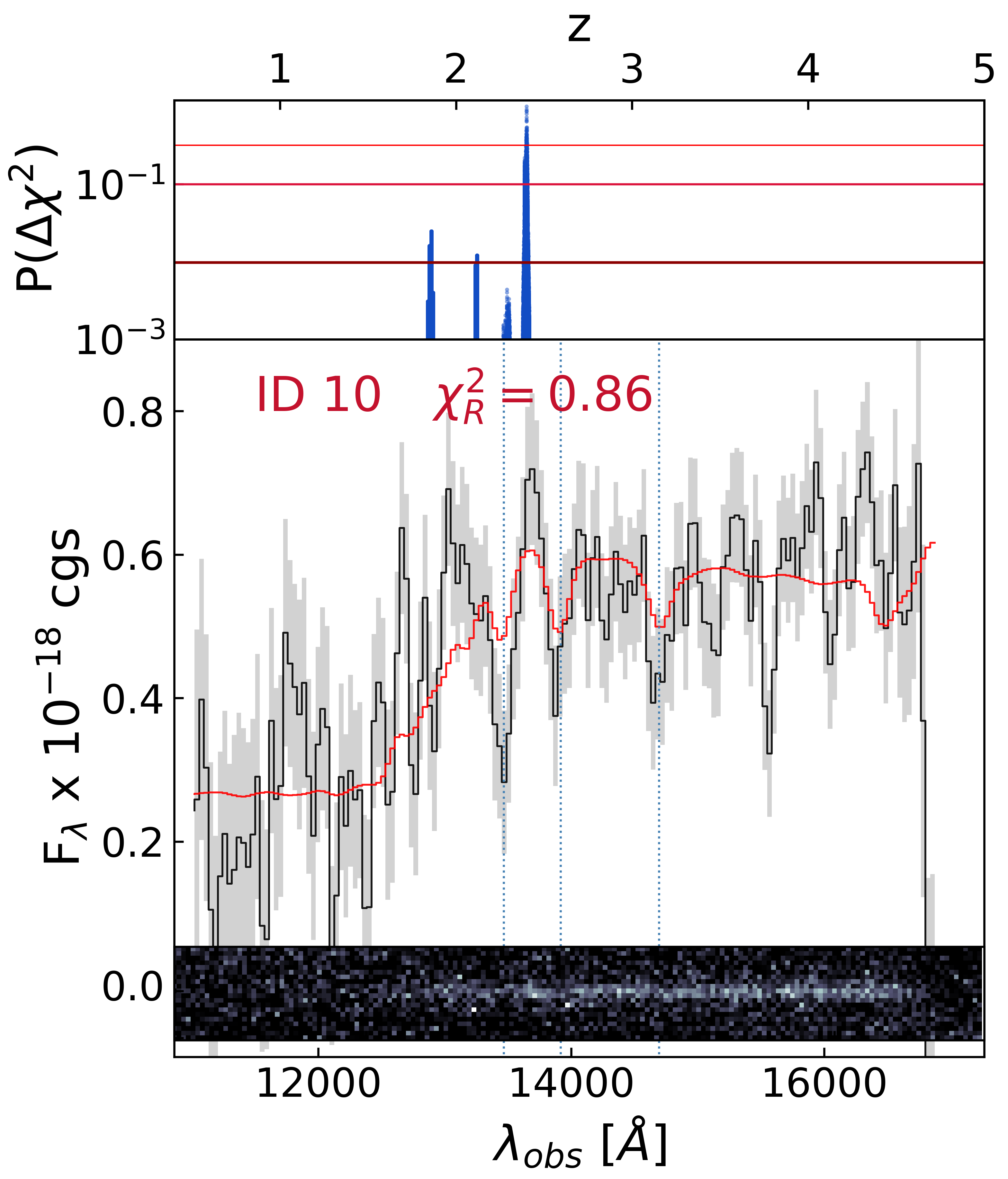}
\end{subfigure}%
\begin{subfigure}{.4\textwidth}
  \centering
  \includegraphics[width=0.9\linewidth]{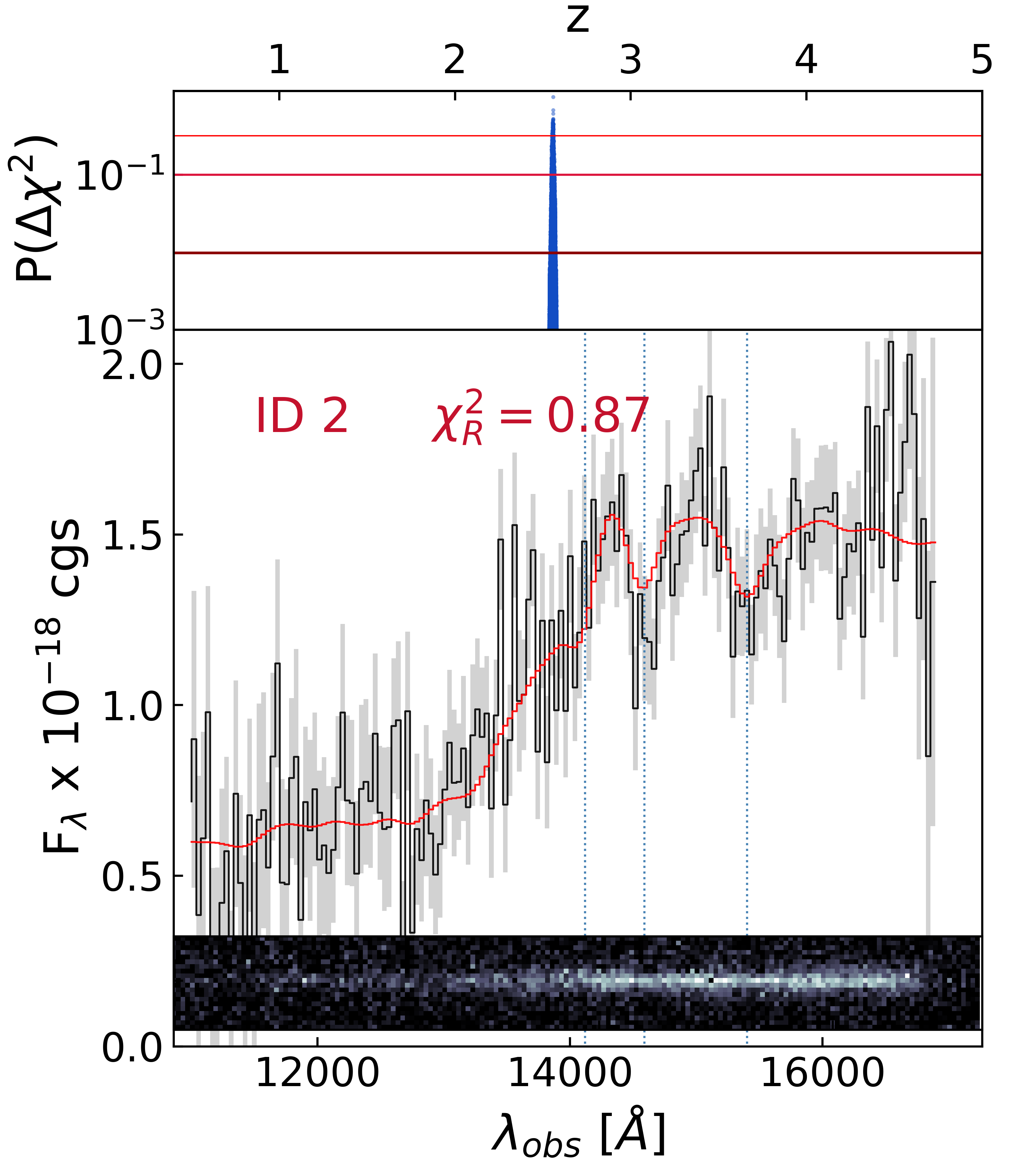}
\end{subfigure}
\begin{subfigure}{.4\textwidth}
  \centering
  \includegraphics[width=0.9\linewidth]{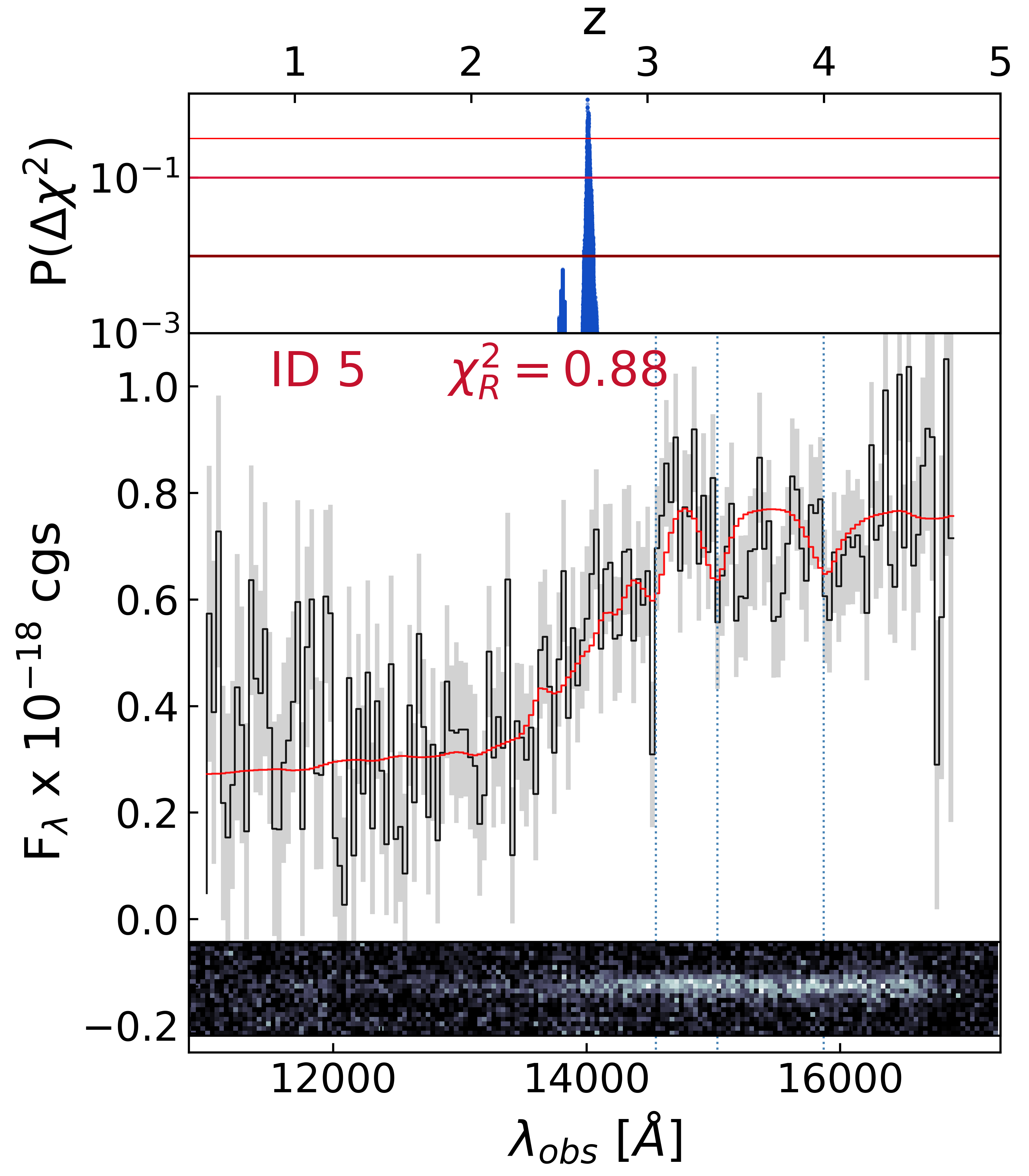}
\end{subfigure}%
\begin{subfigure}{.4\textwidth}
  \centering
  \includegraphics[width=0.9\linewidth]{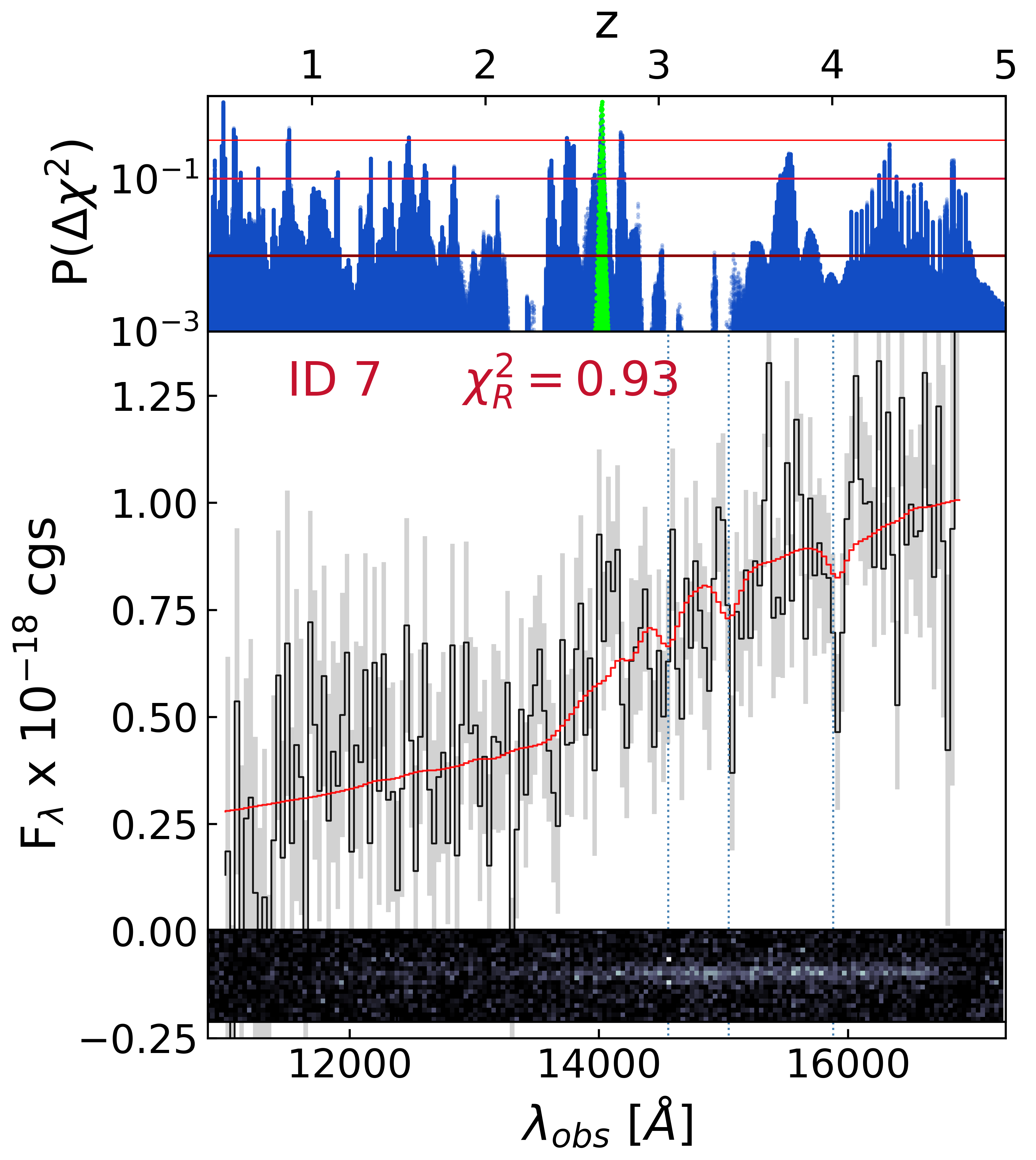}
\end{subfigure}
\begin{subfigure}{.4\textwidth}
  \centering
  \includegraphics[width=0.9\linewidth]{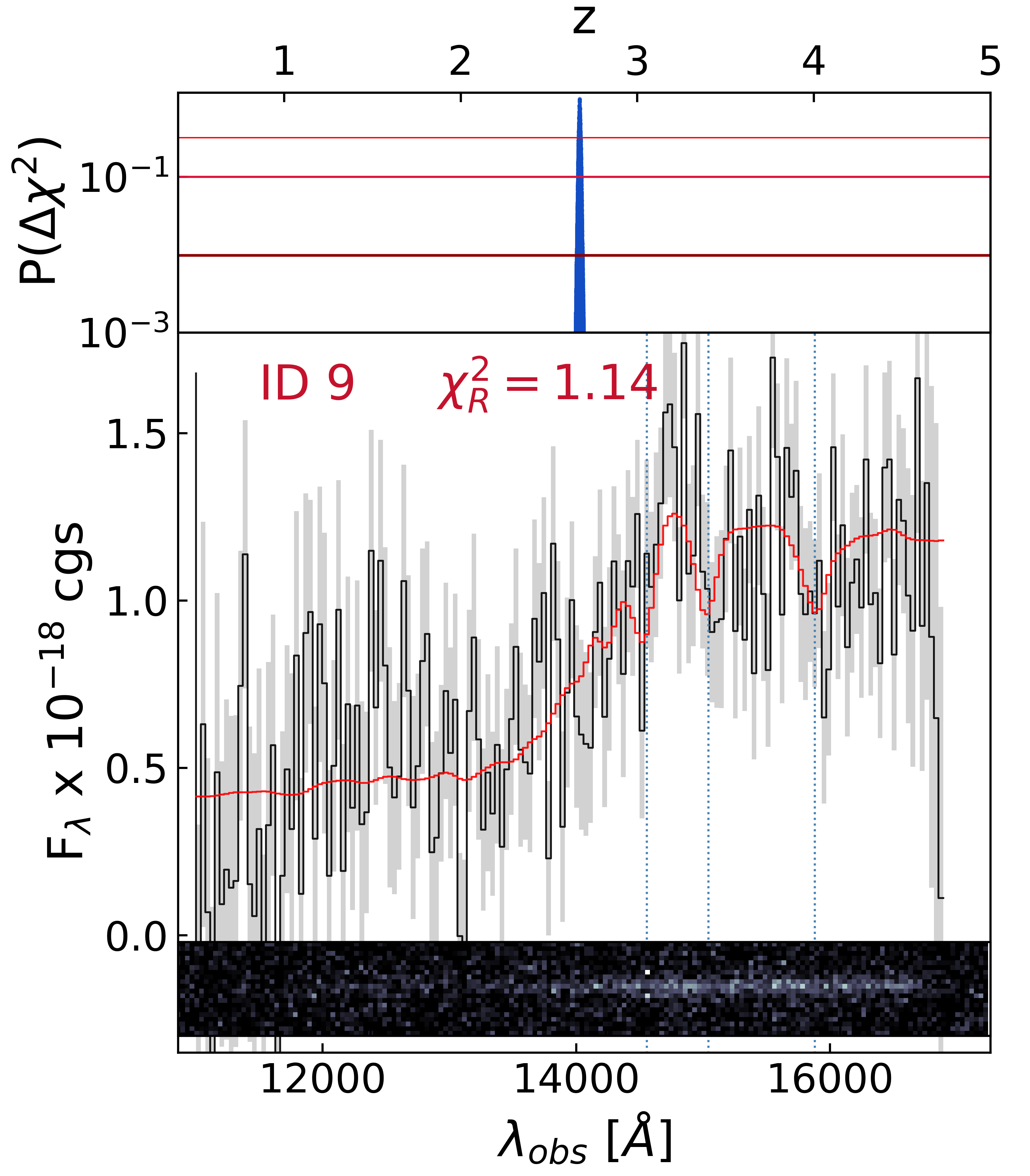}
\end{subfigure}%
\begin{subfigure}{.4\textwidth}
  \centering
  \includegraphics[width=0.9\linewidth]{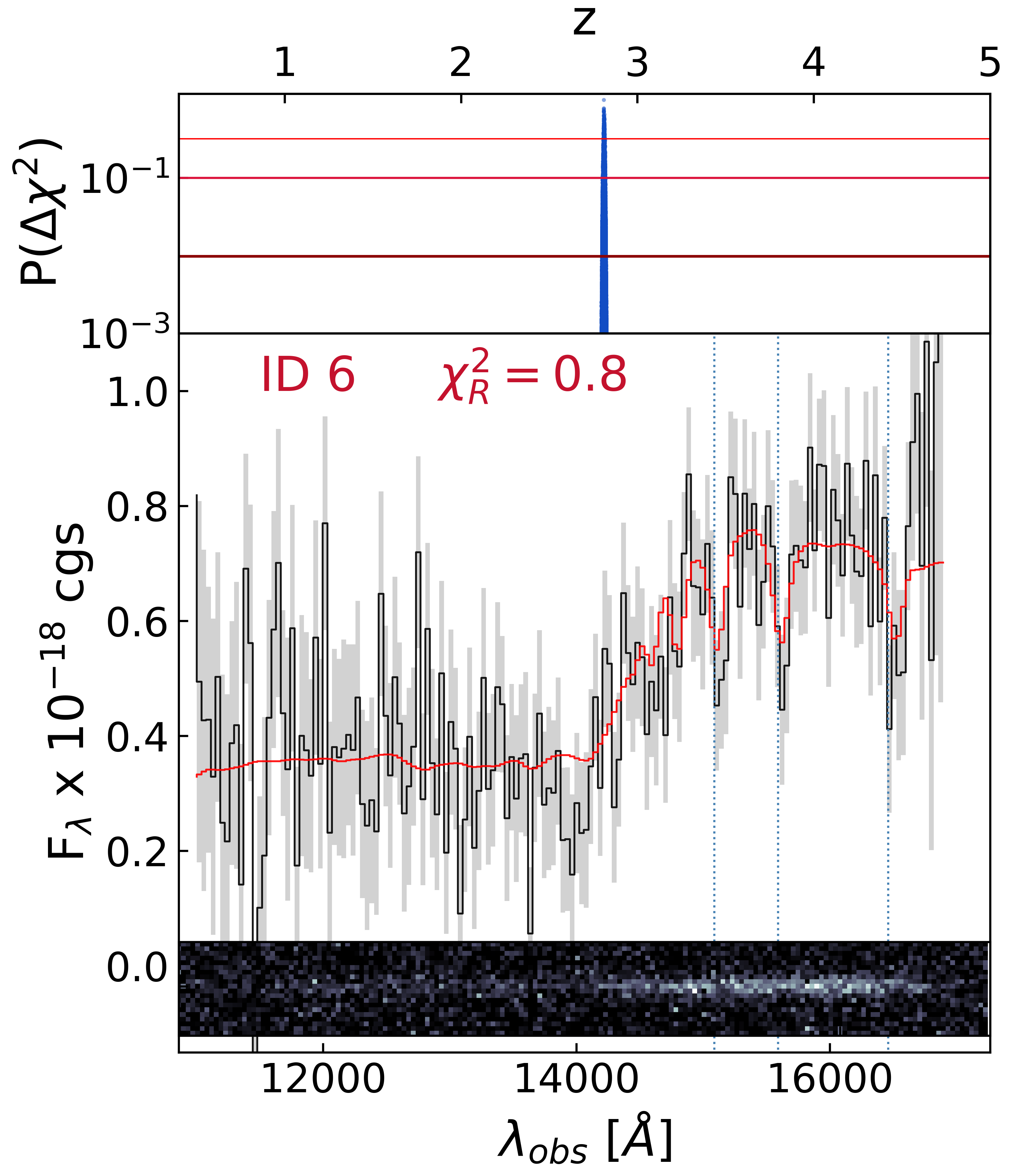}

\end{subfigure}

\caption{Upper panels: Redshift probability distribution for each target. Solid lines from light to dark red mark 1, 2, 3$\sigma$ confidence levels respectively. Green points for ID 7 mark the redshift solutions obtained combining spectroscopy and photometry as described in Sect. \ref{sec:identspecz}. Middle panels: Optimally extracted 1D grism spectra (black) and best-fitting solutions (red) of our targets. The noise vector is shown in grey in each panel. Bottom panels: Corresponding 2D G141 spectra. The color scale is in linear scale. Galaxies are shown in order of increasing redshift.}
\label{fig:test}
\end{figure*}

\begin{figure*}
\ContinuedFloat
\centering
\begin{subfigure}{.4\textwidth}
  \centering
  \includegraphics[width=0.9\linewidth]{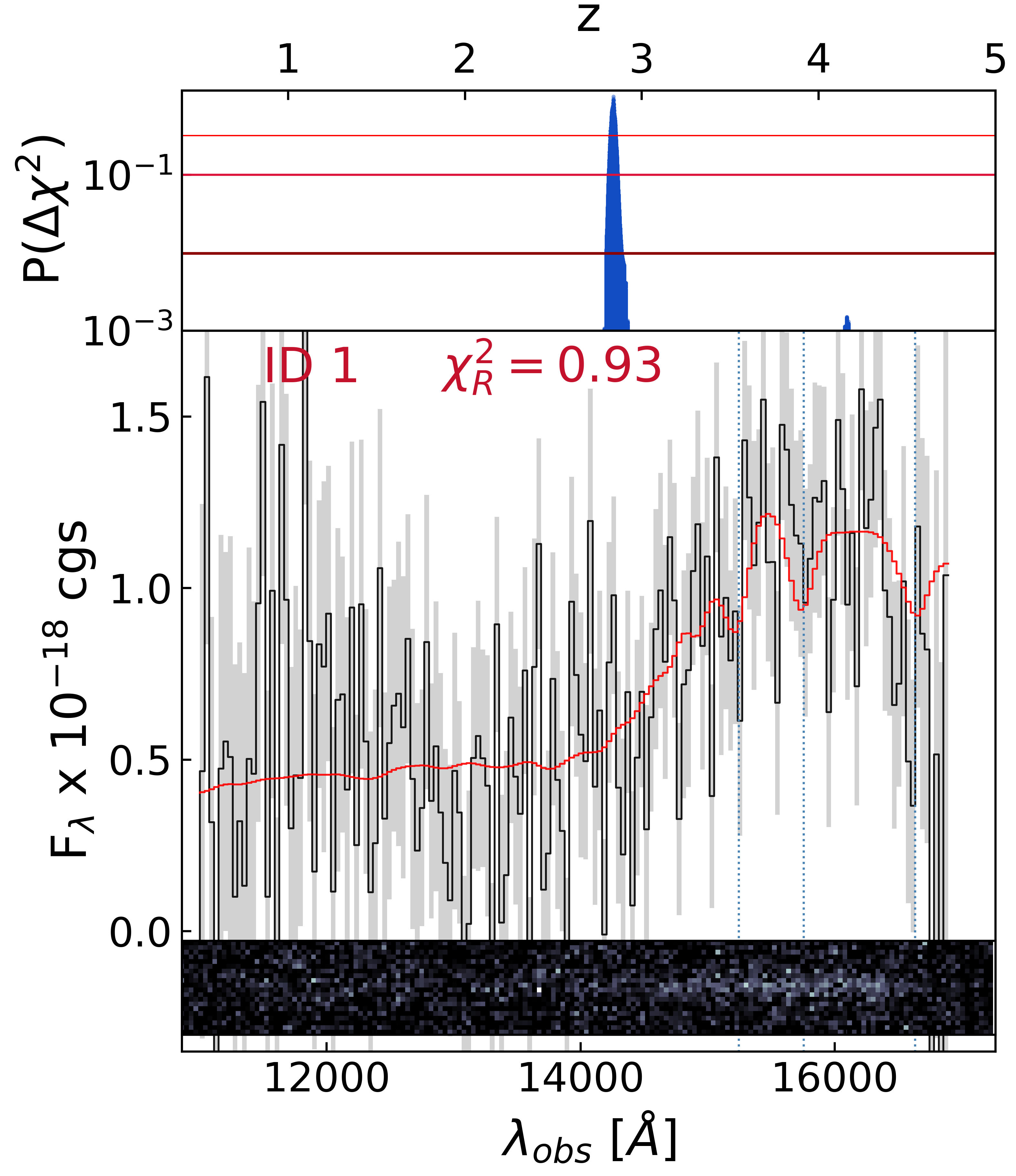}
\end{subfigure}%
\begin{subfigure}{.4\textwidth}
  \centering
  \includegraphics[width=0.9\linewidth]{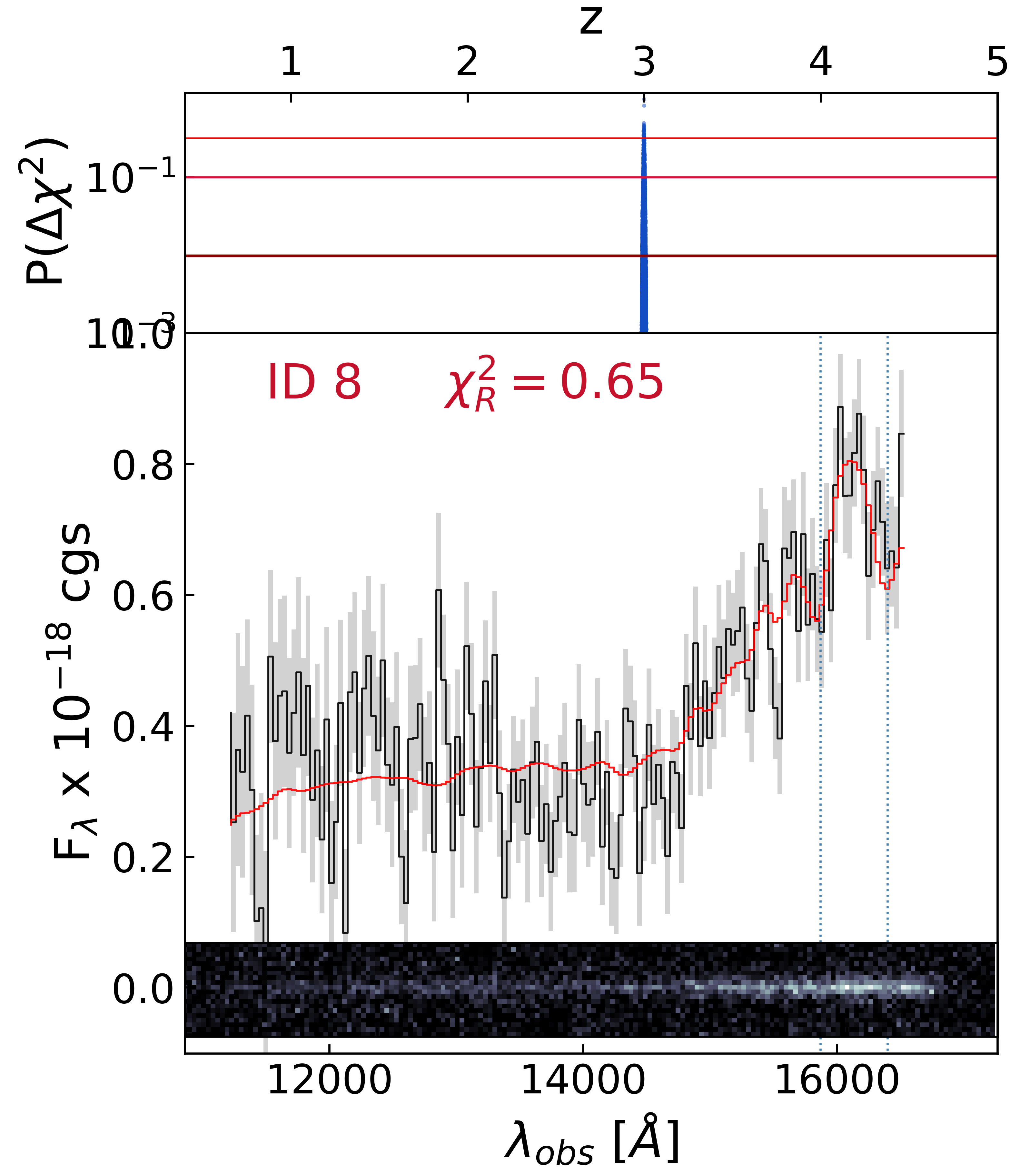}
\end{subfigure}
\begin{subfigure}{.4\textwidth}
  \centering
  \includegraphics[width=0.9\linewidth]{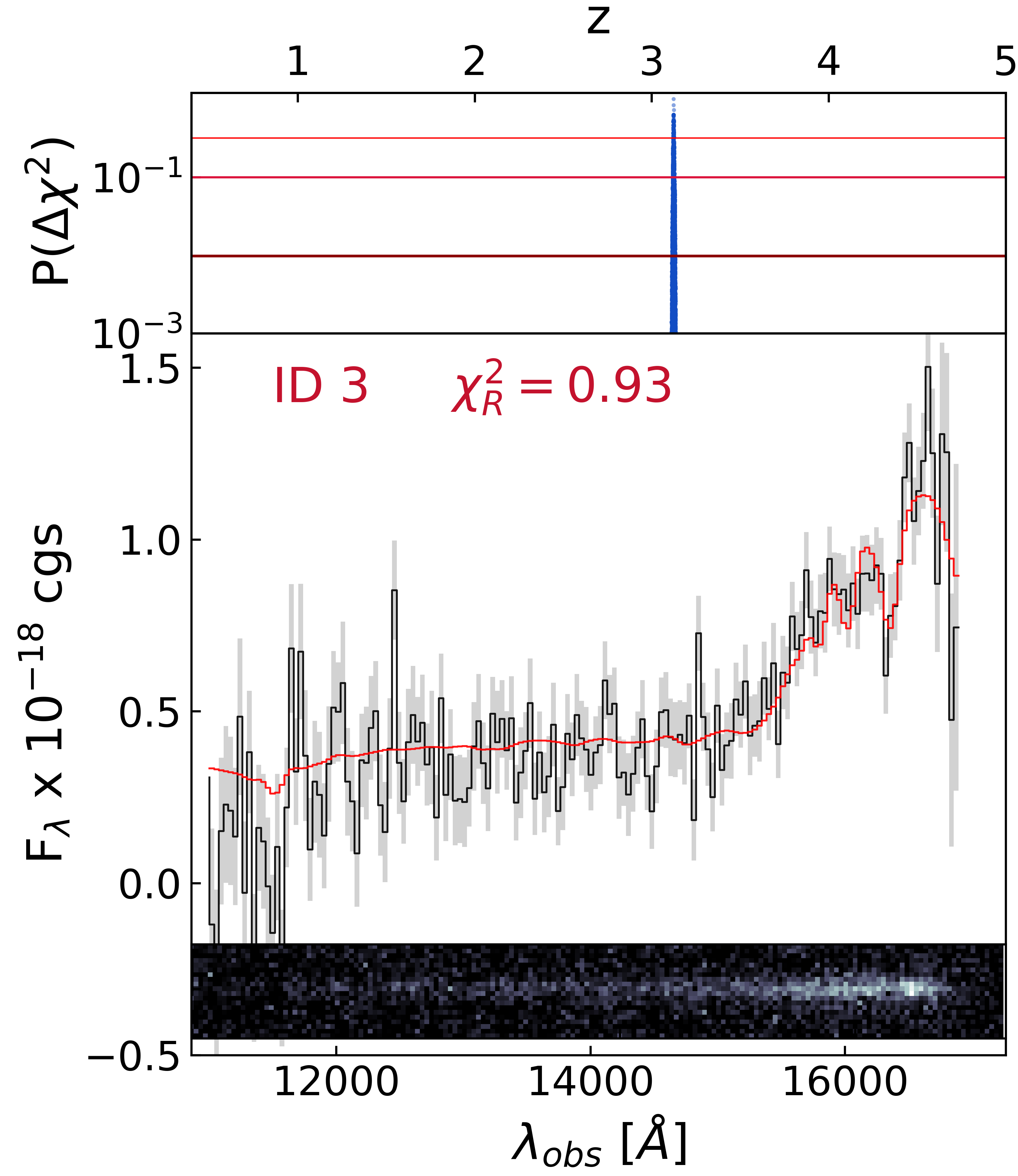}
\end{subfigure}%
\begin{subfigure}{.4\textwidth}
  \centering
  \includegraphics[width=0.9\linewidth]{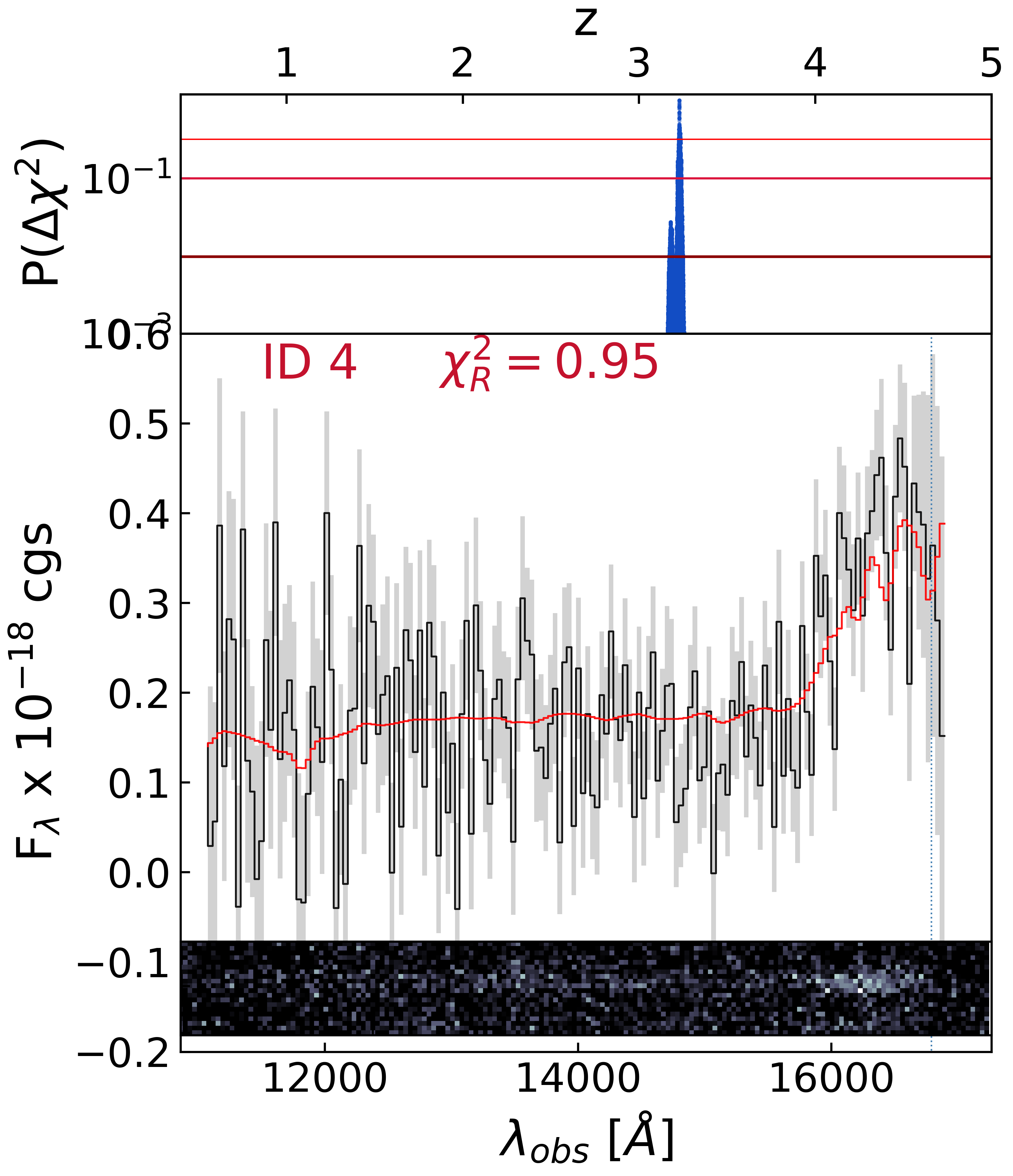}
\end{subfigure}

\caption{Continued}
\label{fig:test2}
\end{figure*}

\subsection{Redshift identification}
\label{sec:identspecz}
Seven of our sources showed unambiguous redshift solutions at a  3$\sigma$ confidence already during the low-resolution pass (dz=0.01). Two show a secondary peak within the same confidence level and one (ID 7) lacks a marked spectral break. Nine galaxies out of ten are consistent with single solutions at 1$\sigma$. Despite the low-resolution of the G141 spectroscopy, a redshift identification is made possible thanks to the presence of prominent spectral breaks and strong absorption lines for most of the targets (see Fig.~\ref{fig:test}). For the highest-redshift sources, namely ID 3, 4 and 8, MgI and MgII 2640 -- 2850 \AA\ absorption lines also enter the spectrograph, albeit with relatively low S/N.
Notably, ID 7 lacks absorption features that are as strong as for the remaining sample. This makes the spectrum formally consistent with multiple redshift solutions ($0.4<z<3.5$) within 1$\sigma$ confidence. We verified the consistency of the lowest redshift solutions with the available COSMOS2015 photometry placing z>1.5 as a lower-limit, confirmed by its SED which rises in flux from J to H band. The final spectroscopic confirmation of this source was obtained performing the combined fit of its photometry and grism spectrum across the $1<z<3.5$ redshift range at high-resolution (dz=0.001). This test confirmed the previously derived best-fit solution and reduced the error bars on z$_{\rm{spec}}$ at all confidence levels. At 1$\sigma$ it yields $z=2.674^{+0.005}_{-0.009}$, whereas at 3$\sigma$ it yields $z=2.674^{+0.021}_{-0.026}$ showing the stability of the best-fit solution compared to the information derived from spectroscopy alone (see Fig.~\ref{fig:test}). In the remaining analysis we will refer to the former confidence level, as done for the rest of the sources.\\

ID 10 shows excess flux at 11870 \AA\ . The secondary peak in the redshift probability distribution is placed at z=2.3, where the fitting routine attempts to reproduce the two most prominent absorption lines with H$\delta$ and H$\gamma$ instead of H$\epsilon$ and H$\delta$ favoured by the best-fitting template. This redshift is not low enough to explain the excess flux with a [OII]$\lambda$3727 emission line. Such line would have to be placed at z=2.18, yet it would not match any of the absorption features present in the spectrum. We therefore interpret it as a spurious noise-driven feature.

\subsection{Performance of photometric redshifts}
\begin{figure}
  \resizebox{\hsize}{!}{\includegraphics{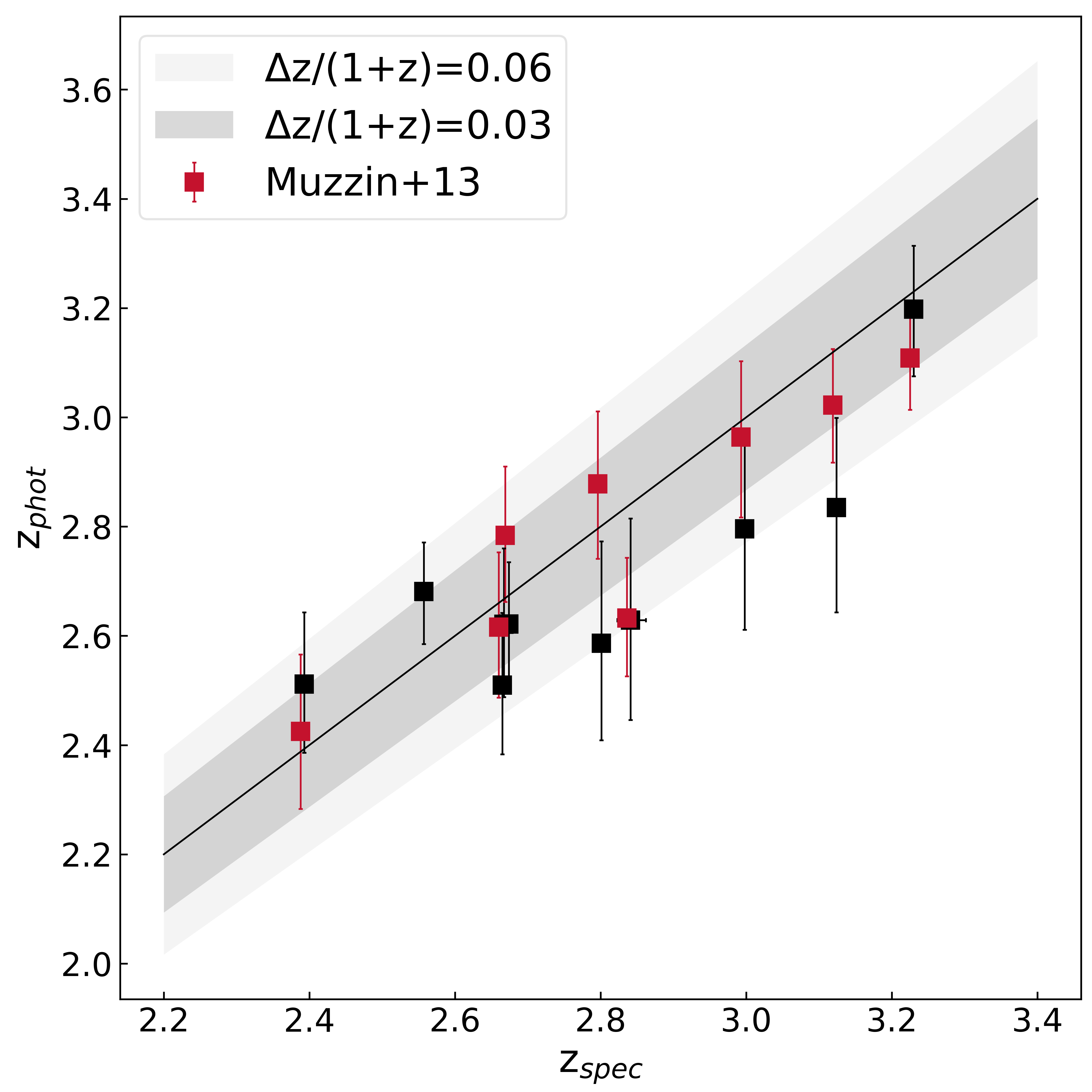}}
  \caption{Comparison of calibrated z$_{\rm{phots}}$ and spectroscopic redshifts. Black squares mark the original calibrated z$_{\rm{phots}}$ used for the selection. Red squares mark z$_{\rm{phots}}$ from \citet{Muzzin13}. The solid black line and relative shaded area mark the 1:1 relation and the nominal dispersions between the original calibrated z$_{\rm{phots}}$ and the newly derived z$_{\rm{spec}}$ (see text). Redshifts from \citet{Muzzin13} have been shifted horizontally by -0.005 for clarity.}
  \label{fig:zphot}
\end{figure}

We compared the resulting z$_{\rm{spec}}$ with the calibrated photometric redshifts used for the sample selection to assess the quality of the latter (Fig.~\ref{fig:zphot}).
The quoted normalized median absolute deviation of the adopted photo-zs\footnote{$\sigma_{\rm{NMAD}}=1.48 \cdot \rm{median}(|z_{\rm{phot}}-z_{\rm{spec}}|/(1+z_{\rm{spec}}))$, Hoaglin et al. 1983} was $\sigma_{\rm{NMAD}}=0.025$, estimated on the spectroscopically confirmed sample at z$\sim$1.5, reducing to 0.018 once galaxies with less reliable $z_{\rm{spec}}$ were excluded. As clarified in \citet{strazz15}, the accuracy is maximum for bright objects (such as those used for spectroscopic confirmation) and decreases for fainter ones (either less massive sources or at higher z). In order to take this into account, in Fig.~\ref{fig:zphot} we show the error bars computed using the $z_{\rm{phot}}$ accuracy estimated by the authors for faint objects, as a function of the K-band magnitude in \citet{mc10}. For objects between K=[20.8, 22] - such as in our case - the expected uncertainty is $\sim$0.040(1+z). The  $z_{\rm{phot}}$ used for the selection of our sample largely agree with the $z_{\rm{spec}}$ derived here, with a small systematic underestimation of 0.04\%. We included the performance of photometric redshifts from \citet{Muzzin13} as well. ID 1 and ID 2 are outside the UltraVISTA area and therefore not present in this latter catalog. 
The nominal $\sigma_{\rm{NMAD}}$ we derived here by comparison with our spectroscopic sample are 0.057 and 0.033 for the two catalogs, respectively. We ascribe this difference in redshift accuracy to the different depths and number of photometric bands of the two catalogs. As noted in \citet{strazz15}, already at z$\sim$1.5 the same photo-z calibration method yielded an accuracy similar to that independently provided by \citet{Muzzin13} ($\Delta z/(1+z)=0.015$) when the photometry of \citet{Muzzin13} was used instead of that of \citet{mc10}.


\section{Combining spectroscopy and NUV-NIR photometry}
\label{sec:combo}

\begin{figure*}
\centering
\includegraphics[width=\textwidth]{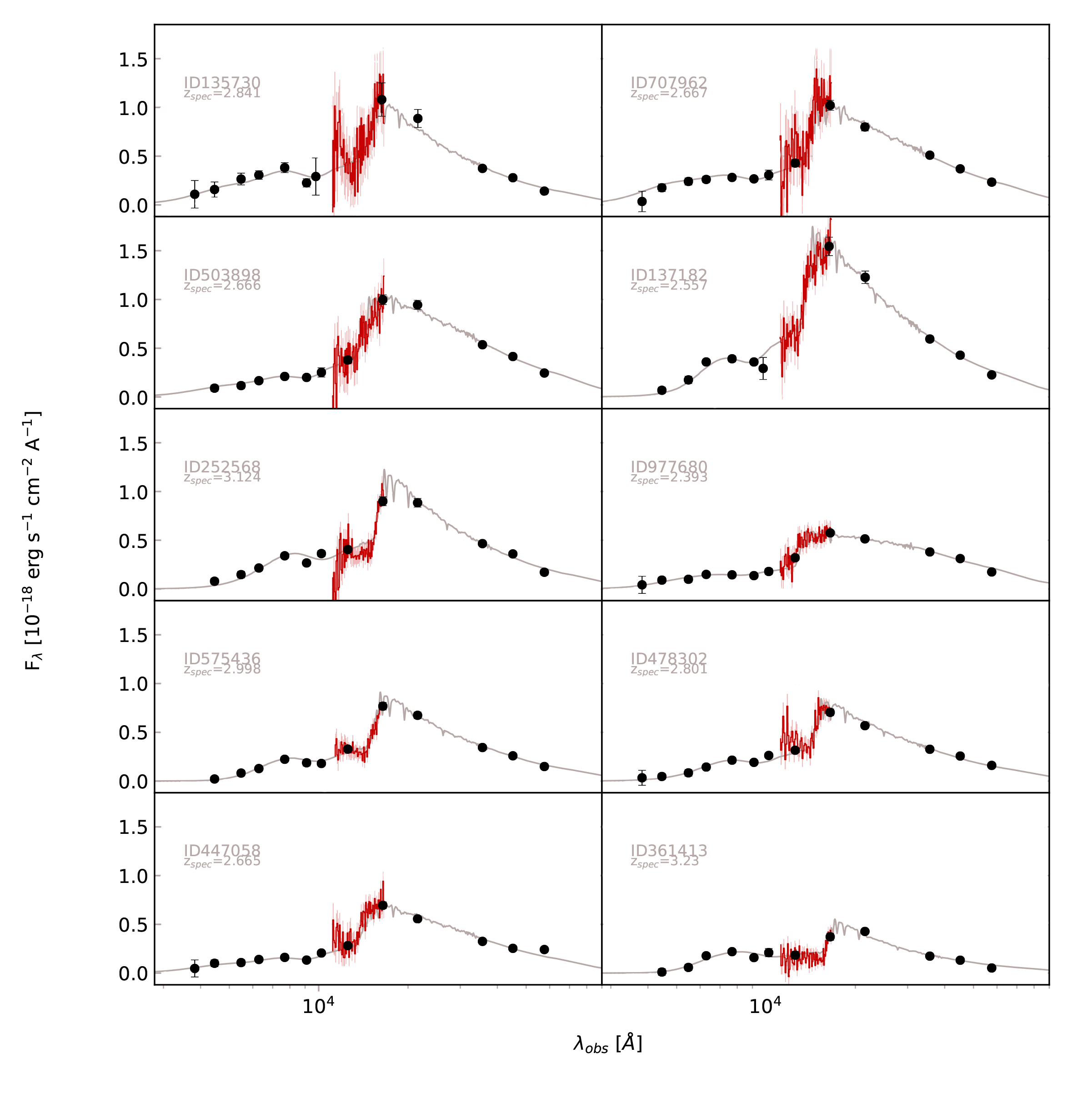}
\caption{Photometry from the catalog of \citet{Laigle16} (black dots). Grey curves show the best-fitting templates smoothed to the G141 resolution. The observed frame grism spectra are superimposed, rebinned for clarity.}
\label{fig:SEDs}
\end{figure*}

In order to characterize the physical properties of the targets, the HST grism spectroscopy was combined with COSMOS2015 broad-band photometry from CFHT/u$^{*}$ to IRAC/5.8$\mu$m\footnote{IRAC/8$\mu$m was excluded from the fit due to higher chances of AGN contamination.} \citep{Laigle16}. A lower limit of 0.05 mag was used for the photometric errors in all the bands. The two data sets were fit separately and were later combined by adding the $\chi^2$ matrices of the two fits. This procedure selects the solution that best-fits both data sets, minimizing the effect of residual mismatches in normalisation between the spectrum and the photometric SED. The same range of model parameters that was used for the spectral fitting is used in Fig.~\ref{fig:SEDs} except for the redshift grid, which was fixed to the 1$\sigma$ range around the best-fitting z$_{\rm{spec}}$. At this stage, the metallicity was fixed to the solar value according to the normalisation and scatter of the local ETG mass-metallicity relation \citep{Thomas10}. Recent clues from HST/G102 grism spectra of $1<z<1.8$ QGs as part of the CLEAR survey seem to support the idea that massive quenched galaxies were enriched to approximately solar metallicity already at $z\sim3$ \citep{EstradaCarp19}. As we describe in Sect. 5.3, the results obtained were tested against the choice of template metallicities.\\ 

\subsection{Photometric zero-point calibrations}
\label{sec:zps}
\begin{figure}[h!]
  \resizebox{\hsize}{!}{\includegraphics{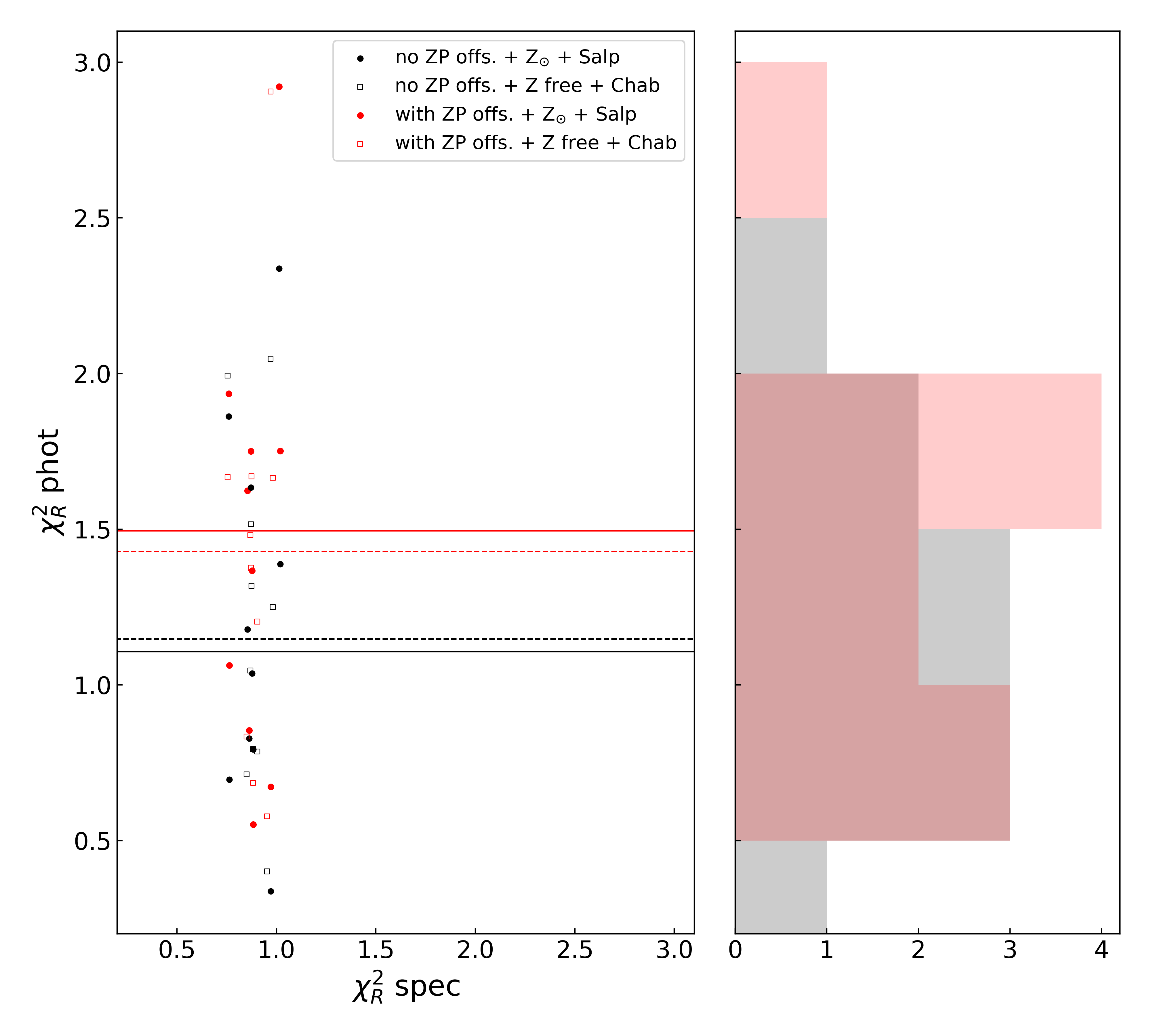}}
  \caption{Distributions of photometric reduced $\chi^2$ before (red dots) and after removing the ZPs recalibration (black dots). Median values are marked by red and black solid lines, respectively. $\chi^{2}_{R}$ values obtained adopting a Chabrier IMF and leaving the metallicity free to vary are shown as empty squares. Their corresponding medians are shown as dashed lines.}
  \label{fig:chired_zp}
\end{figure}

When comparing the mass-weighted ages and dust extinction values obtained from the modeling of the grism data alone with the grism data combined with broad-band photometry (see Sect.~\ref{sec:combo}), we find them being inconsistent at more than 3$\sigma$ in most cases.
Our SED fits to total fluxes resulted in relatively high reduced $\chi^2$ ($\chi^2_R$), as shown in Fig.~\ref{fig:chired_zp}.
Specifically, the probability associated with the total $\chi^2$ is $\sim$0.5\%, given the total degrees of freedom of the photometric fit of the whole sample.
Changing the IMF or leaving the metallicity of templates free did not improve the $\chi^2_R$ distribution.

\begin{figure*}
\centering
\includegraphics[width=\textwidth]{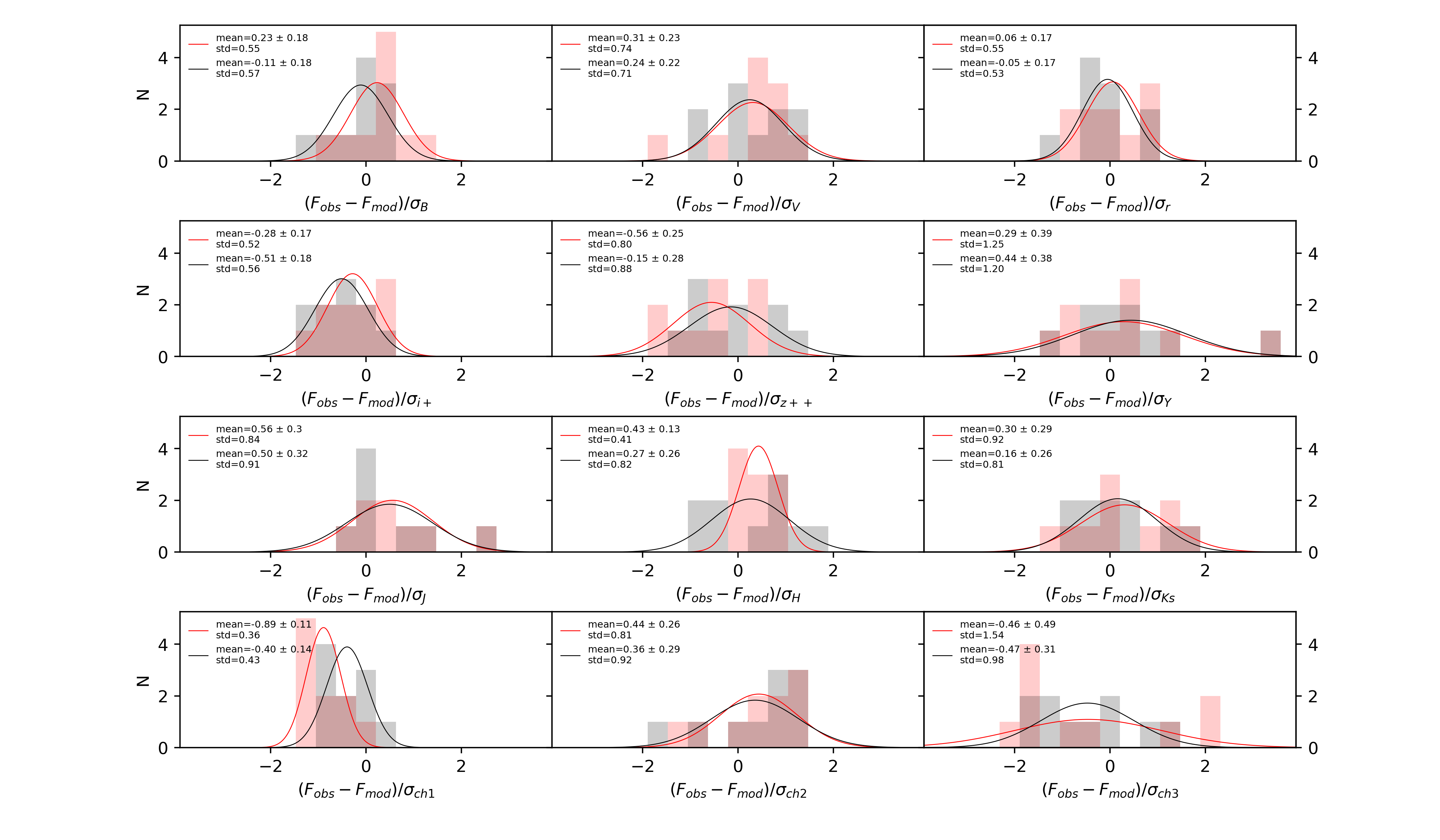}
\caption{Normalized residuals distribution of target galaxies in each photometric band. Red and black histograms show residuals adopting recalibrated and original photometric zero-points, respectively. Solid red and black curves show Gaussian fits to such distributions. Their means, standard mean errors and standard deviations are listed in each panel.}
\label{fig:no_zp_distrib}
\end{figure*}

As shown in Fig.\ref{fig:no_zp_distrib}, the means of the distributions of the normalized residuals in each band appear to suffer from systematic shifts, namely $B$, $V$, $i^+$, $z^{++}$ (marginally), $J$, $H$, IRAC/3.6$\mu$m, IRAC/4.5$\mu$m.
As noted in \citet{Capozzi16}, the procedure of recalibrating photometric zero-points (ZPs) to optimize photometric redshift retrieval (commonly performed when building photometric catalogs) can impact the results of SED fitting. These tweaks can introduce systematics in several bands, since the recalibrations are often based on specific samples at a specific redshift. In the case of COSMOS2015 these were tailored on various samples of spectroscopically confirmed QGs at z<2.5 \citep{Ono12, Krogager14, Stockmann20}, among a much larger number of star-forming galaxies. In addition to this, the templates used to derive these adjustments can also have a role in introducing systematics.
In particular, those used in COSMOS2015 are: two BC03 templates with an exponentially declining SFH with a timescale $\tau$ = 0.3 Gyr and extinction-free templates as in \citet{Ilbert13}; a set of 31 templates including spiral and elliptical galaxies from Polletta et al. (2007) and 12 BC03 templates of young blue star-forming galaxies.

\begin{figure}
  \resizebox{\hsize}{!}{\includegraphics{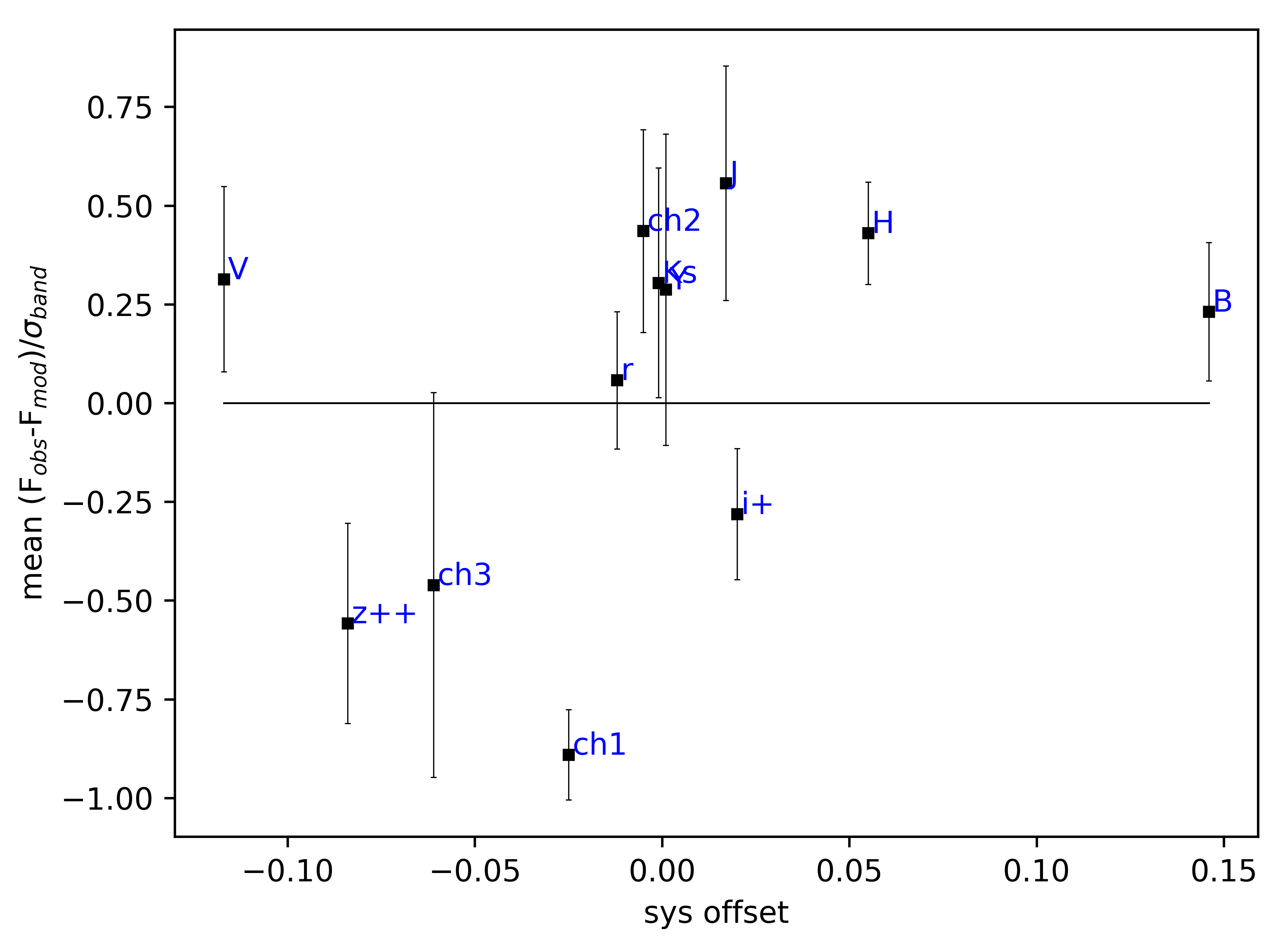}}
  \caption{Mean residuals of photometric bands as a function of the systematic offset to be applied to correct photometric ZPs. Error bars mark the standard error on the mean shown in Fig.~\ref{fig:no_zp_distrib}.} 
  \label{fig:offsets}
\end{figure}

The offsets that we find in our data suggest that model fluxes tend to overestimate observed fluxes when COSMOS2015 ZP corrections subtract flux from the observed signal and vice versa (see Fig.~\ref{fig:offsets}). Dropping ZP corrections reduces most of the wavelength-dependent systematics, producing a better agreement between models and the originally observed fluxes. Figs.~\ref{fig:no_zp_distrib} and ~\ref{fig:chired_zp}, show that dropping ZP offsets has the largest effect compared to changing grid parameters at reducing the median of the $\chi^{2}_{R}$ distribution. 
With this choice the probability associated with the resulting total $\chi^2$ is 9\%.
A high $\chi^{2}_{R}$ could be also flagging systematically low photometric errors. A common practice in this case is to rescale photometric errors to reach a $\chi^{2}_{R}\sim1$. However, given the behaviour highlighted in Fig.~\ref{fig:no_zp_distrib} and Fig.~\ref{fig:chired_zp} such rescaling appears unnecessary. In fact, without using rescaled ZP offsets the median $\chi^{2}_{R}$ is very close to 1.
Lastly, a high $\chi^{2}_{R}$ could be caused by broad residual distributions produced, for example, by the presence of some outliers in the sample. Visual inspections of imaging cutouts relative to each object in each band did not reveal peculiarities. 
We thus conclude that use of the ZP recalibrations derived in \citet{Laigle16} increase the inconsistencies between our spectroscopic and photometric data.
For these reasons, the rest of the analysis was performed on the original COSMOS2015 photometry, namely without making use of any ZP correction.

\subsection{Dust attenuation laws}
\label{sec:dustlaws}
\begin{figure}
  \resizebox{\hsize}{!}{\includegraphics{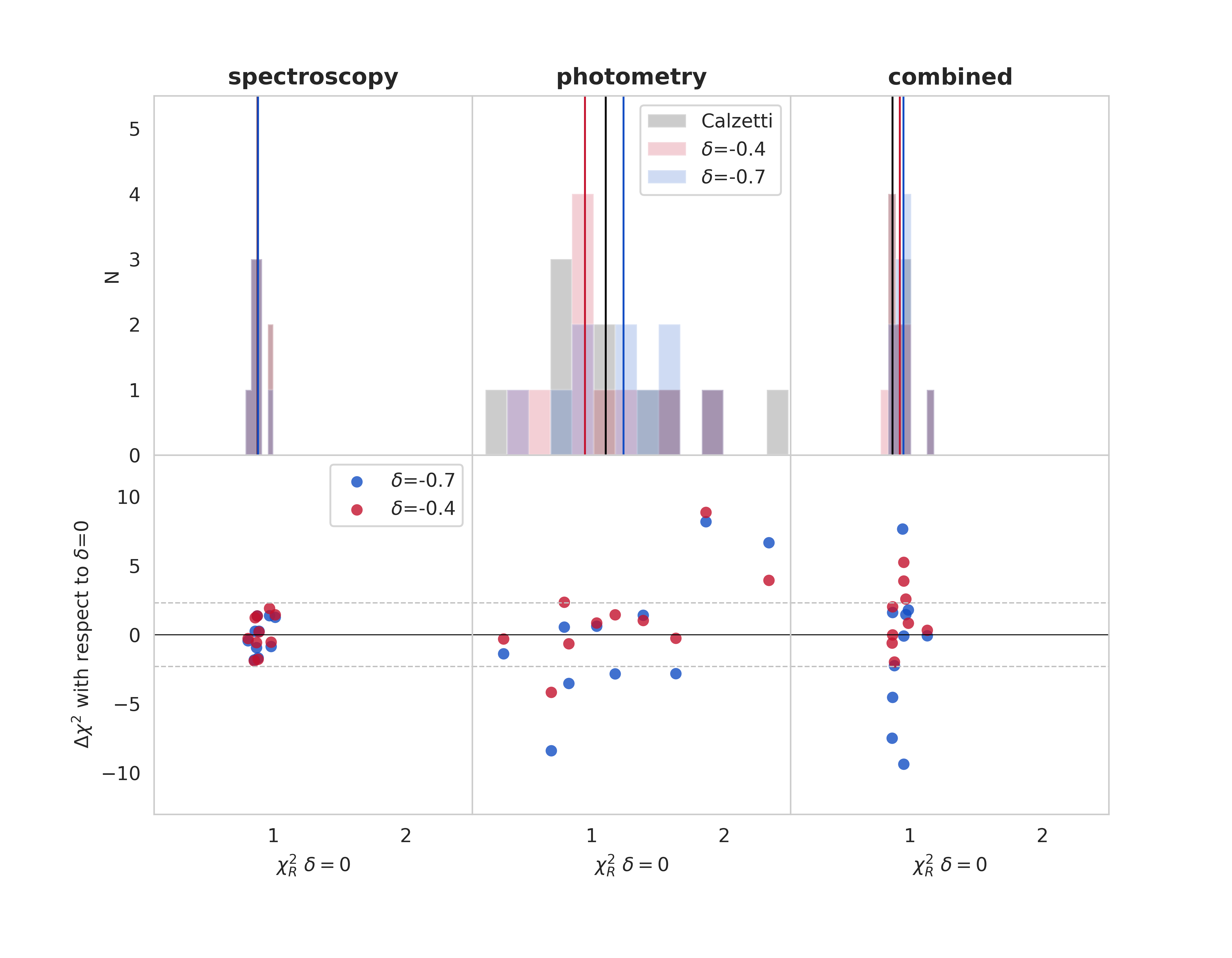}}
  \caption{Top: Reduced $\chi^2$ distributions of the fits to the spectroscopy (left), photometry (middle) and the combined data sets (right panel) as a function of the extinction law. Vertical lines mark the medians of the distributions. Bottom: $\chi^2$ difference of the best-fitting solutions obtained using $\delta=-0.4$ and $\delta=-0.7$ with respect to $\delta=0$. Grey dashed lines mark the levels of $\pm \Delta \chi^2=2.3$.}
\label{fig:deltachilaws}
\end{figure}

After the impact of photometric recalibrations was reduced, we explored whether any inconsistency in terms of $\Delta \chi^{2}$ between the best-fit spectroscopic solution and the combined solution could be ascribed, for example, to an unsuitable attenuation law or to a SFH that was too smooth to simultaneously reproduce the NUV emission from young stellar populations and the NIR emission arising from the bulk of the stellar mass. The use of attenuation laws steeper than Calzetti has been suggested to depend on the SSFRs of galaxies \citep{KriekConr13, Salim18}. The slope of the curve is generally dependent on the grain size distribution and geometry, with steeper curves associated with differential attenuation according to the age of stellar populations. To test the role of dust attenuation recipes, we included in the fitting library alternative attenuation curves in addition to \citet{Calzetti2000}.
We adopted the method proposed in \citet{Salim18}, following from \citet{Noll09}, in which the Calzetti curve is multiplied by a power-law term with exponent $\delta$ which sets the slope of the curve itself. Negative $\delta$ values imply a steeper slope in A$_{\lambda}$/A$_{\rm{V}}$ than in the Calzetti law at rest frame $\lambda < 5500 \AA$. In this formalism, the Calzetti law has $\delta$=0 by definition, whereas $\delta$=-0.4 is similar to the SMC curve. The curve is further modified by introducing the UV bump \citep{fm86}, modeled as a Drude profile D$_{\lambda}$, with fixed central wavelength $\lambda_0$=2187\AA\, , FWHM=274 $\AA$ from the best-fit results in \citet{Noll09} in GMASS galaxies at 1.5<z<2.5 for which the bump was clearly spectroscopically detected. The amplitude of the bump was linked to $\delta$ according to the linear relation found by \citet{KriekConr13} (see their eq. 4 and 5 solving for W$_{H_{\alpha}}$). Hence the law was constrained mainly by one parameter, its slope, which we let vary between 0, -0.4 and -0.7. This latter value was introduced to test the expected behaviour of local quiescent galaxies as shown in \citet{Salim18}. 
In Fig.~\ref{fig:deltachilaws} we show the distribution of the $\chi^2_R$ of spectroscopic, photometric and combined fits (upper panels), as well as the $\chi^2$ difference between their respective best-fit solutions obtained with $\delta=0$ and the best-fit solutions using $\delta$=-0.4 and -0.7 (e.g. $\chi^2_{\rm{min}}(\delta=0) - \chi^2_{\rm{min}}(\delta=-0.4) $, lower panels). The curve with $\delta=-0.4$ appears to reduce the dispersion of the $\chi^2_R$  distribution for the combined fits, as well as showing systematically a smaller $\chi^2$ with respect to $\delta=0$. When testing the overall goodness of fit in terms of the slope, $\delta=-0.4$ was the best-fit solution preferred by the majority of the targets Fig.~\ref{fig:deltachilaws}. Five galaxies out of nine tend to reject $\delta=-0.7$ at a 5\% level (but not at 1\%). However, given the overall similar $\chi^2_R$ distributions of the combined fits it was not possible to reject any of the adopted curves consistently for the entire sample. The probabilities associated with the median combined $\chi^2_R$ are 89, 76 and 67\% for $\delta=0$, $\delta=-0.4$ and $\delta=-0.7$ respectively. Therefore, for the remaining analysis, we let $\delta$ vary, marginalising over it when deriving physical parameters.

\section{Quiescence of individual targets} 
\label{sec:quiescencetest}

We hereby tested the quiescence of our galaxies based on the emission from their stellar component. First we tested the information that can be extracted solely from the grism spectra, then we compared with the results obtained by adding the available broad band NUV-NIR photometry. We further extended the analysis to longer wavelengths to probe possible obscured star formation.

\subsection{Results from the stellar component}

\begin{figure}
  \resizebox{\hsize}{!}{\includegraphics{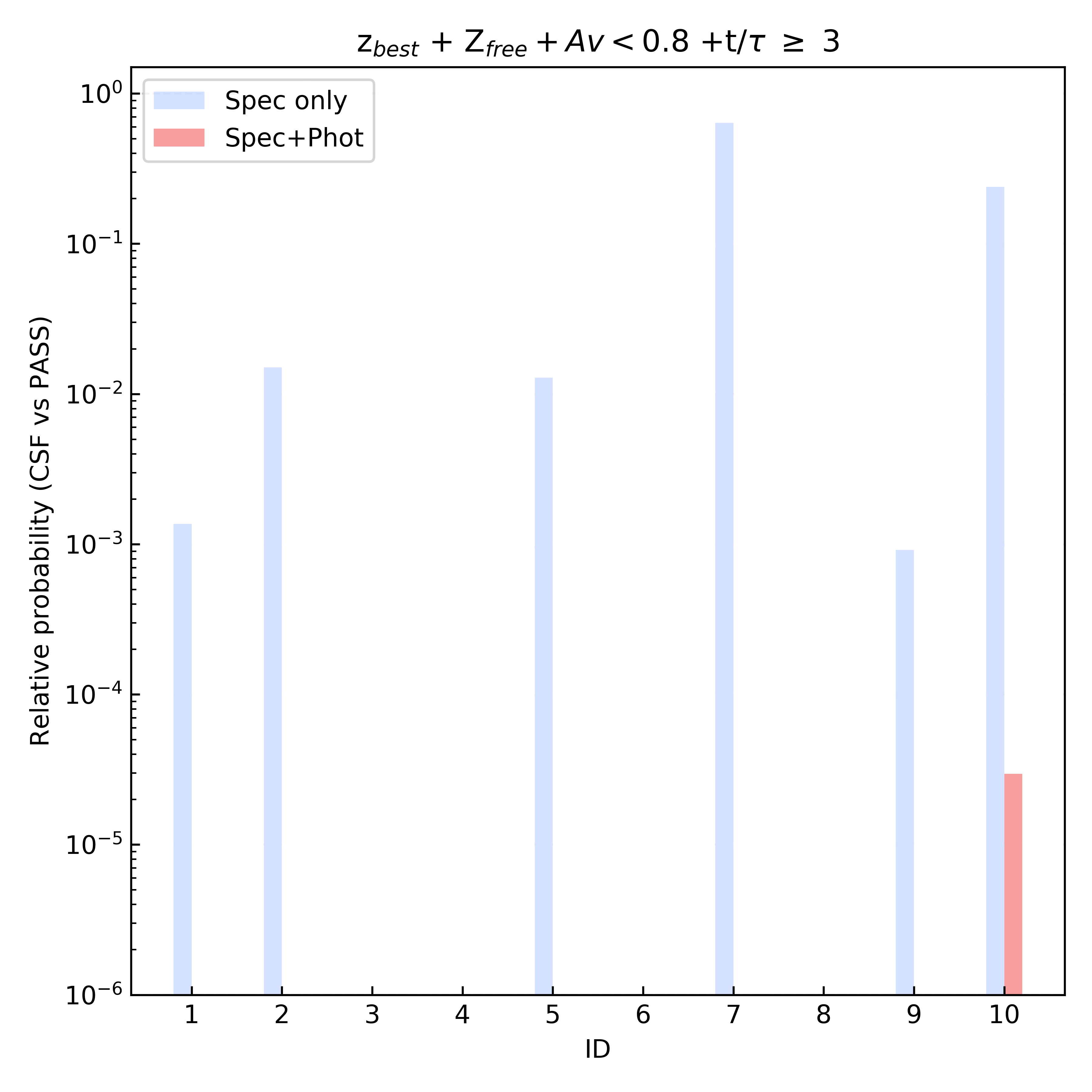}}
  \caption{Probability of constant star-forming solutions relative to that of passive solutions. Blue bars mark the results from spectroscopy alone. Red bars show the probabilities from the combined fit. Bars with relative probabilities lower than 10$^{-6}$ are not reported.}
  \label{fig:qtest}
\end{figure}

We tested the quiescent/dusty star-forming nature of each galaxy by comparing the goodness-of-fit of the best-fitting constant star-forming (CSF) template with a free dust extinction parameter, with that of a solution defined as passive by constraining the best-fitting SFH as follows: t$_{50}\geq$0.3 Gyr, A$_{V}<$0.8 mag and t/$\tau \geq$3 where t$_{50}$ is the lookback time at which the galaxy formed half of its stellar mass (our mass-weighted ages), t is the lookback time at the onset of star formation and $\tau$ is the timescale of the SFH. This ratio corresponds to a drop in SFR of about a factor of 20 with respect to the initial value for an exponentially declining SFH. To classify a galaxy as quiescent, the probability of a CSF solution relative to the passive solution has to show a probability of $<$0.01, as inferred from their $\chi^{2}$ difference. This simple parametrisation is able to discern to a zero-order level the consistency of both the spectrum and the photometry with a heavily dust-attenuated star-forming component, whether it fits better than the passive solution and the confidence level of its consistency.
The test was performed fixing the redshift to z$_{\rm{spec}}$. We first tested the spectra alone and then combined the photometric information by summing the $\chi^{2}$ matrices of the two fits. In Fig.~\ref{fig:qtest} we show the results of the test. We verified that letting z$_{\rm{spec}}$ vary within its 1$\sigma$ confidence range does not impact the probabilities significantly. Letting the metallicity of the templates vary (adopting 0.4, 1 and 2.5 Z$_{\odot}$) has a similarly negligible effect.
Once the redshift of the target can be constrained to a sufficient accuracy, the addition of the photometry, with its wide wavelength coverage and overall quality, can help rejecting a SFH in those cases where spectroscopy alone is not able to robustly distinguish between the two.
The test on ID 7 was performed letting z$_{\rm{spec}}$ vary within its 3$\sigma$ confidence level of the combined fit used for its spectroscopic confirmation (see Sect.~\ref{sec:identspecz}). Given the lack of prominent (emission or absorption) lines, the spectrum alone is not able to reject star-forming solutions. The combined fit, however, rejects such solutions (P=0, $\Delta \chi^{2}$=183, $\chi^{2}_{\rm{R}, SF}$=2.0, $\chi^{2}_{\rm{R}, PASS}$=0.9), even when adopting tighter constraints on the passive solution (e.g. t/$\tau$>10 and t>0.5).
The second object showing the highest consistency with a star-forming template is ID 10. Despite the addition of photometry rejecting star-forming solutions, it should be noted that this object lies in a region of the UVJ diagram where contamination is expected to be more frequent  \citep[see][]{lustig}. Moreover, its best-fitting combined mass-weighted age and Av are still pointing towards a very young and dust-reddened stellar content ($t_{50}=0.3^{+0.3}_{-0.1}$ and $Av=1.6^{+0.1}_{-0.3}$, respectively.)  
We discuss very young sources with MIPS and Chandra detections in the following sections. We anticipate here that we consider ID 10 in particular an AGN host and likely an object close to its quenching phase.
Considering the arguments presented above, we conclude with reasonable confidence that all of our galaxies can be classified as passive on the basis of their UV-to-NIR emission.

\subsection{SFR constraints from multiwavelength data sets}

The stellar component of our galaxies suggests that they are inconsistent with being highly attenuated star-forming galaxies. SFRs estimates based on near-UV/optical tracers can nonetheless be underestimated in the presence of complex dust geometries \citep{PoggiantiWu00}. This is particularly compelling in our case since our galaxies appear to be very recently quenched. We explore here what are the realistic constraints on the possible presence of obscured star formation or AGN activity by taking advantage of the recently released IR to (sub)millimeter super-deblended catalog of \citet{Jin18} (hereafter J18) and of the Chandra COSMOS-Legacy Survey catalog of \citet{Civano16}. First, we convert the flux densities of our 24$\mu$m detections into the SFRs expected from z$\sim$3 similarly massive MS galaxies and compare them to the available FIR constraints. Afterwards, we convert the very same flux densities into hard X-ray luminosities to verify whether they are consistent with being AGN-powered. Finally, we derive individual 3$\sigma$ upper limits on the obscured SFR from VLA 3GHz flux densities.\\
The prior-extraction method used in J18 fits the PSF of MIPS 24 $\mu$m, VLA 1.4 and 3GHz images \citep{smolcic17} at the positions of known K$_{S}$-selected (plus radio 3GHz-selected) sources. This procedure improves faint-source identification with respect to blindly extracted catalogs such as in \citet{Lefloch09} by significantly reducing flux errors (by roughly a factor of two) while also improving source deblending. This allows us to investigate in more detail individual mid- or far-IR detections that could have been missed by previous catalogs or judged of low significance. In particular, we recall here that our sample selection allowed for objects with Spitzer/MIPS 24$\mu$m detections (S/N$\geq$4) from the \citet{Lefloch09} catalog, or of higher S/N in case of SEDs with no acceptable star-forming solutions suggestive of an AGN-driven MIR flux.
We refer the interested reader to Appendix A where we report the multi-wavelength cutouts as well as the available reliable detections and upper limits for our sources. SED fitting attempts were performed only for galaxies with available priors for source deblending. We caution that the best-fitting AGN components are only intended to display the maximum AGN contribution allowed by the 24$\mu$m detections.

\subsubsection{IR}
\label{sec:FIR}
The super-deblended catalog marks five of our galaxies as detections in the MIPS 24$\mu$m band (ID 1, 5, 6, 7 and 10, see Table~\ref{tab:final}). None of them is significantly detected in the FIR. Specifically, all of them have a S/N$_{\rm{FIR+mm}}$<5, where S/N$_{\rm{FIR+mm}}$ is the square root of the quadratic sum of the S/N computed in each band from 100$\mu$m to 1.2mm.
ID 1 formally counts multiple detections in Spitzer/MIPS and \textit{Herschel}/PACS bands but it is flagged as unreliable since it lies in a region of COSMOS affected by incomplete prior coverage and underestimated flux uncertainties. Visual inspection of its MIR and FIR cutouts did not reveal any detection at the source position (see Appendix A). We thus consider the IR detections of ID 1 as unreliable and exclude it from the following test.\\
At the mean redshift of this sample, observed frame 24$\mu$m emission corresponds approximately to 6$\mu$m rest frame. Emission at these wavelengths can arise from a range of processes: star formation (policyclic aromatic hydrocarbons (PAH) emission lines and/or warm dust continuum), a dusty torus obscuring a central AGN, warm diffuse cirrus clouds heated by old stellar populations or circumstellar dust around asymptotic giant branch (AGB) stars \citep{DL07, bethermin15, fumagalli14}. The $\sim$6" FWHM of Spitzer/MIPS PSF is larger than the typical optical projected size of our galaxies (<1" at 5000 \AA\, rest frame) and prevents us from distinguishing whether the emission is extended or centrally concentrated.\\
We computed the individual SFRs expected from the remaining four 24$\mu$m detections under the hypothesis that their emission is powered by star formation at the MS level emitting at the observed flux density. We corrected the flux densities by a factor of 1.7, as recommended in J18 and adopted conversions from \citet{magdis12}. These conversions were driven from template SEDs of MS galaxies whose variation as a function of redshift is mainly driven by the strength of the mean radiation field <U>, which maps the sSFR evolution. Such templates assume a fraction of dust mass into PAH of $q_{\rm{PAH}}=2.5$\% for z>1.5 MS galaxies.\\  
The results can be found in Table~\ref{tab:final} where we also report individual 3$\sigma$ upper limits for undetected sources.
The 24$\mu$m derived SFRs range between $\approx$300 and 900 M$_{\odot}$ yr$^{-1}$. This is the same order of magnitude of z$\sim$3 log(M$_{\star}$/M$_{\odot}$)$\sim$11 MS galaxies albeit somewhat higher, inconsistent with FIR non-detections. These MS galaxies, in fact, typically show FIR \textit{Herschel} flux densities of about a few to 10 mJy \citep{Schreiber2015, Liu18, Jin18} which would be detected in \textit{Herschel}/SPIRE.
ID 10, 6 and 5 appear to tentatively show SPIRE/250- and 350 $\mu$m signal at the source position as revealed by visual inspection. For the latter two, the contamination by nearby projected FIR bright sources due to poor spatial resolution is evident. For ID 10 the J18 catalog formally provides non-detections at 100 and 160$\mu$m and no deblending at longer wavelengths. In postage stamps, the SPIRE/250 $\mu$m signal appears to show emission centered on the source position in the middle between two other 24$\mu$m bright sources. J18 attributed the  250- and 350$\mu$m signal to the severe blending of these two sources within the SPIRE large PSF, larger than the distance of these sources from our target ($\approx$10").
Concerning the sources that remain individually undetected at 24$\mu$m, the 3$\sigma$ upper limits are too shallow to reject milder (but substantial) SFRs. Stacking the rest of the sample at 24$\mu$m results in 0.036 $\pm$ 0.018 mJy which translates into a shallow upper limit of <200 M$_{\odot}$ yr$^{-1}$. We caution that the available $3\sigma$ depth of the super-deblended data from Spitzer/MIPS, \textit{Herschel}/SPIRE, \textit{Herschel}/PACS in COSMOS is not sufficient to securely reject sub-MS levels of obscured star formation on a galaxy-by-galaxy basis.

\subsubsection{X-rays}
\label{sec:Xrays}

ID 10, ID 6 and ID 7 have counterparts in the hard X-ray domain with rest frame luminosities of the order of log(L$_{X, 2-10keV}$[erg s$^{-1}$]) $\sim$ 43.7, 44.3 and 44.3 respectively, assuming a photon index $\Gamma=1.4$ \citep[e.g.][]{Gilli07}.
We tested whether their 24$\mu$m emission is consistent with being powered by an accreting black hole by converting the observed frame 24$\mu$m flux densities into unobscured rest frame X-ray (2-10keV) luminosities, adopting the relation of \citet{fiore09} (see their eq. 1). This relation assumes that the 2–10 keV luminosity, computed directly from the observed fluxes without any correction for intervening absorption, can be considered representative of the intrinsic X-ray luminosity. The relation has a scatter of 0.2 dex and outliers in the case of significant X-ray absorption. The expected L(2-10keV)$_{24\mu m}$ for these three sources are in agreement with the observed ones within the uncertainties (see Table~\ref{tab:final}). Therefore, although we cannot reject the scenario in which some level of star formation would contribute to the rest frame $6\mu$m emission, we conclude that our data are fully explained by an AGN obscured by a dusty absorber.\\
Lastly, ID 5 shows a 0.1 mJy 24$\mu$m detection at 10$\sigma$ significance which is not matched by an X-ray detection. Its spectrum and photometry are both pointing towards a passive nature, therefore we tend to favour the hypothesis for which strong obscuration might be playing a role in hiding X-ray photons from the central engine.\\

\subsubsection{Radio}
\label{sec: radio}
We derived individual 3$\sigma$ upper limits on the obscured SFR from the super-deblended VLA 3GHz flux densities using the FIR-radio correlation of \citet{Delvecchio2020} assuming that radio emission is given by star formation alone. One galaxy is detected at 3GHz at 19$\sigma$ (ID 1). The unphysically high SFR estimated for it ($\sim10^{4}$ M$_{\odot}$ yr$^{-1}$, see Table~\ref{tab:final}), implies that the origin of its radio emission is to be ascribed to AGN radio jets. Otherwise, the inferred upper limits result in $<$120-190 M$_{\odot}$yr$^{-1}$, which does not conclusively rule out sub-MS levels of star formation on a galaxy-by-galaxy basis. Finally, as derived in \citet{Deuge20}, the peak flux density of S$_{3GHz}$=2.72$\pm$0.93 $\mu$Jy obtained by mean-stacking 3GHz-undetected sources, translates into an upper limit on the global obscured SFR of $\sim$50--60 M$_{\odot}$ yr$^{-1}$, a level of star formation a factor of 5--6 lower than the coeval MS.\\

In summary, the average obscured SFR of our sample has been constrained to be below $\sim$50--60 M$_{\odot}$ yr$^{-1}$ by a mean-stack detection at 3 GHz. However, individual 3 GHz radio upper limits to the obscured SFR are $<$120-190 M$_{\odot}$yr$^{-1}$, therefore not sufficient to reject, on a galaxy-by-galaxy basis, star formation at 1$\sigma$ below the estimated value for the MS at z$\sim$2.8 corresponding to our stellar masses. Our sample contains four secure MIPS 24$\mu$m detections (f$_{24\mu m}\sim$ 0.1--0.2mJy) with no FIR counterparts given also the shallow upper limits at these redshifts. Three of these detections are consistent with being AGN-powered judging from their luminous X-ray counterparts. The combined passive stellar emission for the remaining MIPS source suggests that this galaxy is likely an obscured AGN host. The lack of individual strong constraints on the residual obscured SFR at these redshifts, combined with very young mass-weighted ages and high dust extinction values for some of our targets, renders the classification on a galaxy-by-galaxy basis somewhat ambiguous. We argue that extended spectral coverage (e.g. covering H$\alpha$ and [NII]) could be of help on this matter and that conclusive evidence on the nature of these high-z QGs can only be obtained with targeted deep mm observations.

\section{Age determination}
\label{sec:age}

\subsection{Spectral fit}
\begin{figure*}[h!]
 \centering
 \includegraphics[width=\textwidth]{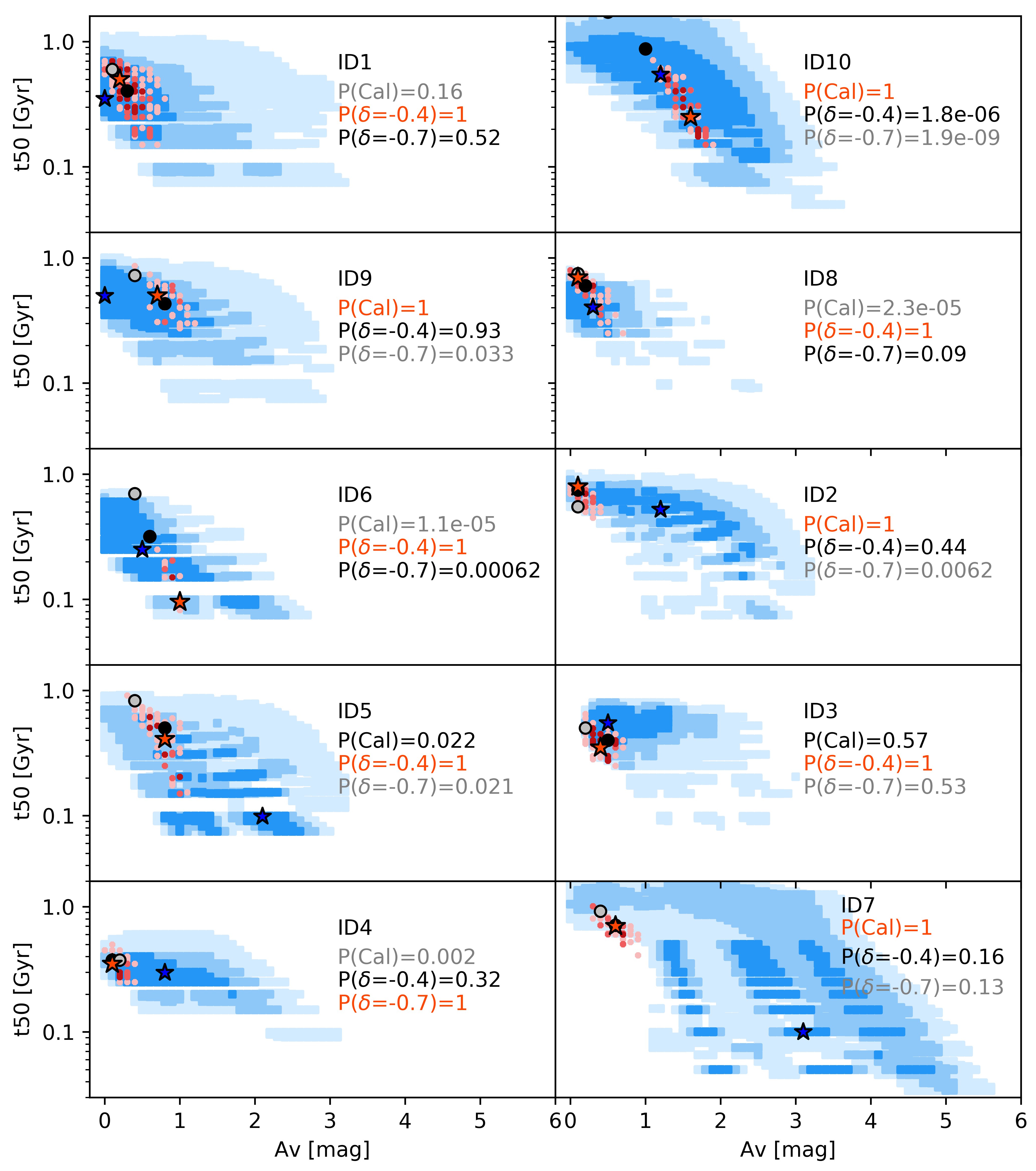}
 \caption{Mass-weighted ages and dust extinction values for our targets. Blue squares mark the confidence regions (3 to 1 $\sigma$ going from light to dark points) extracted from the spectra alone. Red points show the solutions of the combined fit. Blue and red stars in each panel mark the best-fit solution from the spectroscopic fit and the combined fit respectively. The relative probabilities of the extinction laws} are shown for each galaxy. As a reference, we report the best-fitting combined solution at fixed extinction law as black and grey dots, their color coding follows that of the aforementioned relative probabilities.
 \label{fig:t50Av}
 \end{figure*}

Recent works suggest that, when it comes to estimating age and optical dust extinction, relatively simple parametrisations of the SFH perform similarly as more flexible ones and, all in all, behave in a relatively stable way at high redshift \citep{Belli19, Valentino20}. We conservatively marginalized over the different SFHs adopted here to render the best-fitting values and their uncertainties more robust. 
Fig.\ref{fig:t50Av} shows the resulting t$_{50}$ and A$_{\rm{V}}$ extracted from the grism spectra (blue points) compared to those derived including COSMOS2015 photometry (red points). Light to dark shading marks 3,2 and 1$\sigma$ level confidence values respectively, obtained following \citet{Avni76} with two interesting parameters. The degeneracy between t$_{50}$ and A$_{\rm{V}}$ appears to be strongly mitigated by the addition of the photometry once the redshift is constrained with sufficient accuracy to the spectroscopic value. The bulk of our targets are consistent with having formed half of their stellar mass relatively recently, systematically having t50 below 1 Gyr. In some cases, such as ID 6 and 10, the best-fitting combination suggests dust enshrouded young stellar populations. Intriguingly these two galaxies are also detected in X-rays and 24 $\mu$m, as discussed in Sects. 6.1 and 6.2, and might be galaxies which just entered into their quiescent phase or with residual SF below the levels probed our FIR data.

\subsection{Relative strength of spectral breaks}

\begin{figure*}[h!]
 \centering
\includegraphics[width=0.9\textwidth]{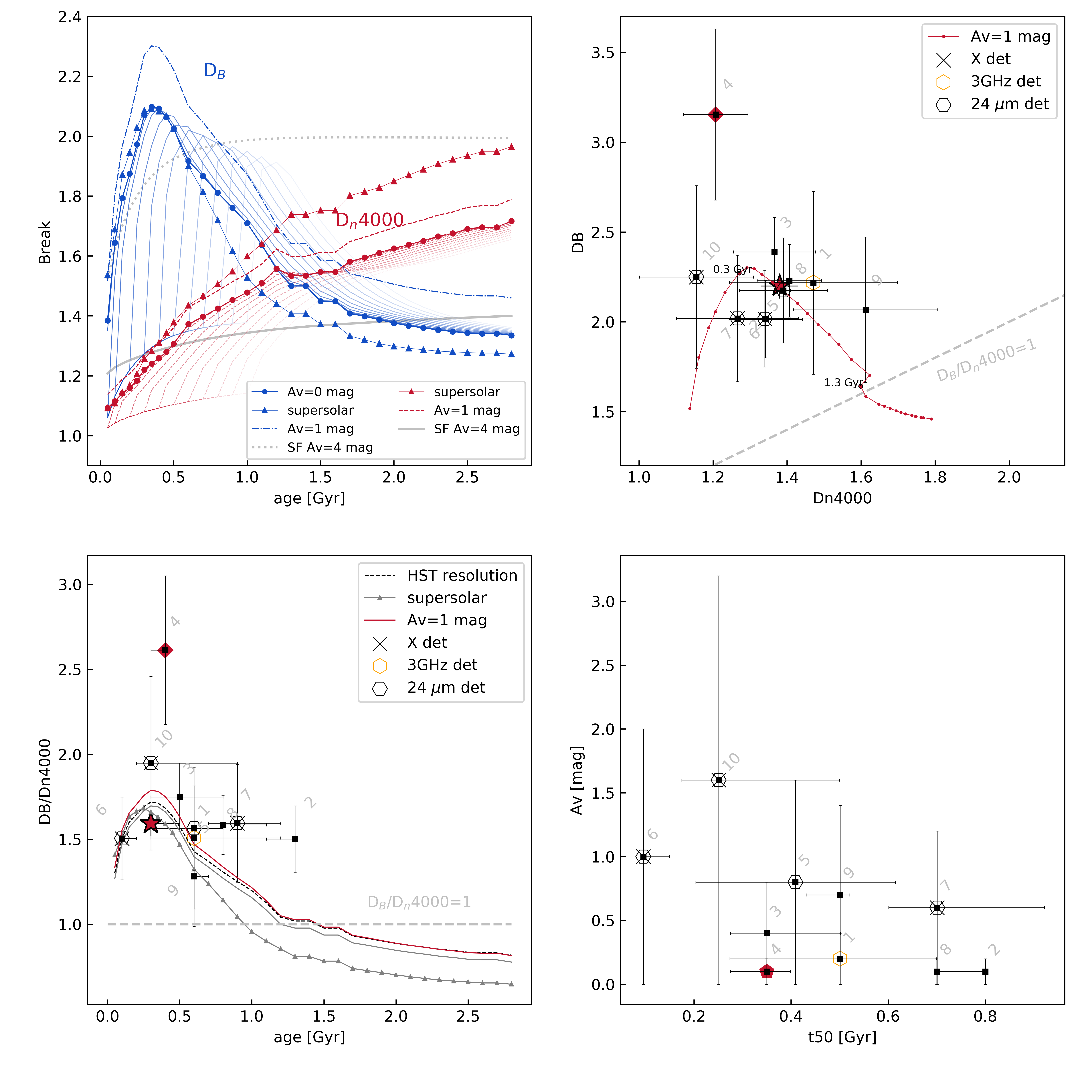}
\caption{\textit{Upper left}: Variation of the Balmer and 4000 \AA\ breaks for CSPs as a function of their age using BC03 templates at solar metallicity. Lighter curves mark the evolution for a truncated SFH with increasing duration of star formation. The effect of adopting templates of different metallicity is displayed for the values that produce the largest variation, i.e. 2.5Z$_{\odot}$ (triangles). Dotted and solid grey curves show the behaviour of a constant SFH for the two breaks respectively.
\textit{Upper right:} Individual values of D$_{B}$ and D$_{n}$4000 for our targets. The red track shows their evolution with age and is attenuated by 1 mag. The dashed grey line marks the 1:1 relation. \textit{Lower left:} Variation of the index ratio as a function of age for an SSP-like template. The effect of smoothing templates to the HST resolution is shown by a dashed black curve. The effect of a A$_{\rm{V}}$=1 mag attenuation is shown instead by the red curve. The full transition between a Balmer-dominated and a 4000\AA\, -dominated spectrum is flagged when D$_{B}$/D$_{n}$4000=1 (dashed grey line), which occurs around 1.3 Gyr of passive evolution. Dark grey triangles mark the evolution for 2.5Z$_{\odot}$ templates. \textit{Lower right:} Best-fit values for the dust attenuation and mass-weighted age from the combined fit. Chandra X-ray, VLA 3GHz and Spitzer/MIPS 24 $\mu$m detections are marked by black crosses, yellow and black hexagons respectively.}
\label{fig:PSBratio}
\end{figure*}

The choice of SFH to infer evolutionary parameters intrinsically carries a degeneracy with the functional form adopted. A more direct approach is to quantify the light-weighted contribution of recent star formation by measuring the relative contribution to the integrated stellar spectrum of short-lived massive stars with respect to long-lived lower-mass stars. Balmer absorption lines reach their maximum strength in A-type stars with a spectral break at 3646 \AA\,. Stars of lower mass and lower effective temperature produce metal absorption lines (CaII H \& K, Fe and Mg) which result in a sharp spectral break at 4000 \AA. Moreover, the underlying continuum changes shape with time, progressively losing emission in the NUV/blue spectral range while flattening in the NIR. The different evolutionary rates of the stars producing the lines and their fractional contribution to the optical light at fixed mass, make it possible to trace the evolutionary stage of a galaxy. In Fig~\ref{fig:PSBratio} (upper panels) we show the evolution of the spectral break measured through the D$_{B}$ definition \citep{Kriek06} and the D$_{n}$4000 definition \citep{Balogh99} respectively, as well as the relative strength (the ratio) between the two (lower left panel). Lighter shaded curves show the variation with increasing duration of star formation. The ratio is shown as a function of age of composite templates built with a short truncated SFH. The ratio is only mildly dependent on dust reddening because the two indices share a similar wavelength range. Additionally, the two indices are fairly robust against low-resolution. The ratio varies strongly during the first 1 Gyr or so, reaching its maximum around 0.3-0.5 Gyr. Eventually, it drops below 1 when the light-weighted contribution from A-type stars fades away. Constant star formation results instead in a ratio of $\sim$1.1 rather constant with time. Varying the metallicity of the input templates has the effect of anticipating the transition to D$_{B}$/D$_{n}$4000$>$1. This effect is strongest when supersolar metallicity (2.5Z$_{\odot}$) is adopted. In this case the transition is reached at 0.9 Gyr. We suggest that this ratio could be used to spot post-starburst galaxies with high dust attenuation along and across the UVJ diagram when high-resolution spectra are unavailable.\\
In Fig.~\ref{fig:PSBratio} (upper right) we show the two indices computed on our targets. The dashed grey line highlights the transition where the post-starburst ratio equals 1. The mean error in each side band was divided by the square root of the number of pixels within it. For ID 4, whose rest frame spectrum does not cover the entire wavelength range required to compute the D$_{n}$4000 red sideband, the average flux density was taken as the mean of the best-fitting template in the same range. The error was computed as the mean of the noise spectrum taken on the last 10 spectral bins. We flagged this galaxy with a red diamond. The red star marks the values obtained on the average spectrum in \citet{Deuge20}. Despite the large uncertainties driven by the S/N of our spectra, the indices all lie well above the 1:1 relation, thus marking the presence of young stellar populations in all of our targets. This supports the results of the spectral modeling, highlighting that some of the most massive high-z QGs appear to be only recently quenched \citep{Stockmann20, Valentino20, Forrest2020b}. An overview of the physical parameters derived for our target galaxies can be found in Table \ref{tab:final}.

\section{Tracing AGN activity}
\label{sec:agn}
Here we investigate the incidence and strength of BH activity on newly quiescent galaxies, in the framework of SMBH-galaxy coevolution. In particular we assessed the level of mechanical feedback on our galaxies by studying the rest frame 1.4 GHz luminosity averaged over the entire sample; and the incidence of radiatively efficient accretion by constraining the fraction of X-ray detected galaxies and their rest frame hard X-ray luminosity respectively.
\subsection{Radio}
\begin{figure}[h!]
\resizebox{\hsize}{!}{\includegraphics{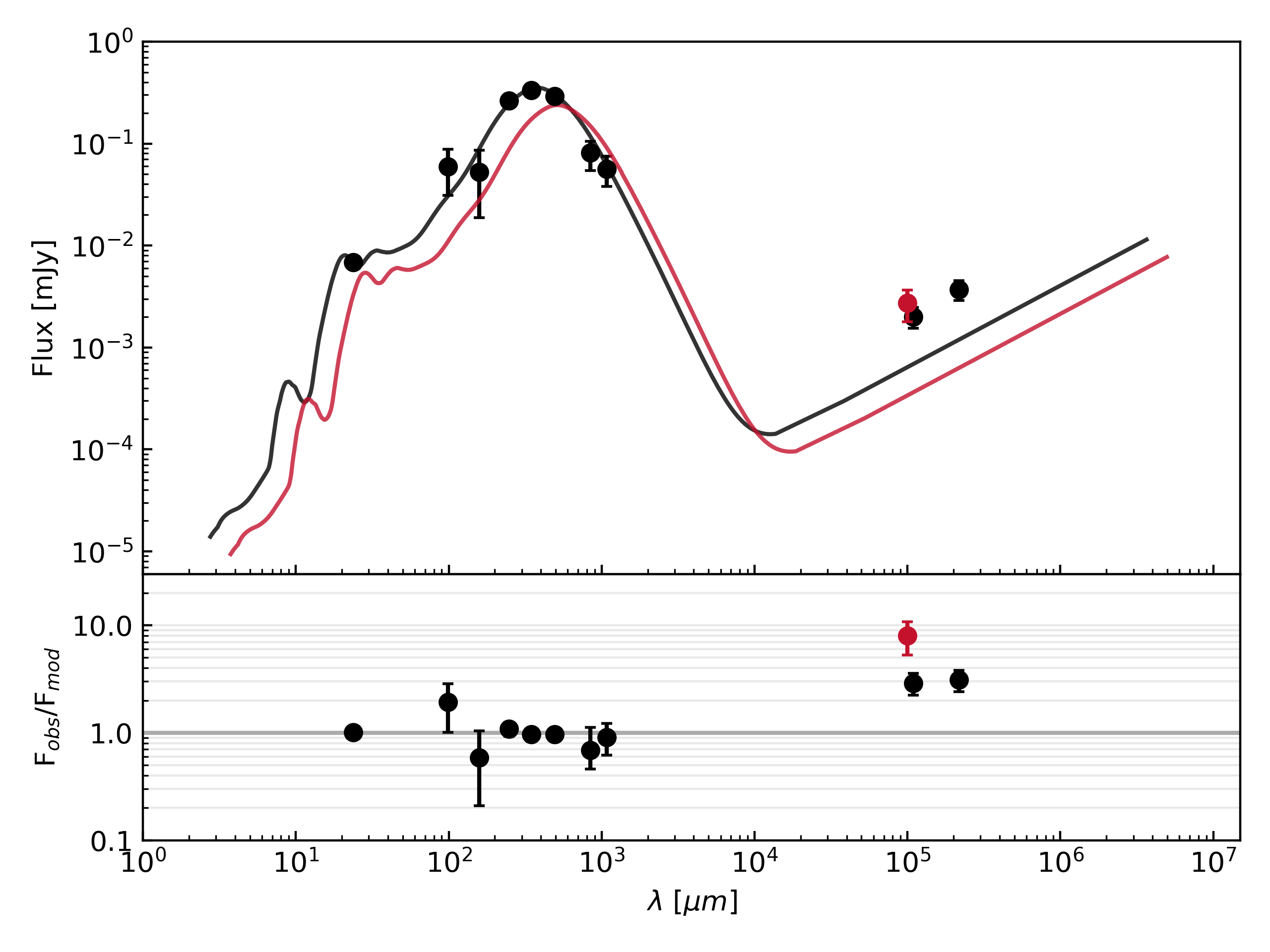}}
\caption{Top panel: Observed 3GHz flux for our sample (red dot) compared to the observed FIR SED for $<M_{\star}>\sim1.1 \cdot 10^{11} \,\, M_{\odot}$ galaxies at z$\sim1.8$ in \citet{Gobat18} (black dots). The best-fitting template of \citet{Gobat18} is shown as a black curve. The same template but rescaled to our redshift and stellar mass is shown as a red curve. Lower panel: Observed flux normalized to the respective model at the corresponding wavelength.}
\label{fig:excess}
\end{figure}

As mentioned, one galaxy (ID 1) is securely detected at 3GHz. Given the unphysically high SFR estimated for it ($\sim10^{4}$ M$_{\sun}$ yr$^{-1}$), we ascribe the origin of its radio emission to AGN radio jets.
Stacking the radio-undetected sources, \citet{Deuge20} obtained a peak flux density of S$_{3GHz}$=2.72$\pm$0.93 $\mu$Jy. This translates into a K-corrected rest frame luminosity of L(1.4 GHz)=$(2.0 \pm0.7) \cdot 10^{23}$ W/Hz, which we interpret in this section as arising from low-luminosity AGN activity. 
Fig.~\ref{fig:excess} shows the observed 3GHz flux density of our targets, compared to the observed stacked SED of z$\sim$1.8 analogues presented in \citet{Gobat18} (hereafter G18). Under the assumption that z$\sim$2.8 quiescent galaxies share this FIR-to-radio SED similar to that of z$\sim$1.8 analogues, our observed flux density appears to be a factor of 2.7 higher than the best-fit model in G18 rescaled to our redshift and average SFR. The 1.4 GHz luminosity expected from residual star formation according to the FIR-radio correlation is $L_{mod}=1.64\cdot10^{22}$ W/Hz. Our observed L(1.4 GHz) is 12 times higher than $L_{mod}$, implying an excess of $1.9\cdot10^{23}$ W/Hz. This excess, in turn, is a factor of 3.8 higher than the excess found in z$\sim$1.8 similarly massive QGs ($5\cdot10^{22}$ W/Hz, G18).

The low statistics implied by our sample size prevents us from making meaningful considerations on the overall duty cycle of AGN activity. It is worth mentioning, however, that the 0.66 duty cycle estimated in G18 from z=1.4 to z=2.5 implied a burst duration of 1.2 Gyr which is 1.6 times larger than our observational window (the cosmic time spanned by our sample is 0.72 Gyr). Together with the ensemble radio detection, this might imply that we are sampling an epoch when low-level AGN activity is almost always on in newly quiescent galaxies, with a stronger radio AGN activity with respect to what inferred for lower-z massive analogues.
\subsection{X-ray}
\begin{figure}[h]
\resizebox{\hsize}{!}{\includegraphics{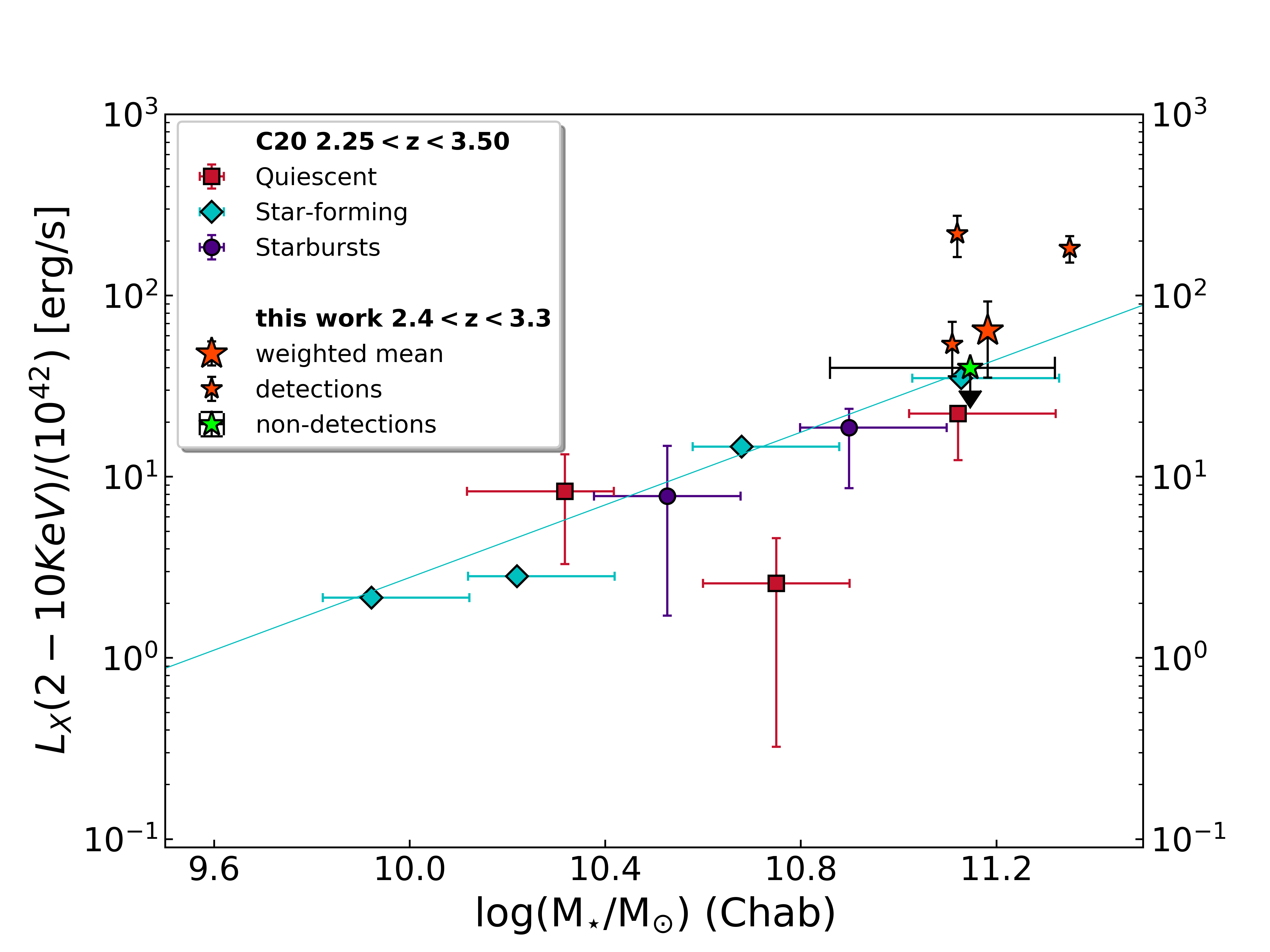}}
\caption{X-ray luminosity in the 2-10 keV band as a function of stellar mass for quiescent (red squares), star-forming (cyan diamonds) and starburst galaxies (violet circles) at $2.25<z<3.50$. Adapted from Carraro et al. 2020.}
\label{fig:Lx}
\end{figure}

L$_{X}$ can be viewed as a tracer of the typical rate of black hole growth in a given galaxy sample. A recent stacked analysis of quiescent galaxies in COSMOS has constrained the average level of rest frame hard X-ray emission to be $L_{X}=2\cdot10^{43}$ erg s$^{-1}$ \citep[hereafter C20]{Carraro20}\footnote{Our stellar masses were converted to a Chabrier IMF for consistency}. While our non-detections are consistent with C20, our mean rest frame L(2-10)keV obtained as $$ L(2-10)keV=\frac{L_{\rm{X, uplim}}\times N_{\rm{nondet}}+ \Sigma_{i=1}^{N_{\rm{det}}} L_{\rm{X, i}}}{N_{\rm{nondet}}+N_{\rm{det}}} $$ is higher by a factor of 2--3 at face value (see Fig.~\ref{fig:Lx}). The error bar on the average is computed as the error on the weighted mean.

\begin{figure}[h]
\resizebox{\hsize}{!}{\includegraphics{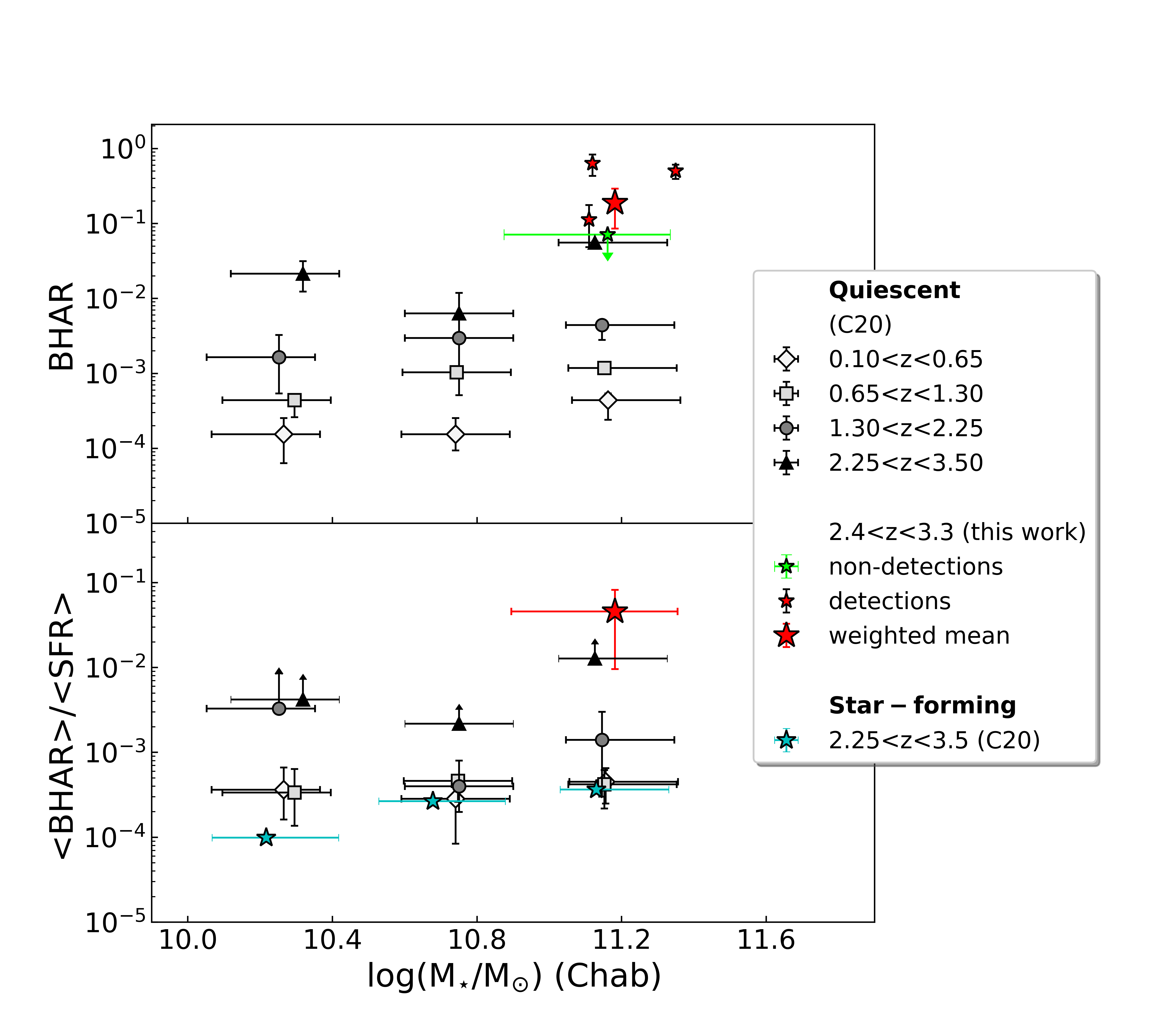}}
\caption{BHAR (top) and BHAR per unit star formation rate (bottom) as a function of stellar mass for quiescent galaxies (greyscale points) in COSMOS. Different symbols mark different redshift bins. Main sequence galaxies in the same redshift range as studied in this work are added in the bottom panel as cyan stars. Error bars on the average BHAR reflect the dispersion of the weighted mean on the rest frame L(2-10keV) of the sample. The average SFR$_{\rm{[OII]}}$ was converted to a Chabrier IMF. Adapted from \citet{Carraro20}.}
\label{fig:BHAR}
\end{figure}

Assuming the \citet{Lusso12} bolometric corrections and a $M_{BH}=M_{\star}/500$ conversion as in \citet{HR04}, we computed Eddington ratios for each target, defined as the bolometric X-ray luminosity (or its 3$\sigma$ upper limit) divided by the Eddington luminosity expected at the stellar mass of the galaxy. We obtain Eddington ratios of $\lambda_{EDD} \sim 2-11$\% for detected sources and 3$\sigma$ upper limits lower than 1\% for all the undetected ones.
We repeated the test using 24$\mu$m derived L(2-10)keV, obtaining $\lambda_{EDD}$ values that were a factor of 2 higher on average.\\
This $\lambda_{EDD}$ translate into black hole accretion rates (BHAR) \footnote{BHAR(M$_{\star}$, z)=(1-$\epsilon$)$\cdot$ L$_{2-10keV}\cdot$ kbol(M$_{\star}$, z)/($\epsilon$ c$^{2}$)=$\lambda_{\rm{EDD}}\cdot L_{\rm{EDD}}\cdot 10^{-45.8}$ M$_{\sun}$ yr$^{-1}$, where the efficiency of mass conversion is $\epsilon$=0.1} which are largely in agreement with C20 at z$\sim$3. Dividing the mean <BHAR> by the average <SFR$_{\rm{[OII]}}$> estimated from the 5$\sigma$ [OII] detection from the average spectrum of 9 of our targets \citep[see][]{Deuge20}, we obtain an increase by a factor of $\sim$30 with respect to $1.3<z<2.25$ QGs at the same mass (see Fig.~\ref{fig:BHAR}), consistent with the lower limit for massive $2.3<z<3.5$ QGs inferred from the same paper. 
Our mean $L_{X}$ is marginally consistent with that star-forming galaxies in the same mass and redshift range, whereas the [OII]-derived dereddened SFR lies around $\sim$60 times below the MS. This translates into $<BHAR>/<SFR>$ being a factor of $\sim$60 higher than the high-mass end of the MS at z$\sim$3. We recall that this ratio would be even higher if even part of the oxygen ionisation were powered by AGN activity rather than actual star formation. This supports the idea that, while the stellar mass growth of the host galaxy has already ceased, the BH mass growth in high-z QGs takes longer to fade away, as already pointed out in C20, and might have had a role in quenching.\\

The stochastic nature of detectable AGN activity implies that it is usually only observed in a small fraction of galaxies at a given time. \citet{aird19} reported that high-z massive QGs exhibit enhanced AGN fractions compared to low-z SF galaxies hosting an equivalent SFR, suggesting that AGN activity in QGs might be fuelled and sustained by stellar mass loss rather than the availability of cold gas. Their fraction of highly accreting quiescent galaxies ($\lambda_{\rm{sBHAR}}>0.1$, i.e. at more than 10\% the Eddington limit) reaches 2-3\% around z$\sim$3. 
The fraction of normally accreting AGN ($\lambda_{\rm{sBHAR}}>0.01$) reaches 20-30\% in the quiescent population with SFRs of order of 0.5-1 M$_{\odot}$ yr$^{-1}$. 

\citet{Schreiber18} find 18\% of X-ray detections in 3.2<z<3.7 massive UVJ-quiescent galaxies, plus an additional 30\% in the young-quiescent (lower left) region of the UVJ diagram.
Several of their young-quiescent SEDs show similarities with our ID 5 and 6 in terms of SED shape and possibly young age. 
\citet{olsen13} reported a 19\%$\pm$9\% luminous AGN fraction in a mass-complete sample of massive UVJ-selected quiescent galaxies at $1.5<z<2.5$, with a total low-luminosity AGN fraction up to 70\%-100\%, advocating in favour of episodic AGN activity to maintain low SFRs in quiescent galaxies. 
Including a possibly Compton thick source, our sample likely contains a 40\% fraction of luminous AGN. We currently do not know in which direction the incompleteness due to the sample selection will affect the AGN fraction, since we are focusing on high-mass quiescent galaxies while excluding the strongest 24$\mu$m detections.

\begin{figure}[h!]
\resizebox{\hsize}{!}{\includegraphics{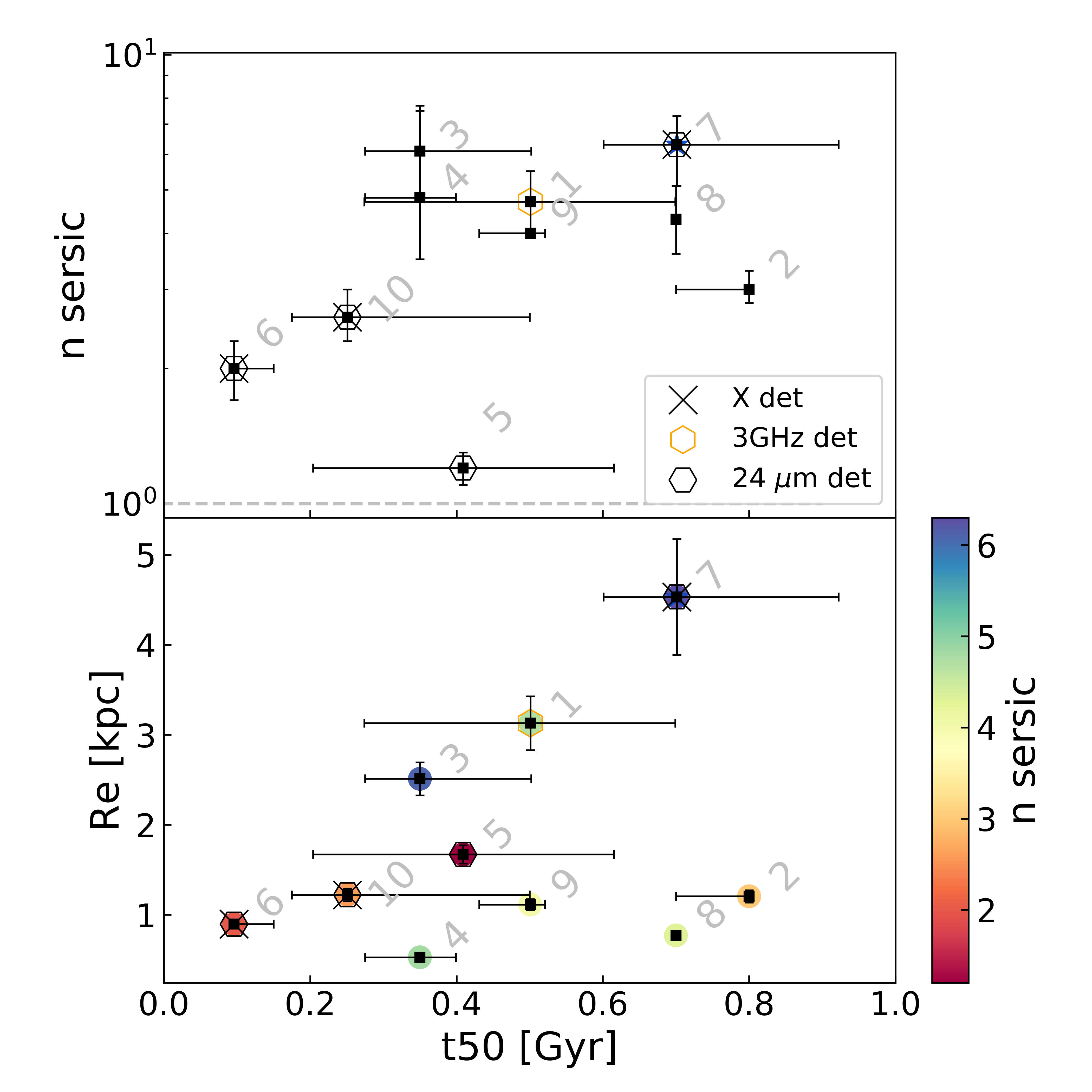}}
\caption{Comparison between our mass-weighted ages and the morphological parameters derived in \citet{lustig}, namely Sérsic index n and effective radius R$_{e}$ at 5000 \AA\ rest frame.}
\label{fig:morphology}
\end{figure}

As noted in \citet{aird19}, stellar mass-loss and AGN feedback tend to be disfavoured mechanisms for causing relatively high-accretion rates and high AGN fractions in sub-MS and quiescent galaxies. The former could sustain the accretion onto the central BH by providing a relatively stable supply of low-angular momentum gas but is expected to result in relatively low accretion rates. Moreover, stellar mass loss is most efficient soon after star formation (2-5 Myr) and declines exponentially afterwards, making its contribution likely not sufficient to explain the highest $\lambda_{EDD}$ measured for some of our objects (unless non-negligible SF is occurring). Instead, AGN feedback assumes that the gas supply, once used by the galaxy to sustain SF, is accreted by the central SMBH. However, even the youngest ages shown by our stellar populations would imply a stability of the radiatively efficient AGN feedback of order of hundreds of Myr, whereas radiatively efficient accretion is expected to be stable on timescales of 0.1 Myr. One other mechanism proposed by \citet{aird19} could be the build-up of a compact bulge, which would increase the stellar density of the host galaxy, hence increasing the rate at which the AGN is triggered by infalling gas, as also supported by observations \citep{Barro13}. Fig.~\ref{fig:morphology} shows that our likely AGN hosts are among the most compact ones in the sample. However, they also show the lowest Sérsic indices (n=1.2-2.6) implying that no clear-cut connection between radiative AGN feedback and central stellar density can be established with the present sample. Globally, the 30-40\% fraction of AGN that are tentatively associated with young objects (seen also young quiescent galaxies in \citet{Schreiber18}) might suggest that episodic AGN feedback might be triggered before a completely passive stellar core is settled.



\section{Discussion}
\label{sec:discussion}
In the present paper we have characterized one of the largest representative samples of 10 high-z massive QGs in terms of spectroscopic confirmation and age estimation. We showed that HST observations are able to probe the (H-band) brightest end of the massive QG population providing access to clear spectral breaks in the majority of the objects. These breaks can be used to quantitatively compare the amplitude of the Balmer absorption lines with respect to metal lines, hence quantifying the light-weighted contribution of young versus old stars at fixed stellar mass in early QGs. In other words, with relatively few HST orbits it is possible to both perform spectroscopic confirmation of QGs up to z$\sim$3.2 and start quantifying the incidence of newly quenched objects within the population. 
The particular configuration of spectral breaks characterising our sample (strong D$_{B}$ and weak D$_{n}$4000) is due to the presence of luminous A-type stars. These often flag a relatively recent shutdown of star formation, however, they have been also linked to galaxies hosting substantial amounts of obscured star formation contaminating the rest frame colors that are commonly used to select high-z quiescent galaxies \citep{PoggiantiWu00, lemaux}. The quiescence of our targets was first tested and confirmed through their combined spectral (rest frame NUV/optical) and photometric (rest frame UV-to-NIR) emission, rejecting the presence of catastrophic photometric errors or prominent emission lines. They are however consistent with very young mass-weighted ages, which makes their final interpretation less clear.  
Given the shallow upper limits on the obscured SFR placed by the current Spitzer/MIPS and Herschel/PACS and SPIRE data, we relied on the mean-stacked shallow detection at 3 GHz to constrain the potential obscured SFR to be on average below 50--60 $M_{\odot}$ yr$^{-1}$ \citep{Deuge20}, hence 5--6 times lower the coeval MS at most. 
The individual 24$\mu$m flux densities of three 5-10$\sigma$ detections in the J18 super-deblended catalog were converted into hard X-ray luminosities, compared to Chandra COSMOS Legacy X-ray detections, and judged to be consistent with being AGN-powered. Converting individual 3 GHz 3$\sigma$ upper limits into SFRs results in $<$120-190 M$_{\odot}$yr$^{-1}$, which are not conclusive to exclude substantial obscured star formation on a galaxy-by-galaxy basis. 
We also note that the origin of the mid-IR emission is in principle unclear since it can arise from multiple phenomena such as a dusty AGN torus, star formation, hot circumstellar dust around AGB stars and/or the presence of diffuse cirrus clouds heated by hot stellar populations \citep{fumagalli14} or a combination thereof.
From the broad agreement between the prominent Balmer breaks and the mass-weighted ages of our galaxies we conclude that they have quenched relatively recently prior to observation. However, dedicated mm observations are required for 40\% of our objects to conclusively assess the level of residual star formation. These objects might potentially represent examples of rapidly transitioning galaxies that underwent a sharp truncation of their star formation, possibly through AGN feedback.

\subsection{The emergence of massive quiescent galaxies at high-z}

\begin{figure}[h!]
\resizebox{\hsize}{!}{\includegraphics{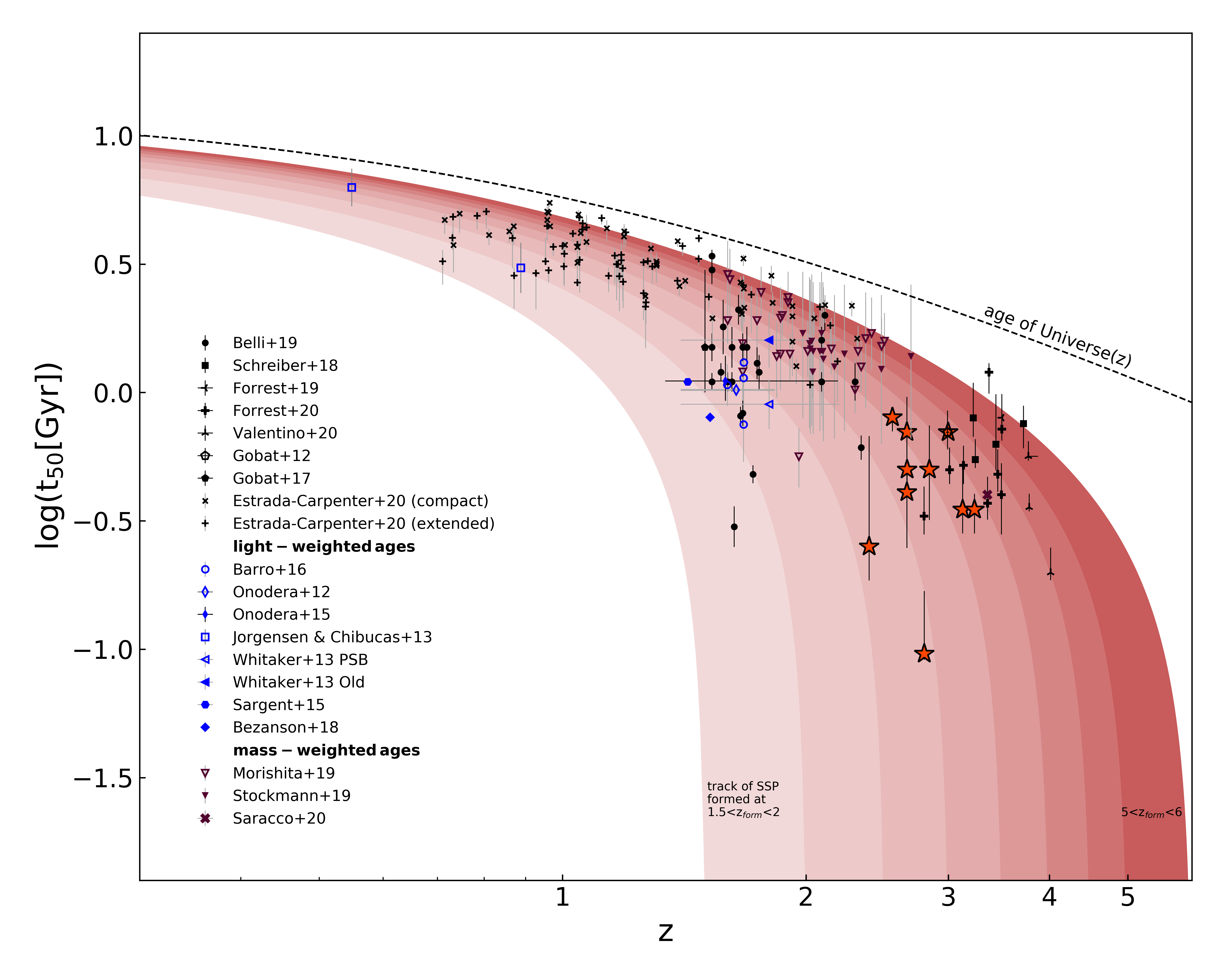}}
\caption{Stellar ages as a function of observed redshift of log(M$_{\star}$/M$_{\odot}$)$>$10.5 quiescent galaxies. Adapted from \citet{Onodera15}. Here mass-weighted ages from the literature differ from t$_{50}$ in that they do not specifically refer to the lookback time at which \textit{50\%} of M$_{\star}$ is formed. The four targets selected from \citet{Schreiber18} correspond to those recently followed up by \citet{Esdaile}. Solid grey lines show, from thin to thick, the age of simple stellar populations made at a z$_{form}$ from 1.5 to 5. The dotted red line marks the age of the Universe as a function of redshift.}
\label{fig:ages}
\end{figure}

Despite the overall stability of the spectral fitting mentioned above, we recall that age estimates still rely upon the spectral fitting scheme adopted, on the assumptions made in the choice of template libraries and ultimately on the shape of the SFH used. Keeping such caveats in mind, we can start making meaningful statements on the ages of our QGs through relative comparisons of the mass-weighted ages within our sample. The bulk of our targets is consistent with having suppressed their star formation very recently, between 300-800 Myr prior to observation. The median value is 0.5 Gyr with a dispersion of 0.2 Gyr. Two outliers are present, ID 6 and ID 10, showing younger ages than the bulk of the sample, 0.1 and 0.25 Gyr, respectively. Despite the fact that for ID 10 the D$_{B}$/D$_{n}$4000 ratio appears to be the strongest (hence suggesting the highest contribution of A-type stars with respect to the underlying stellar population), the S/N of our spectra prevents us from identifying significant differences among our galaxies. Moreover, the downward trend of the D$_{B}$/D$_{n}$4000 ratio at $t_{50}\leq0.1$ Gyr precludes the possibility of testing the age inferred from the spectral fit, such as for ID 6. 
Assuming an exponentially declining SFH and our best-fit $t_{50}$, our galaxies are consistent with having formed half of their stellar mass around z$_{\rm{form}}\sim3.5$ at a SFR$\sim$1800-3000 M$_{\odot}$ yr$^{-1}$, similarly to what was reported in \citet{Valentino20}, but shifted at a later epoch. We caution, however, that such simple representations of SFHs are unlikely to be representative of the peak SFR if the true SFH were more complex, such as in the case of multiple phases in the SFR \citep{Barro16a} or mergers (which imply a degeneracy with mass assembly, hence in lower SFRs split between the progenitors). Interestingly, the common selection of high-z QGs preferentially selects bright blue UVJ quiescent objects where "dust-poor" PSBs often lie. In some cases, it extends to a bluer region outside the standard quiescent boundaries (either in the UVJ or in the NUVrJ selection) where compact transitioning galaxies are thought to lie along their fast drop in SFR \citep{Belli19, Schreiber18, Valentino20, Forrest20a}. This implies that high-z quiescent galaxies are selected more or less in the same evolutionary phase, namely after O and B stars exited the turn-off and before the same happens for A-type stars. This appears to be manifesting through similar distributions of (mass-weighted) ages among the highest-z samples (perhaps unsurprisingly, see Fig.~\ref{fig:ages}). This also means that the magnitude cut necessary for spectral acquisition biases the selection of high-z QGs against dusty PSBs or galaxies more slowly transitioning into quiescence \citep{Belli19}. Moreover, considering the cosmic time between z=2.8 and z=1.8, the mass-weighted ages inferred for our targets appear to be broadly consistent with passive evolution into old QGs at intermediate redshifts \citep{Whitaker13}. This is not the case for higher-z massive QGs, confirming that the high-z selection directly probes the continuous injection of new compact quiescent galaxies into the passive population.\\
The enhanced fraction of PSBs \citep[60\%-70\%,][]{Deuge20, lustig} among photometrically selected log(M$_{\star}$/M$_{\odot}$)$\geq$11 QGs at $z\sim3$, is linked to the progressive migration of the red sequence towards bluer colors with increasing redshift. PSBs represent a increasing fraction of the whole QGs population with redshift \citep{Wild16}. In particular, also because the transition from Balmer to CaII absorption lines is fast compared to the overall lifetime of a galaxy at low redshift (at the epochs spanned in this work, instead, such a phase naturally represents a much larger fraction of a galaxy lifetime). When the growth of the red sequence is considered, the observed fraction of PSBs in massive QGs remains low <1-3\% from z=0 to z$\sim$0.5 \citep{Tran2003} and increases between z$\sim$1-2 to $\sim$20\%--50\%, with percentage variations mirroring different selection criteria \citep{leborgne06, Whitaker13, Wild16}. The mounting fraction of massive QGs with evidence of recent quenching already accounts for half of the population at z$\sim$2, when their number densities start to match \citep{Whitaker12}. This is further supported by the growth rate of log(M$_{\star}$/M$_{\odot}$)$\geq$10.8 PSBs of less than 1 Gyr at z$\sim$2 which accounts for half that of the whole quiescent population \citep{Belli19}. The decrease in cosmic time allows to study the width of the distribution in quenching times of QGs: the emerging picture is one in which a continuous injection of objects into the quenched population manifests into the fast increase in the number density of young quiescent galaxies starting at $z\sim 1.5-2$ \citep{Whitaker11, Whitaker13} with a reversal of their relative contribution to the red sequence with respect to old galaxies by $z\sim$3. This trend appears to continue towards higher redshifts, where relatively old quiescent galaxies appear to be still unobserved \citep{Forrest2020b, Marsan2020}. The latest spectroscopic constraints at z$\sim$3.5 appear to find quasars or star-forming redshift interlopers among the reddest objects in the passive UVJ region, however the long integrations required limit both the number and the quality of these spectra \citep{Forrest2020b}. The question of whether PSBs scatter also into the reddest area of the passive UVJ region\footnote{E.g. (U-V)+(V-J)>2.8} due to photometric errors or intrinsic properties such as dust or metallicity (which can also be read as whether or not the z$\sim$3 population is missing the descendants of any z$\sim$4-5 massive quiescent galaxies) can only be addressed with very expensive targeted observations. The enhanced sensitivity of JWST will enable mapping the full distribution of PSBs on the UVJ diagram to the highest-redshifts. Establishing a detailed demographics of z$\sim$3 (or higher) QGs to the faintest magnitudes will help clarify the distribution of dust attenuation in spectroscopically confirmed extremely red objects and in turn provide insights on the global star formation history of high-z QGs. 

\subsection{On the morphology of quiescent galaxies at z$\sim$3}
We here leverage the results obtained by \citet{lustig} on the H-band imaging of our targets to interpret their plausible evolutionary stage considered their ages and morphology. 
Excluding our only object with a Sérsic index n=1.2 (ID5), 9 (8) of our galaxies have n>2 (2.5), even those objects with an axis ratio of about 0.5--0.6. This indicates that a centrally peaked, bulge-like stellar component is already established in these recently quenched galaxies. Similarly, \citet{Almaini17} showed that photometrically selected log(M$_{\star}$/M$_{\odot}$)$>$10.5 PSBs at $1<z<2$ consist of compact proto-spheroids, with Sérsic indices consistent with the established passive galaxy population at the same epoch and significantly larger than those of star-forming galaxies. They reported evidence for PSBs being more compact on average than fully passive galaxies at the same epoch. Based on these elements, they argued that for z$>$1 PSBs structural transformation preceded (or accompanied) the quenching event, leaving a compact remnant that later grew in size. Although the small number statistics prevents us from finding any significant correlation between morphology (either R$_{e}$ or n) and t$_{50}$ on a galaxy-by-galaxy basis, the data at our disposal show that our galaxies have median Sérsic indices and axis ratios of 4.5$_{-1.4}^{+0.3}$ and 0.73$_{-0.12}^{+0.06}$, with spectra consistent with being very recently quenched. Additionally, as noted in \citet{lustig} (see their Fig. 6 and 8) our galaxies have,  on average, significantly higher Sérsic indices with respect to the coeval star-forming population and effective radii about a factor of 3 smaller. Assuming that our mass-weighted ages are not significantly underestimated and that post-starburst features are not driven by minor rejuvenation effects leading to small variations in stellar mass, the young ages retrieved for our galaxies indicate that their average formation epoch is around z$_{\rm{formation}}\sim$3.5. Under the assumption that typical star-forming galaxies at our z$_{\rm{formation}}\sim$3.5 follow the extrapolation of the size evolution in \citet{vdw14} and rescaling at half of our stellar mass, our galaxies would still be more compact by a factor of 2. Although the intrinsic scatter of the mass-size relation for late-type galaxies would make the two populations marginally consistent, the evolution of Sérsic indices with redshift for star-forming galaxies points to the prevalence of exponential profiles (n$\sim$1-1.5) at least in the rest-frame UV \citep{Shibuya15}. These aspects support the idea that the development of a compact bulge-dominated structure is involved in creating quiescent systems and that this process precedes or is at least concomitant with the star formation rate suppression. 
The large uncertainties on our median axis ratio make our sample consistent with both quiescent galaxies at lower redshift and star-forming samples at $2<z<3$. Photometric samples at lower redshift point at a general flattening of the massive QG population towards z$\sim$2 or higher, judging from their rest frame optical projected axis ratio distributions \citep[e.g.][]{vdW11}. This flattening has been interpreted as an increasing fraction of disk-dominated QGs as opposed to classical triaxial spheroids \citep[60\% at z$\sim$2 at log(M$_{\star}$/M$_{\odot}$)$\gtrsim11.1$][]{Chang13b, Chang13a}). We note, however, that the distinction between bulge-dominated and disk-dominated systems is not clear-cut and it is often dependent on the tracer used. Axis ratios, for example, do not fully take the presence of a bulge component into account. The Sérsic indices in these works are mostly n$>$2-2.5 in the rest frame optical, which indicates that at least a spheroidal stellar component is already present in massive quiescent galaxies at z$\sim$2. Evidence in this regard is also presented in \citet{Belli2017, Almaini17, Stockmann20} at z$\sim$2. Despite the small number statistics which affects our sample, our data are inconsistent with the majority of massive quiescent galaxies being \textit{pure disks} at z$\sim$3: the probability associated with finding only 1 object with n$\sim$1 (ID 5) in a sample of 10 galaxies is 1\% assuming a 50\% fraction of pure disks and assuming that these morphologies are evenly distributed on the passive UVJ diagram. Along the same line, the fraction of pure disks that is consistent at 1$\sigma$ with our data is lower than 30\%.  Interestingly, spectroscopically confirmed QGs z$\sim$3-3.5 for which detailed morphology is available mostly show n=3-4 and with varying q \citep[see][]{Esdaile, Marsan15, Saracco20}, with the exception of \citet{Gobat12}. One possibility could be that the magnitude cut imposed for the spectroscopic confirmation biases spectroscopic samples towards high Sérsic index objects. This would reinforce the idea that recently quenched systems generally undergo fast quenching leaving a bulge-dominated remnant, likely without much dust reddening, in most cases. Another possibility could be that the rest frame near-UV/optical sampled by HST/F160W at z$\sim$3--3.5 is more centrally peaked than the rest frame 5000\AA\, sampled at z$\sim$2. High-resolution NIR imaging at longer wavelengths than those sampled by the present work will help clarify this point. Lastly, for ID 8 \citet{lustig} measured q=0.33$_{-0.03}^{+0.03}$ and n=4.3$_{-0.07}^{+0.08}$. These values could be interpreted as due to a strong bar in a fast-rotating compact S0 galaxy lacking an extended and bright outer disk, or alternatively, due to a compact spheroid surrounded by a lower-mass edge-on disk. On a more general note, the question of whether recently quenched galaxies at high-z are compact, bulge-dominated objects with residual rotational support is riveting \citep{Belli2017, Newman18b}. It appears unlikely that high-z quiescent galaxies closely resemble low-z slow rotators, given the shorter cosmic time interval for minor mergers to fully cancel any rotational component. However, the assessment of any rotational support is currently not feasible with the data at our disposal. 


\section{Summary and conclusions}
\label{sec:summary}

We have obtained HST WFC3/G141 grism spectra for one of the first representative samples of ten log$(M_{\star}/M_{\odot})>10.8$ quiescent galaxies at high redshift ($2.4<z<3.2$). HST observations efficiently provided us with the largest sample of QGs with a full continuous coverage in the Balmer/4000 \AA\ spectral region at these redshifts. This allowed us to perform spectroscopic confirmation of QGs up to z$\sim$3.2 and, thanks to widespread prominent Balmer breaks, we were able to start quantifying the incidence of newly quenched objects within the massive quenched population against contamination by lower-redshift interlopers. The quiescence of our targets was tested by means of the combined information of our newly acquired rest frame NUV/optical spectra and COSMOS2015 UV-to-NIR photometry. In addition, we also considered mid-IR, far-IR and radio detections from recently released super-deblended photometry \citep{Jin18}.

Our main conclusions can be summarized as follows:
\begin{itemize}
    \item[$\bullet$] Successful spectroscopic confirmation was achieved for the full sample, confirming the quality of the original photometric selection;
    
     \item[$\bullet$] The joint analysis of our newly acquired rest frame NUV/optical spectra and COSMOS2015 broad-band UV-to-NIR photometry confirms the quiescent nature of all our targets; 
    
        \item[$\bullet$] Although IR-based constraints on the obscured SFRs of our individual targets are weak with the available data (<120-190 M$_{\odot}$ yr$^{-1}$), the quiescent nature inferred from grism spectra and optical/NIR SEDs is globally supported from the 3GHz stack of the sample yielding an obscured SFR < 50 M$_{\odot}$ yr$^{-1}$ \citep{Deuge20};
    
     \item[$\bullet$] The use of photometric zero-point recalibrations proposed in \citet{Laigle16} appears to be disfavoured by our data. These corrections were derived on a set of spectroscopically confirmed quiescent among a much larger number of star-forming galaxies at intermediate redshifts and might not be necessary to extract the SEDs of QGs at high-z;
     
    \item[$\bullet$] An attenuation curve with slope $\delta=-0.4$, thus steeper than Calzetti, tends to reduce the median and the dispersion of the $\chi^2_{R}$ distribution for the photometric fits, as well as showing lower $\chi^2$ in general. Nonetheless, our data do not allow us to securely distinguish among the different slopes adopted;
    
    \item[$\bullet$] Marginalising the spectrophotometric fit over different attenuation curves and SFHs, the typical mass-weighted ages inferred for our objects range from 300-800 Myr, pointing at a recent rapid suppression of their SFR. Their global strength was quantified and compared to that of the 4000 $\AA$ break by means of the D$_{B}$/D$_{n}$4000 ratio which is systematically higher than 1. This spectroscopically confirms on a galaxy-by-galaxy basis the post-starburst nature of massive bright QGs, already pointed out by means of stacking \citep{Deuge20} and by individual high-resolution spectra \citep{Forrest2020b}. More observational efforts are required to explore to which extent strong Balmer absorption lines are spread among objects with a lower mass-to-light ratio;
    
    
    
    \item[$\bullet$] Interpreting our mid-IR and X-ray individual detections and the radio-stack shallow detection as a signature of AGN activity, our results are consistent with a widespread radio AGN activity a factor of 4 stronger than in similarly massive QGs at intermediate redshifts, and a 30-40\% incidence of luminous AGN in which the BH mass growth is substantially enhanced with respect to $z\sim2$ quiescent analogues ($\times30$) and to coeval star-forming galaxies at the same stellar mass ($\times60$). This is in agreement with the recent results of \citet{Carraro20} for the stacked emission of photometrically selected populations of QGs at $2.25<z<3.50$;
    
    \item[$\bullet$] Our galaxies are globally characterized by a bulge-dominated, compact morphology \citep{lustig}. Although no clear trend between mass-weighted ages and R$_{e}$ or Sersic indices could be found in the sample, the young ages yielded by their grism spectra and broad-band photometry suggest that structural transformation may precede or be concomitant with quenching in similarly selected high-z galaxies.
    
\end{itemize}

The fast evolutionary phase probed by the magnitude-limited color selection seems the one in which the majority of z>2.5 QGs are caught, systematically selecting newly quenched (quenching) objects that enter the quiescent population. We expect the number density of spectroscopically confirmed PSBs (which  already matches that of old systems at $z\sim2$ based on photometric selections \citep{Whitaker13}) to fully dominate the QGs mass-function at $z\sim3$. JWST and ALMA will be crucial to trace the full demographics of the quiescent population in an effort to map the distribution in quenching times, dust and molecular gas content, getting further insights into the global SFH of the emerging massive quenched systems.

\begin{sidewaystable}
\begin{threeparttable}
\renewcommand\arraystretch{1.5}
\resizebox{\textwidth}{!}{\begin{tabular}{cc ccc ccc ccc cccc}

\hline
ID  & z$_{\rm{spec}}$ & log(M$_{\star}$/M$_{\odot}$)& t$_{50}$ & A$_{\rm{V}}$ & $\delta$ & DB & D$_{n}$4000 & f$_{24 \mu \rm{m}}$  & logLx & logLx & S$_{3GHz}$ & SFR$_{24\mu \rm{m}}$  & SFR$_{3 \rm{GHz}}$ \\

&  &  & [Gyr] & [mag]& & & & [mJy]  &obs. &  24$\mu$m & [mJy] &  [$M_{\odot} yr^{-1}$] &  [$M_{\odot} yr^{-1}$] \\
\hline 
\hline 

1  &  2.841$^{+0.021}_{-0.018}$ & 11.37 & 0.5$^{+0.2}_{-0.2}$  &  0.2$^{+0.3}_{-0.1}$   & [-0.7,0] & 2.22 $\pm$ 0.51 & 1.47 $\pm$ 0.23 & - & $<$ 43.6 & - & 0.576 $\pm$0.03 &  867 &  $\sim$12930\\

2 &  2.557$^{+0.005}_{-0.005}$  & 11.50 & 0.8$^{+0.0}_{-0.1}$   & 0.1$^{+0.1}_{-0.0}$    & [-0.4,0]      & 2.02 $\pm$ 0.22 & 1.34 $\pm$ 0.10 &  -  &  $<$ 43.16  &  -  &     -   &  -   &  -  \\

3 &  3.124$^{+0.003}_{-0.003}$  & 11.55 & 0.4$^{+0.2}_{-0.1}$   & 0.4$^{+0.2}_{-0.2}$   & [-0.7,0]      & 2.39 $\pm$ 0.19 & 1.37 $\pm$ 0.11 &   0.041 $\pm$ 0.018   &  $<$ 43.6  &     43.73    &     0.011  $\pm$0.004     & <540   &  <307\\

4 &  3.230$^{+0.007}_{-0.006} $   & 10.98 & 0.4$^{+0.1}_{-0.1}$    & 0.1$^{+0.1}_{-0.0}$   & [-0.7,-0.4] & 3.15 $\pm$ 0.48 & 1.21 $\pm$ 0.09 &  0.000 $\pm$ 0.024  &  $<$ 43.6  & 43.60 & 0.003 $\pm$0.003    & <967   &  <242\\

5 &  2.665$^{+0.003}_{-0.007}$  & 11.34 & 0.4$^{+0.2}_{-0.2}$   & 0.8$^{+0.2}_{-0.2}$   & [-0.4] & 2.18 $\pm$ 0.29 & 1.39 $\pm$ 0.12 &   0.171  $\pm$ 0.017   &  $<$ 43.6  & 44.02 & 0.000 $\pm$0.002    & 644 &  <122\\

6 &  2.801$^{+0.005}_{-0.002}$  &11.36 & 0.1$^{+0.1}_{-0.0}$    & 1.0$^{+0.0}_{-0.1}$    & [-0.4]    & 2.02 $\pm$ 0.27 & 1.34 $\pm$ 0.12  &   0.100 $\pm$ 0.021   &  44.34$\pm$0.32 & 43.91 & 0.002 $\pm$0.003 & 508 &  <197\\

7 & 2.674$^{+0.005}_{-0.009}$\tnote{a} & 11.55& 0.7$^{+0.2}_{-0.1}$  & 0.6$^{+0.1}_{-0.2}$ &[-0.7,0] & 2.02 $\pm$ 0.35 & 1.27 $\pm$ 0.17 &   0.111  $\pm$ 0.011   &  44.26 $\pm$0.36 & 43.89 & 0.004 $\pm$0.003 &  419     &  <183\\

8 &  2.998$^{+0.002}_{-0.003}$  & 11.40 & 0.7$^{+0.0}_{-0.0}$  & 0.1$^{+0.0}_{-0.0}$   & [-0.7,-0.4] & 2.23 $\pm$ 0.2 & 1.41 $\pm$ 0.09 &   0.000 $\pm$ 0.018  &  $<$ 43.6  &  43.42 & 0.002 $\pm$0.002  & <394 &  <144\\

9 & 2.667$^{+0.015}_{-0.002}$   & 11.53 & 0.5$^{+0.0}_{-0.1}$  &  $0.7^{+0.1}_{-0.0}$   & [-0.4,0]      & 2.07 $\pm$ 0.41 & 1.61 $\pm$ 0.20 &   0.032 $\pm$ 0.023  &  $<$ 43.6 & 43.50 & 0.005 $\pm$0.003 &  <261 &  <183\\

10 &  2.393$^{+0.011}_{-0.000} $  & 11.33 & 0.3$^{+0.3}_{-0.1}$  & 1.6$^{+0.1}_{-0.3}$    & [0]               & 2.25 $\pm$ 0.51 & 1.15 $\pm$ 0.15 &   0.146 $\pm$ 0.020  &  43.73$\pm$0.29 & 43.87     &     0.005 $\pm$0.003    & 289      &  <155\\

\hline
\end{tabular}}
 \begin{tablenotes}
        \item[a] Obtained with the addition of NUV-to-NIR broad-band photometry as described in Sect.~\ref{sec:identspecz}.
    \end{tablenotes}
\end{threeparttable}
\caption{Best-fit values and their 1$\sigma$ uncertainties. Observed 24$\mu$ m and 3GHz flux densities. Upper limits are given at 3$\sigma$. Those for the observed bolometric X-ray luminosities were derived median-stacking all undetected sources. Stellar masses are taken from \citet{lustig} and converted to a Salpeter IMF. ID 2 is absent from the J18 catalog due to the absence of K$_{s}$ and VLA 3 GHz priors.}
\label{tab:final}
\end{sidewaystable}

\begin{acknowledgements}
We wish to thank the anonymous referee for their constructive and helpful comments. We are grateful to G. Brammer for assistance with the data reduction, to C. Vignali for providing X-ray spectra and M. Salvato for additional redshift constraints and helpful discussions. C.D. is grateful to C. Gomez-Guijarro for helpful discussions. V.S. acknowledges support from the ERC-StG ClustersXCosmo grant agreement 716762. I.D. acknowledges the European Union's Horizon 2020 research and innovation program under the Marie Sk\l{}odowska-Curie grant agreement No 788679. A.C. acknowledges the support from the grants PRIN-MIUR 2017and ASI n.2018-23-HH.0. Based on data products from observations made with ESO Telescopes at the La Silla Paranal Observatory under ESO programme ID 179.A-2005 and on data products produced by TERAPIX and the Cambridge Astronomy Survey Unit on behalf of the UltraVISTA consortium (Laigle cat.). This paper made use of Astropy,\footnote{http://www.astropy.org} a community-developed core Python package for Astronomy \citep{astropy13}, Matplotlib \citep{Hunter07} and Numpy \citep{numpy}.
\end{acknowledgements}


\bibliographystyle{aa}
\bibliography{references.bib}


\begin{appendix} 
\section{Additional plots}
We here add the multi-wavelength cutouts (from K$_{s}$ band to 20 cm) of our sources together with their SED fits when available. SEDs for IDs 1 and 2 are not present because ID1 lies in a region that is subject to unreliable IR fluxes and uncertainties, while ID 2 is not present in the J18 catalog due to lack of UltraVISTA K$_{s}$ and VLA 3 GHz radio priors. 
We recall that the plots on the right are only intended to display the available data and to show the maximum AGN component allowed by our 24 $\mu$m detections.
For completeness: SED fitting was carried out by fixing the redshift to the best-fitting grism value of each galaxy and includes two components: a stellar component from BC03 templates (black curve), and a mid-IR AGN torus from \citet{Mullaney11} (red curve). The downward arrows show the 2$\sigma$ upper limit at a given wavelength.\\

\begin{figure*}
\centering
\begin{subfigure}{.4\textwidth}
  \centering
  \includegraphics[width=0.85\linewidth]{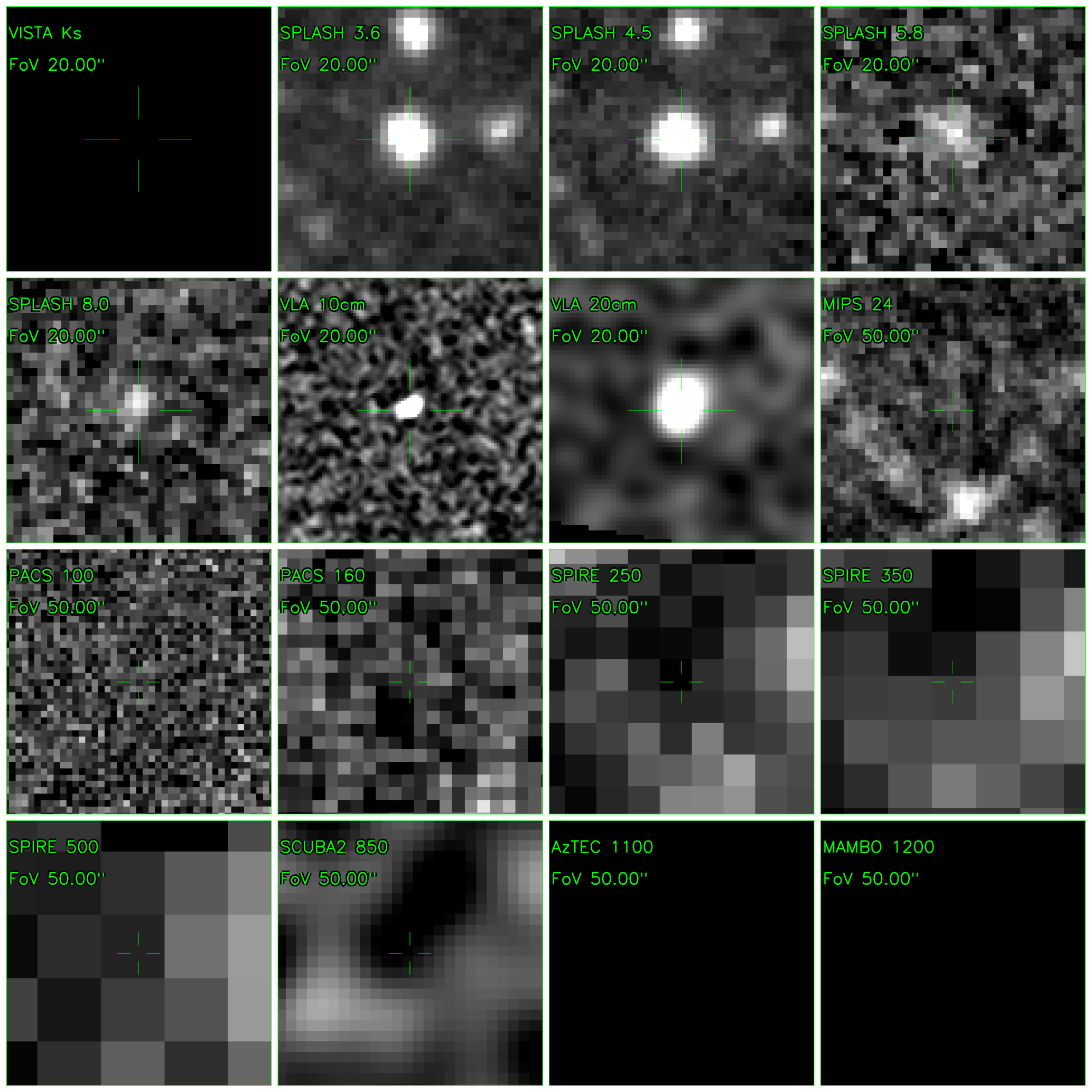}
\end{subfigure}%
\begin{subfigure}{.4\textwidth}
  \centering
  \includegraphics[width=0.85\linewidth]{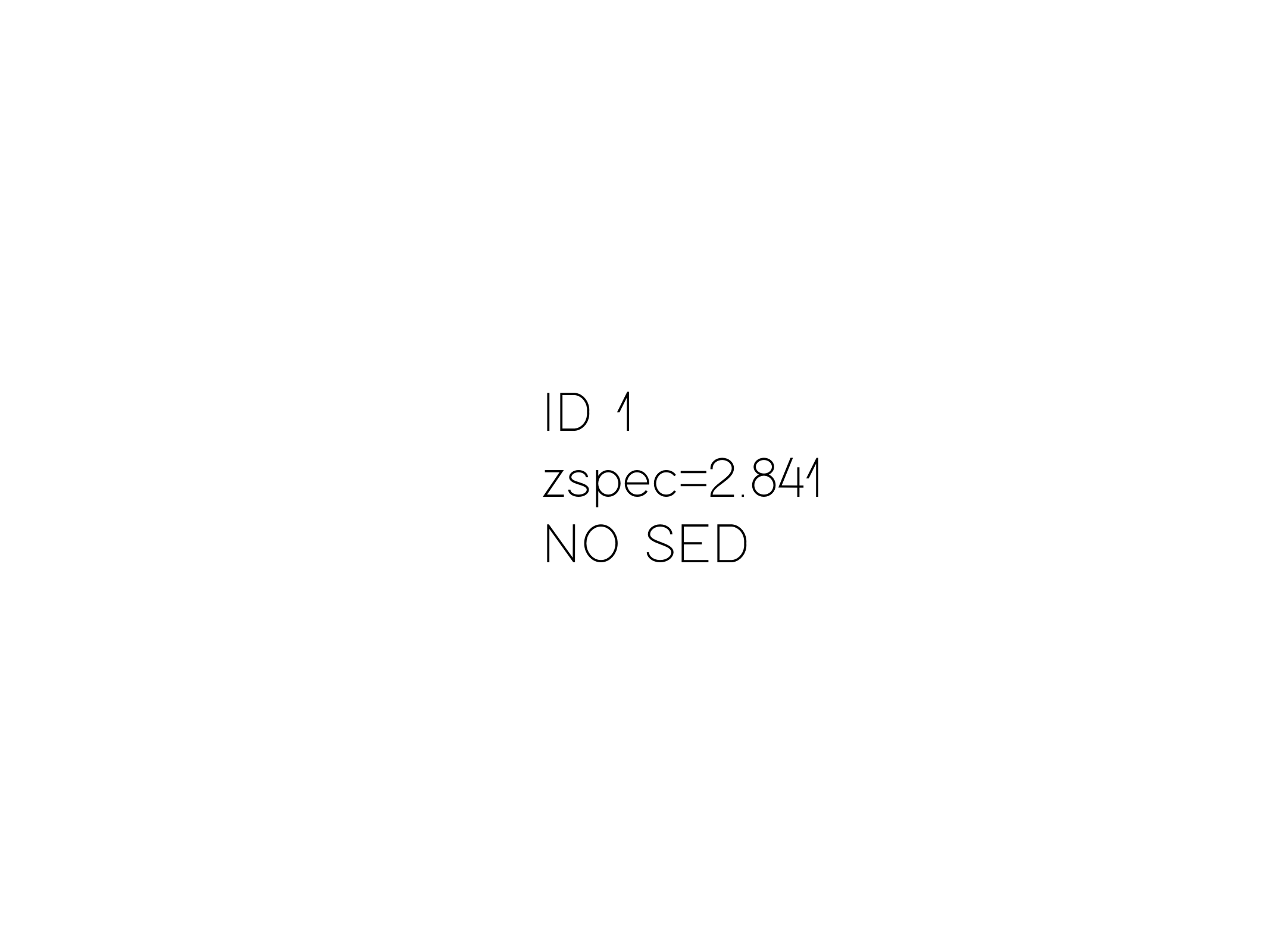}
\end{subfigure}

\centering
\begin{subfigure}{.4\textwidth}
  \centering
  \includegraphics[width=0.85\linewidth]{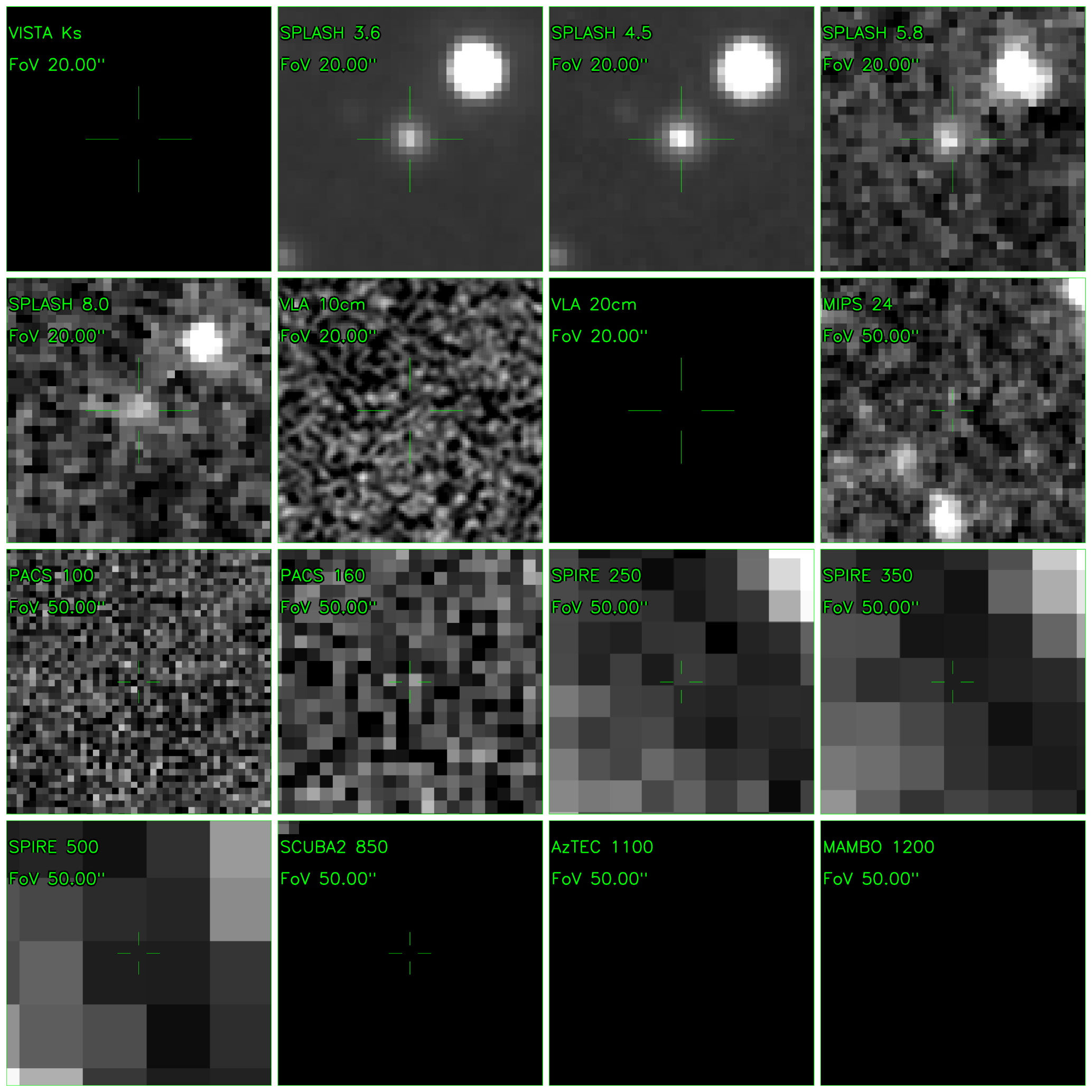}
\end{subfigure}%
\begin{subfigure}{.4\textwidth}
  \centering
  \includegraphics[width=0.85\linewidth]{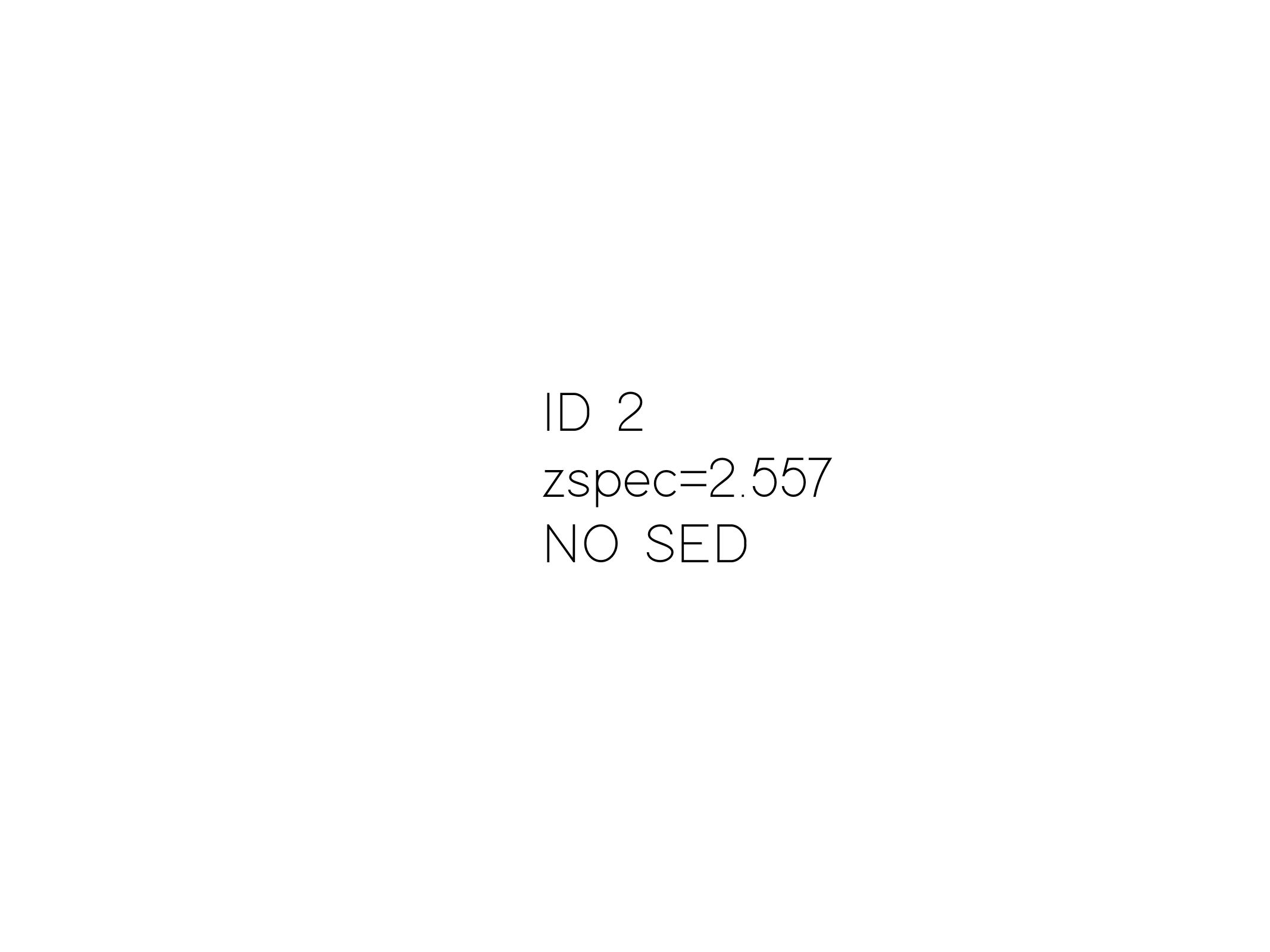}
\end{subfigure}

\centering
\begin{subfigure}{.4\textwidth}
  \centering
  \includegraphics[width=0.85\linewidth]{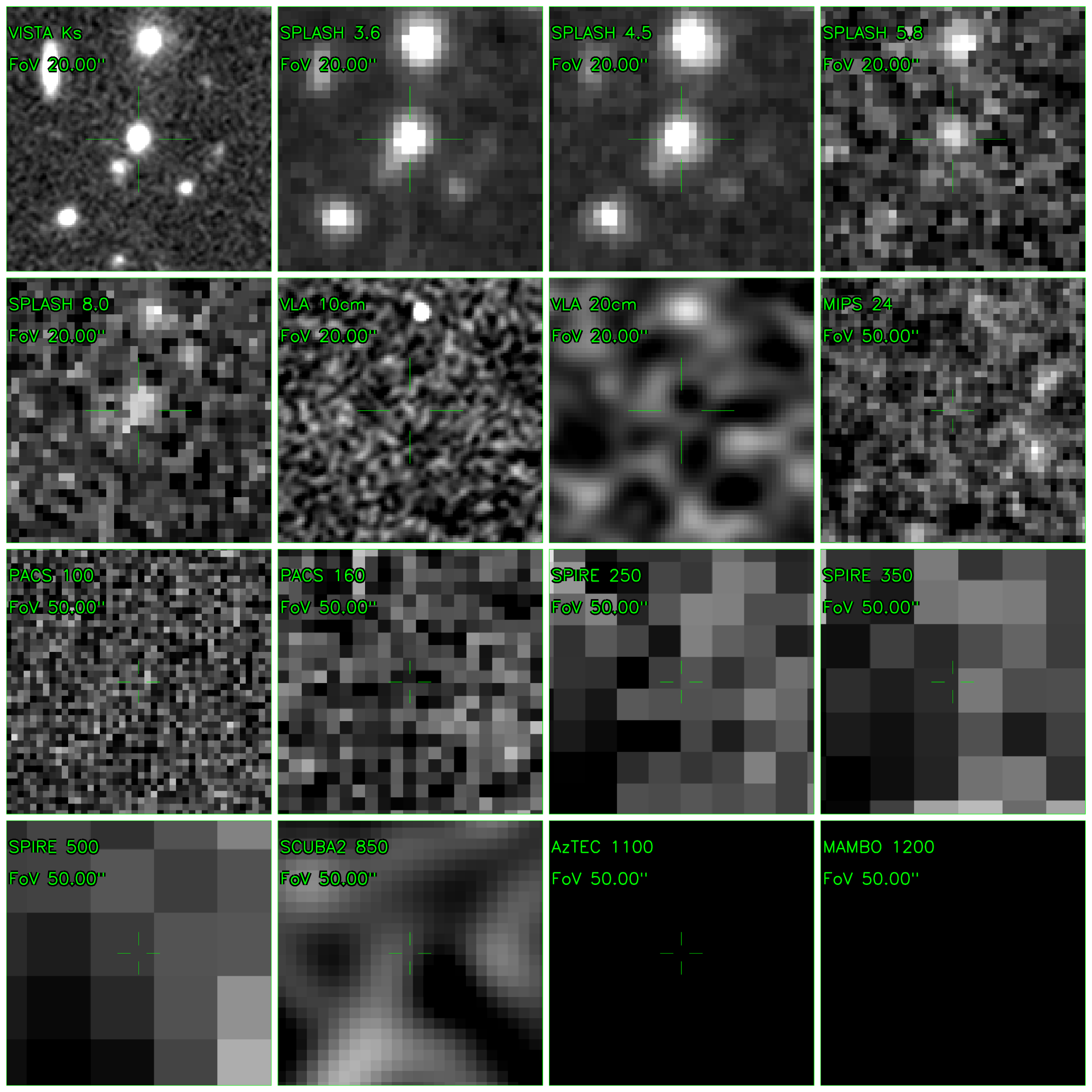}
\end{subfigure}%
\begin{subfigure}{.4\textwidth}
  \centering
  \includegraphics[width=0.85\linewidth]{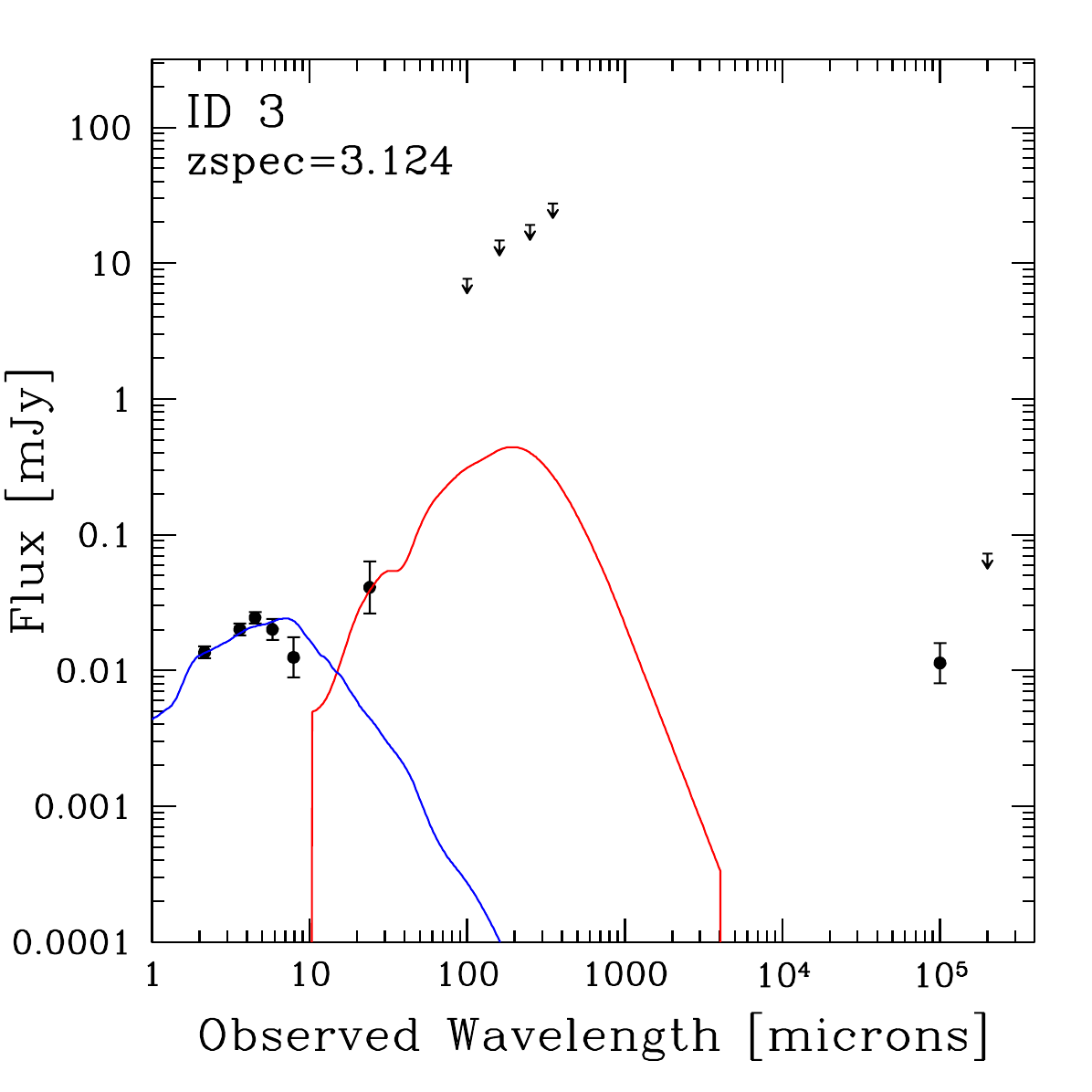}
\end{subfigure}%

\caption{Left panel: Multiband cutouts of our targets. The green text marks the instrument, the observed wavelength in units of $\mu$m and the size of the field of view. Right panel: Fits to the SEDs of our galaxies. We do not report the SED of ID1 because it lies in an area of UltraVista affected by unreliable flux uncertainties. ID2 is not present in the catalog of \citet{Jin18} due to the lack of K$_{\rm{s}}$ and VLA 3GHz priors. The SEDs are fitted with a stellar component (blue curve; Bruzual \& Charlot 2003) and an AGN torus component \citep[red curve;][]{Mullaney11}. The fits where fixed to the grism redshifts derived as in the main text. Upper limits are at 2$\sigma$.}
\label{fig:cl12301}
\end{figure*}

\begin{figure*}

\centering
\begin{subfigure}{.4\textwidth}
  \centering
  \includegraphics[width=0.85\linewidth]{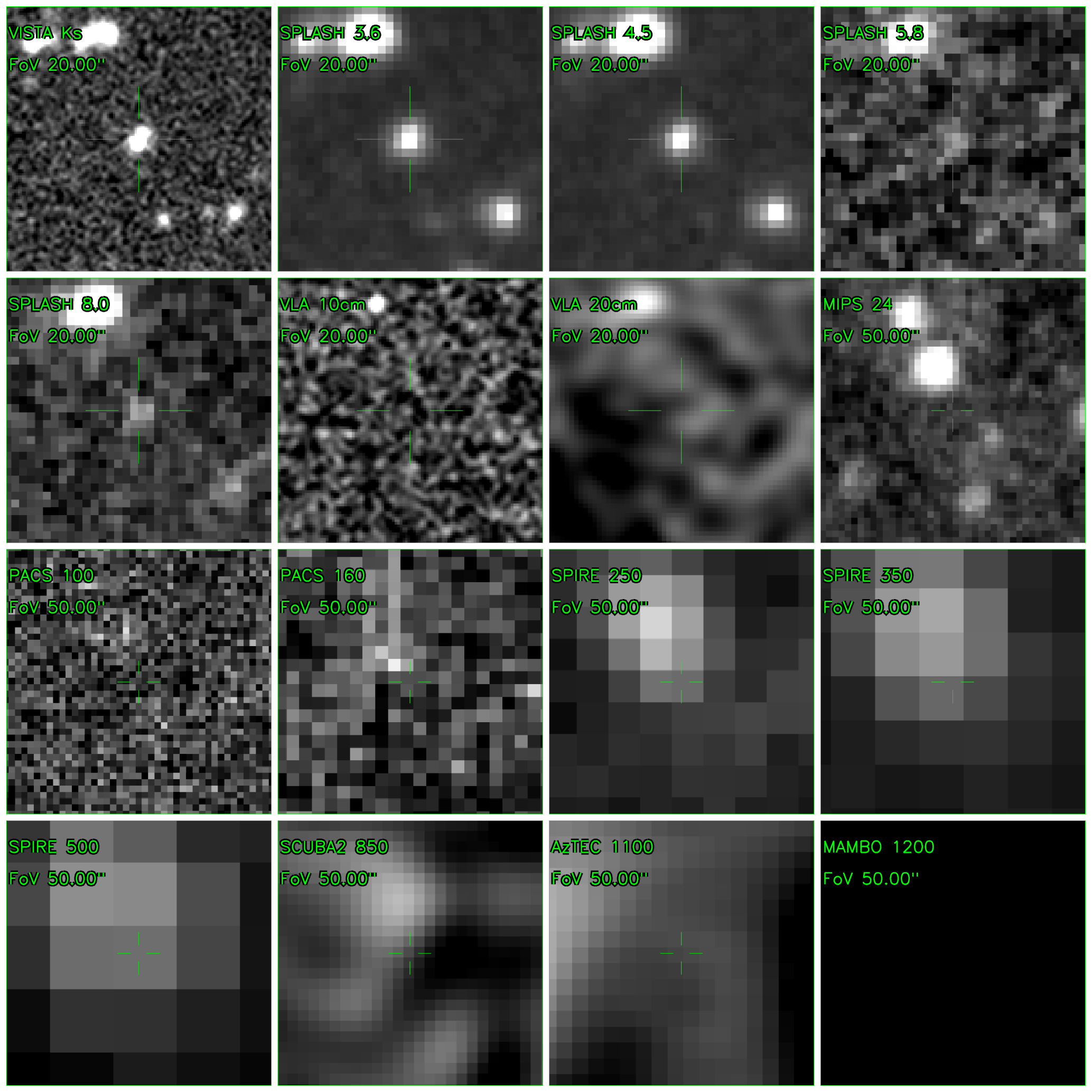}
\end{subfigure}%
\begin{subfigure}{.4\textwidth}
  \centering
  \includegraphics[width=0.85\linewidth]{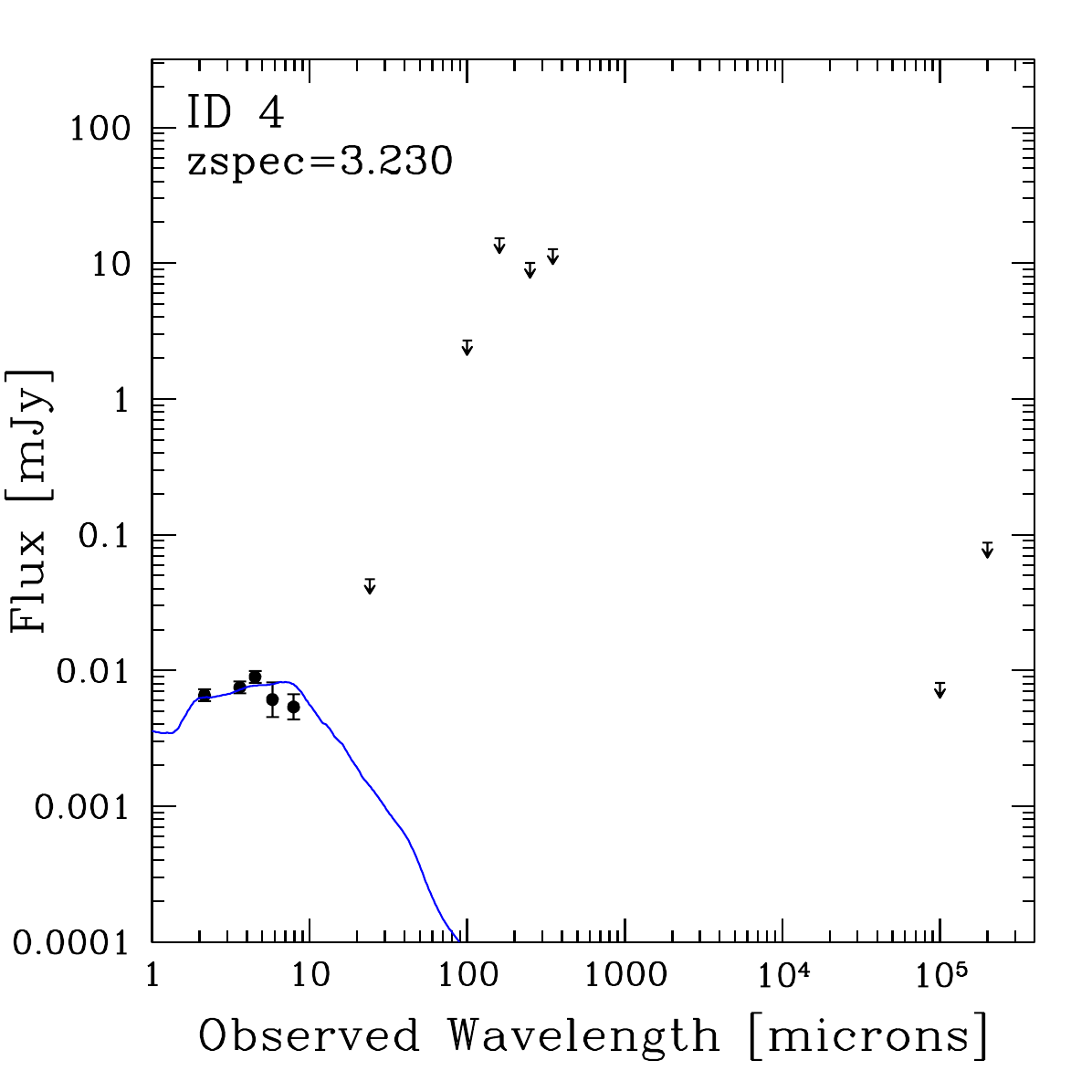}
\end{subfigure}%

\centering
\begin{subfigure}{.4\textwidth}
  \centering
  \includegraphics[width=0.85\linewidth]{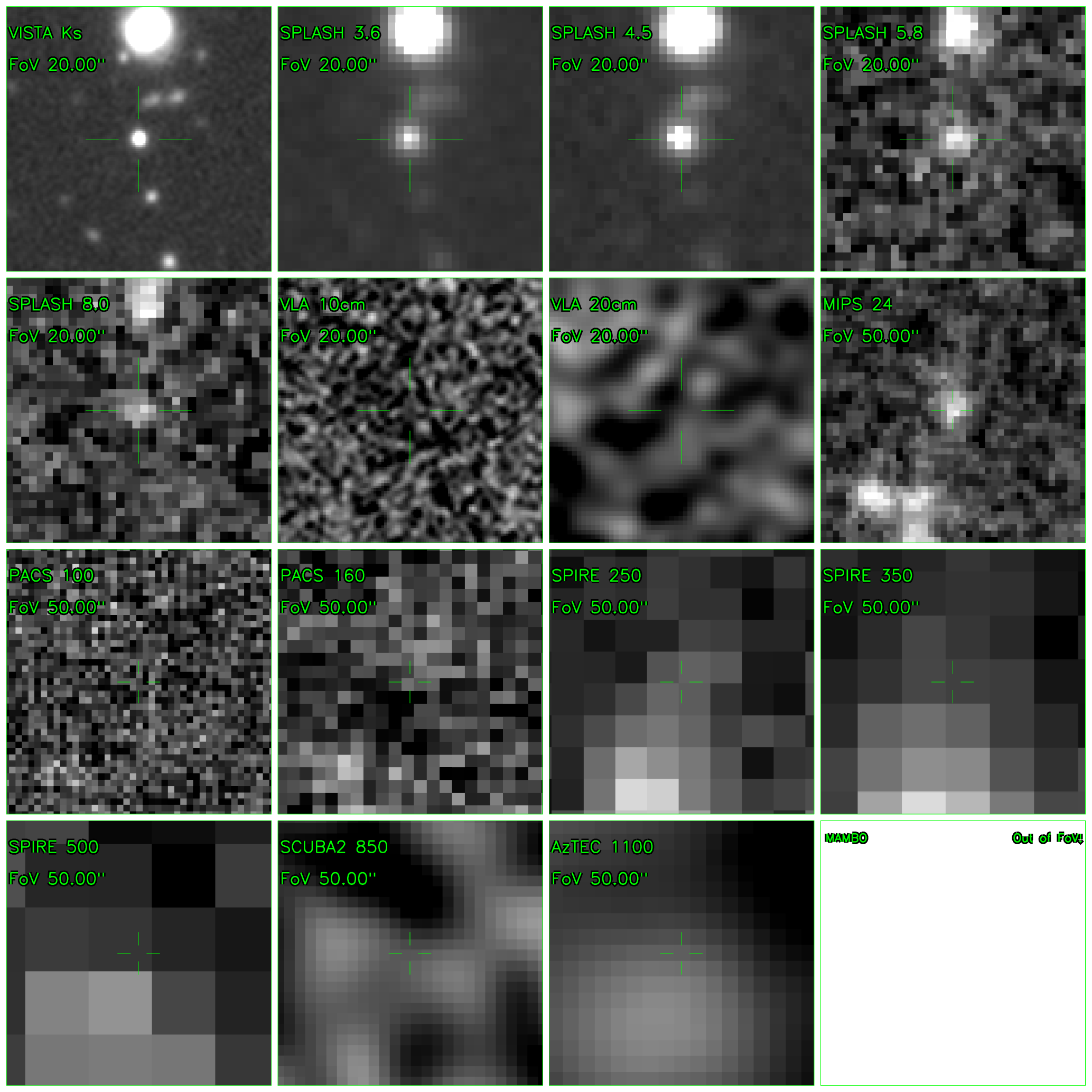}
\end{subfigure}%
\begin{subfigure}{.4\textwidth}
  \centering
  \includegraphics[width=0.85\linewidth]{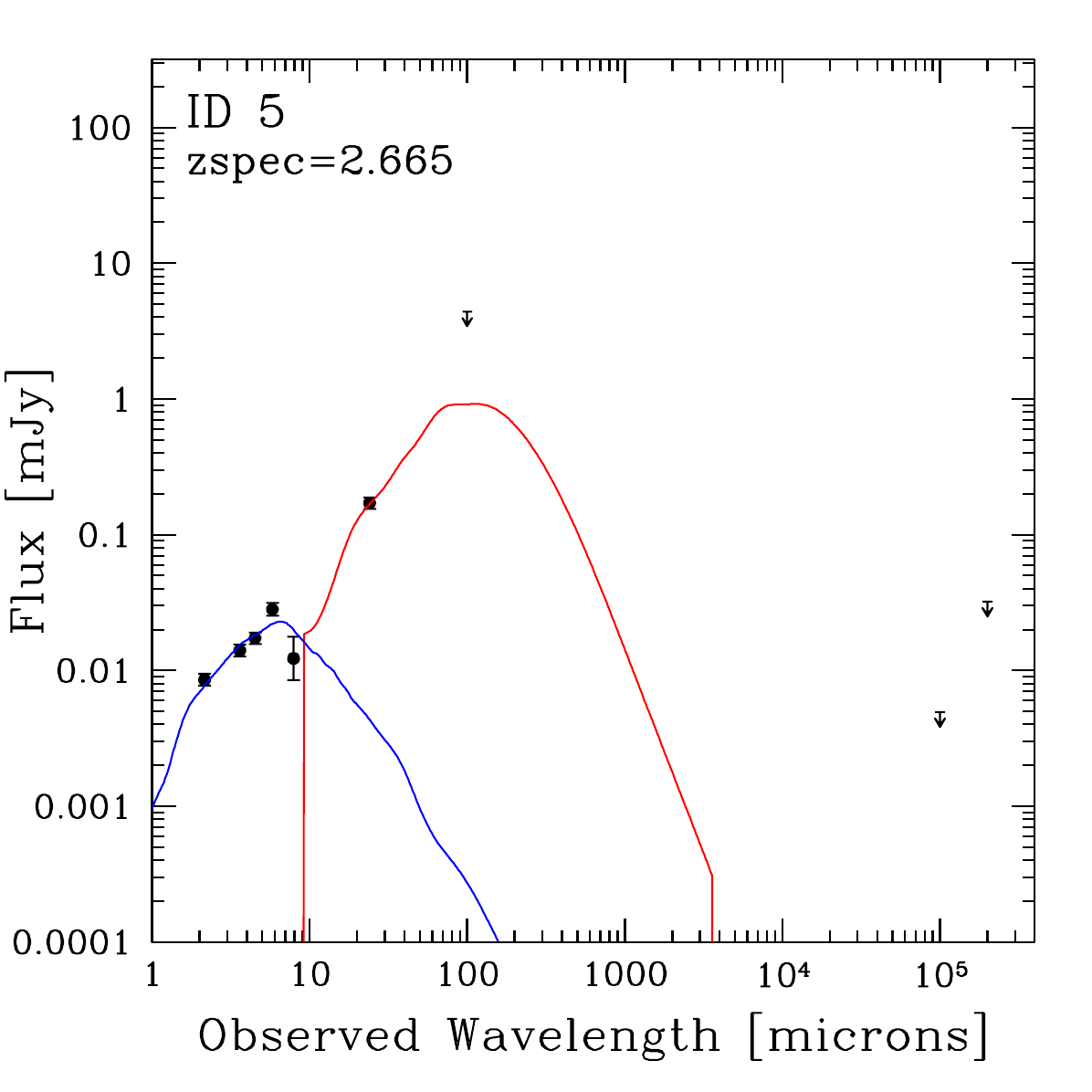}
\end{subfigure}%

\centering
\begin{subfigure}{.4\textwidth}
  \centering
  \includegraphics[width=0.85\linewidth]{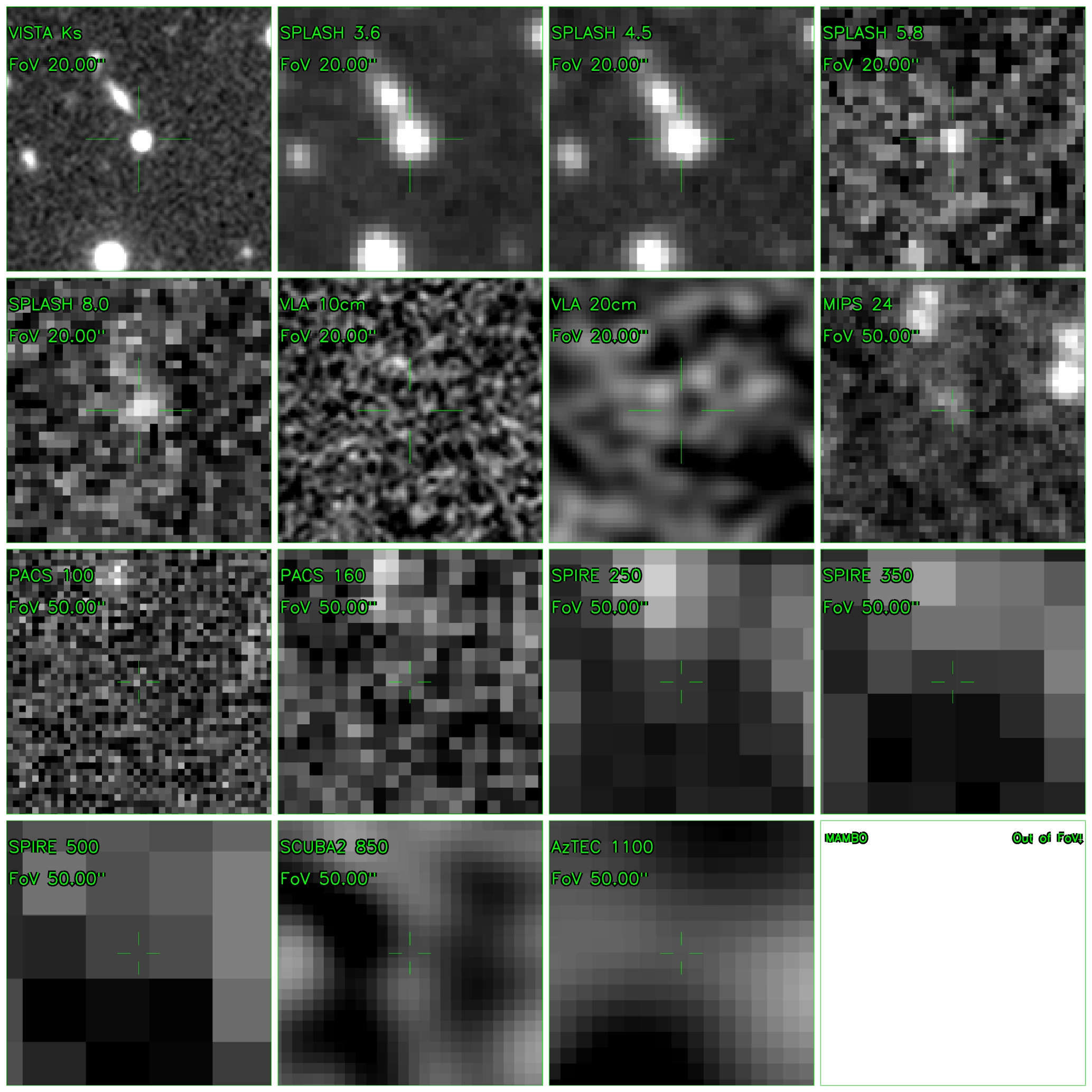}
\end{subfigure}%
\begin{subfigure}{.4\textwidth}
  \centering
  \includegraphics[width=0.85\linewidth]{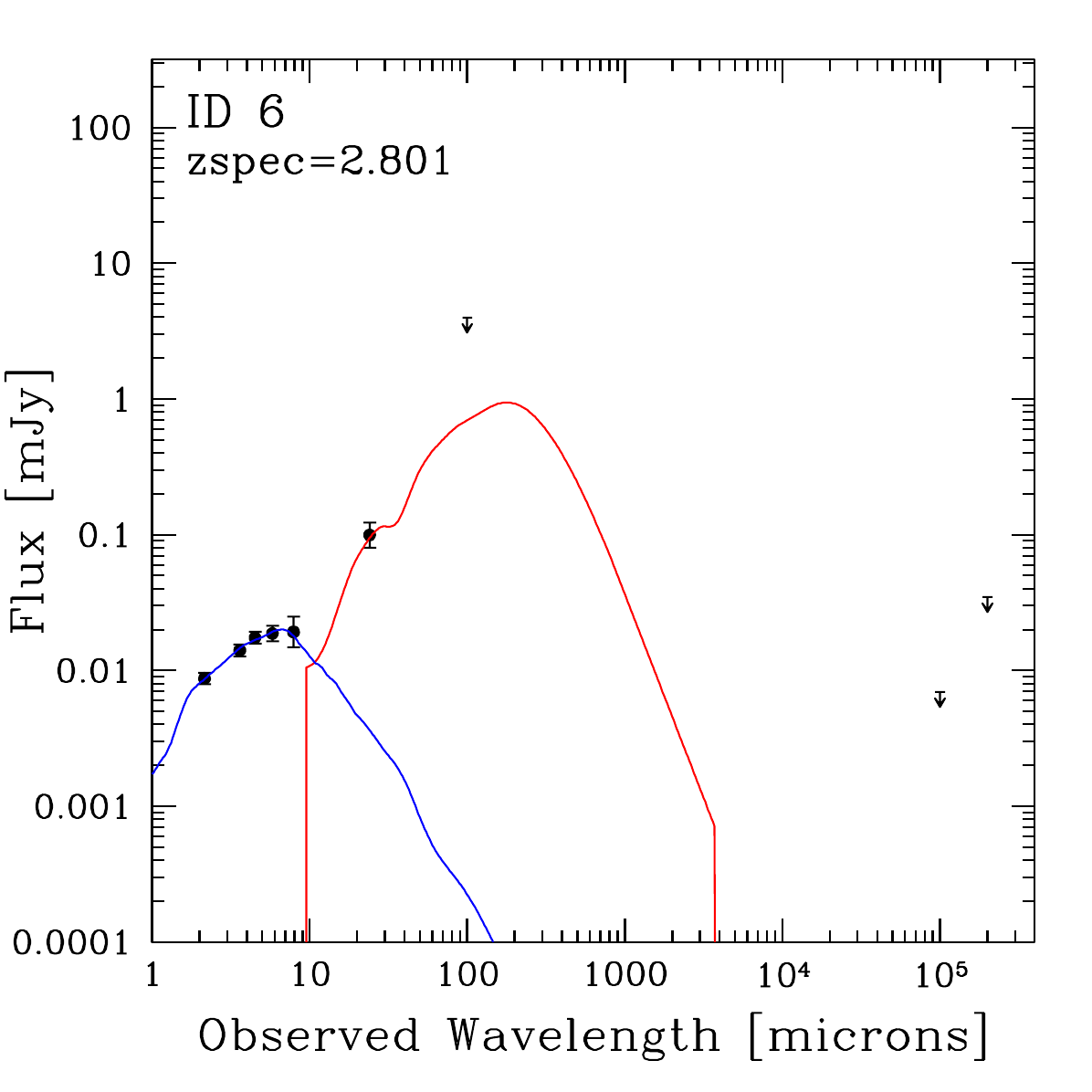}
\end{subfigure}%

\caption{Continued.}

\end{figure*}

\begin{figure*}

\centering
\begin{subfigure}{.4\textwidth}
  \centering
  \includegraphics[width=0.85\linewidth]{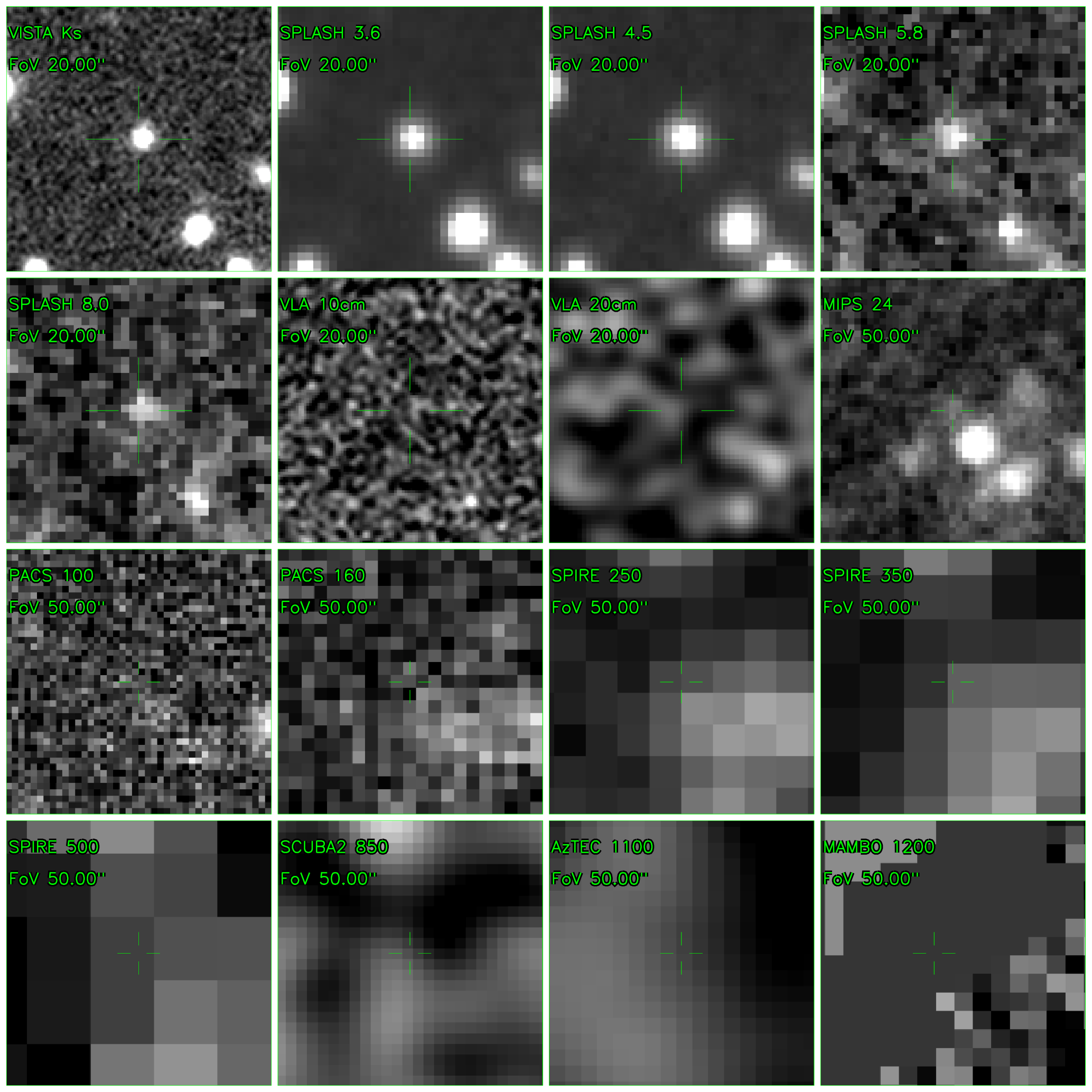}
\end{subfigure}%
\begin{subfigure}{.4\textwidth}
  \centering
  \includegraphics[width=0.85\linewidth]{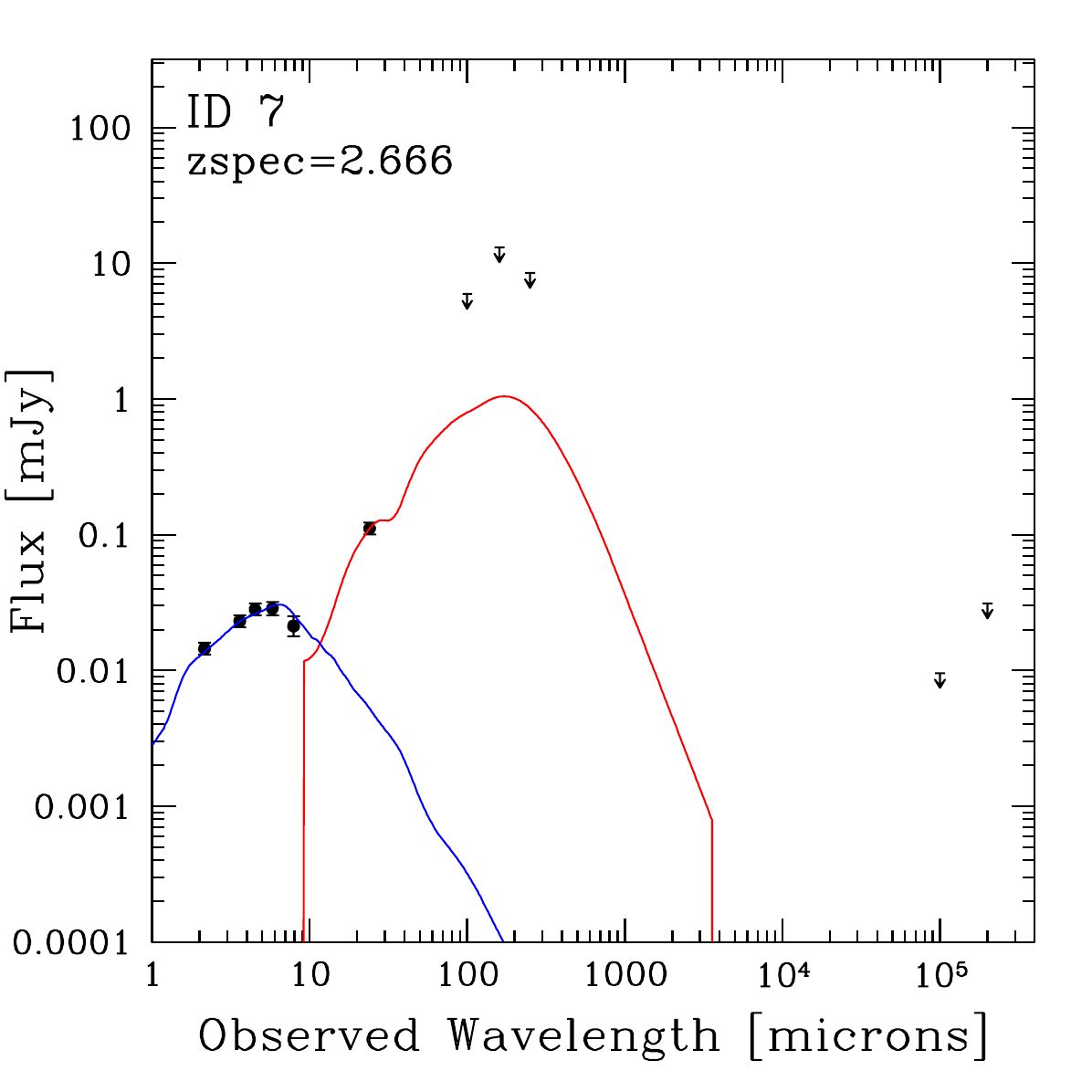}
\end{subfigure}

\centering
\begin{subfigure}{.4\textwidth}
  \centering
  \includegraphics[width=0.85\linewidth]{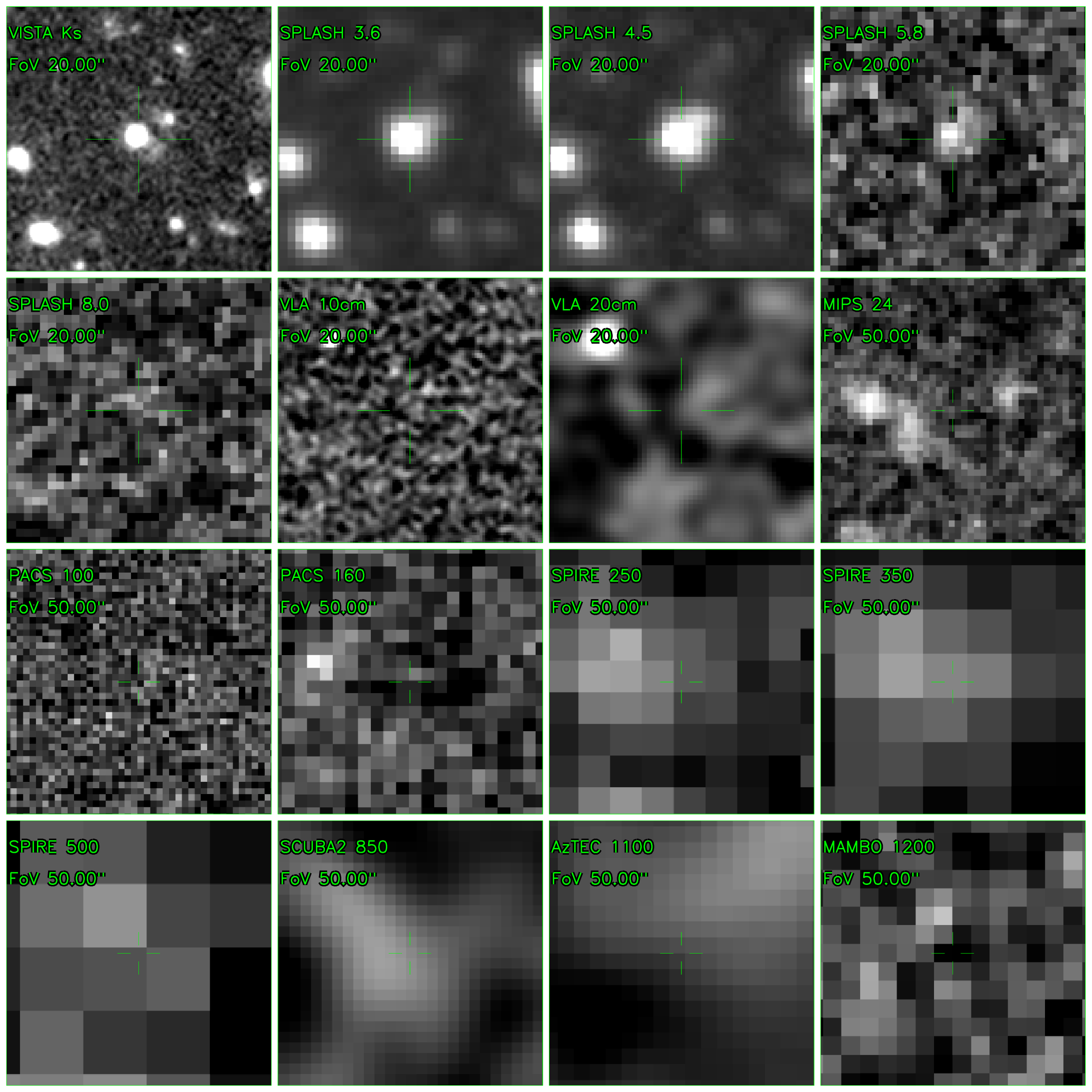}
\end{subfigure}%
\begin{subfigure}{.4\textwidth}
  \centering
  \includegraphics[width=0.85\linewidth]{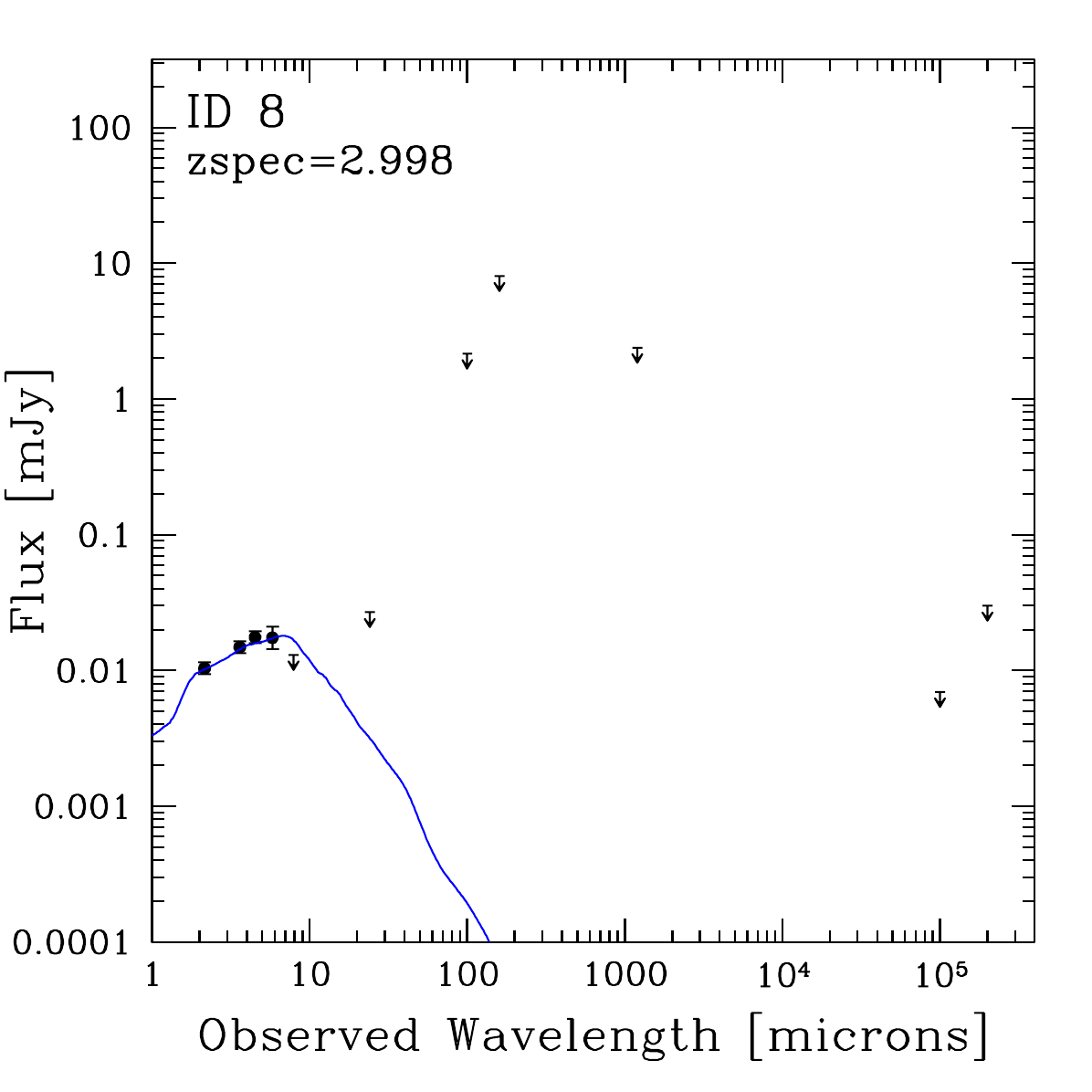}
\end{subfigure}%

\centering
\begin{subfigure}{.4\textwidth}
  \centering
  \includegraphics[width=0.85\linewidth]{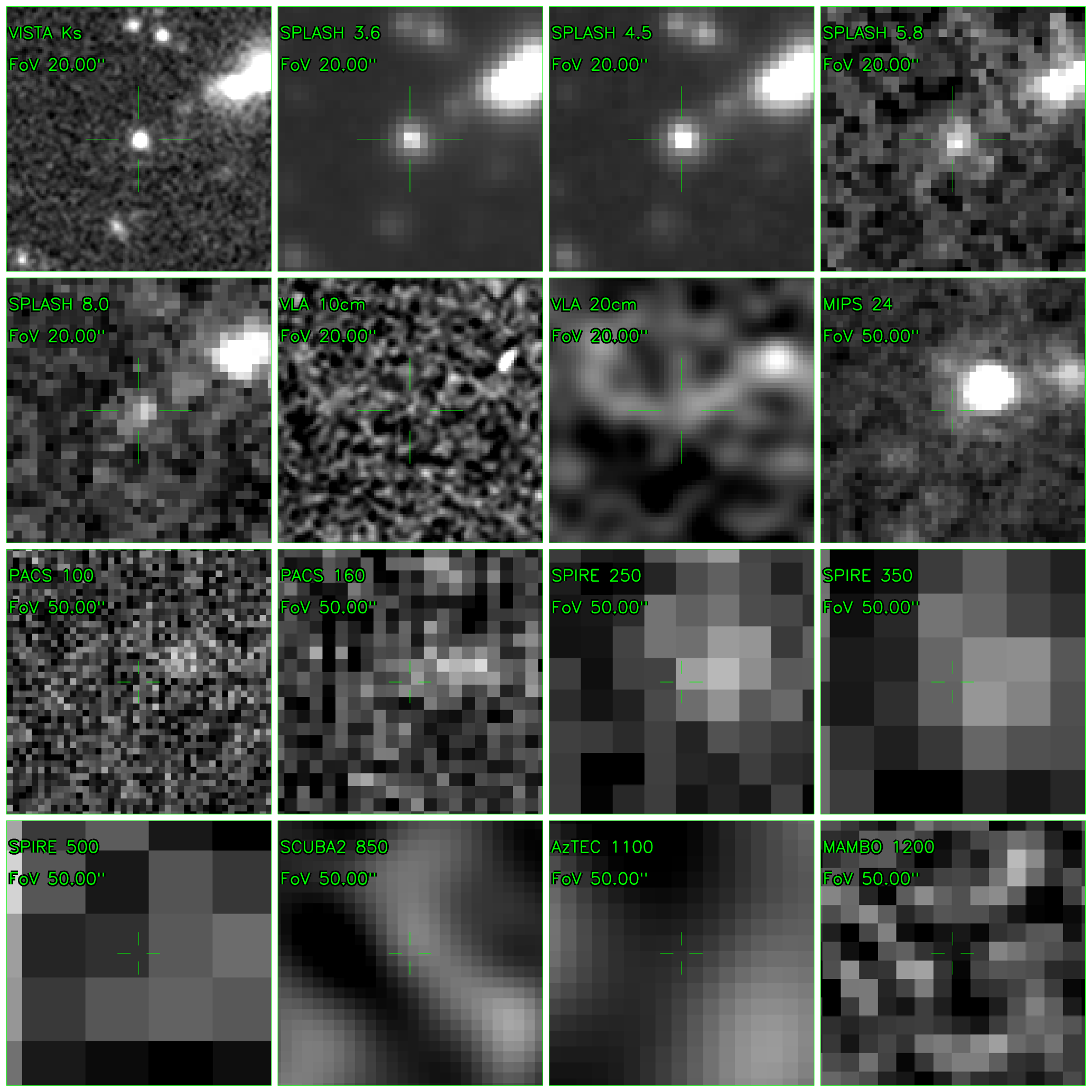}
\end{subfigure}%
\begin{subfigure}{.4\textwidth}
  \centering
  \includegraphics[width=0.85\linewidth]{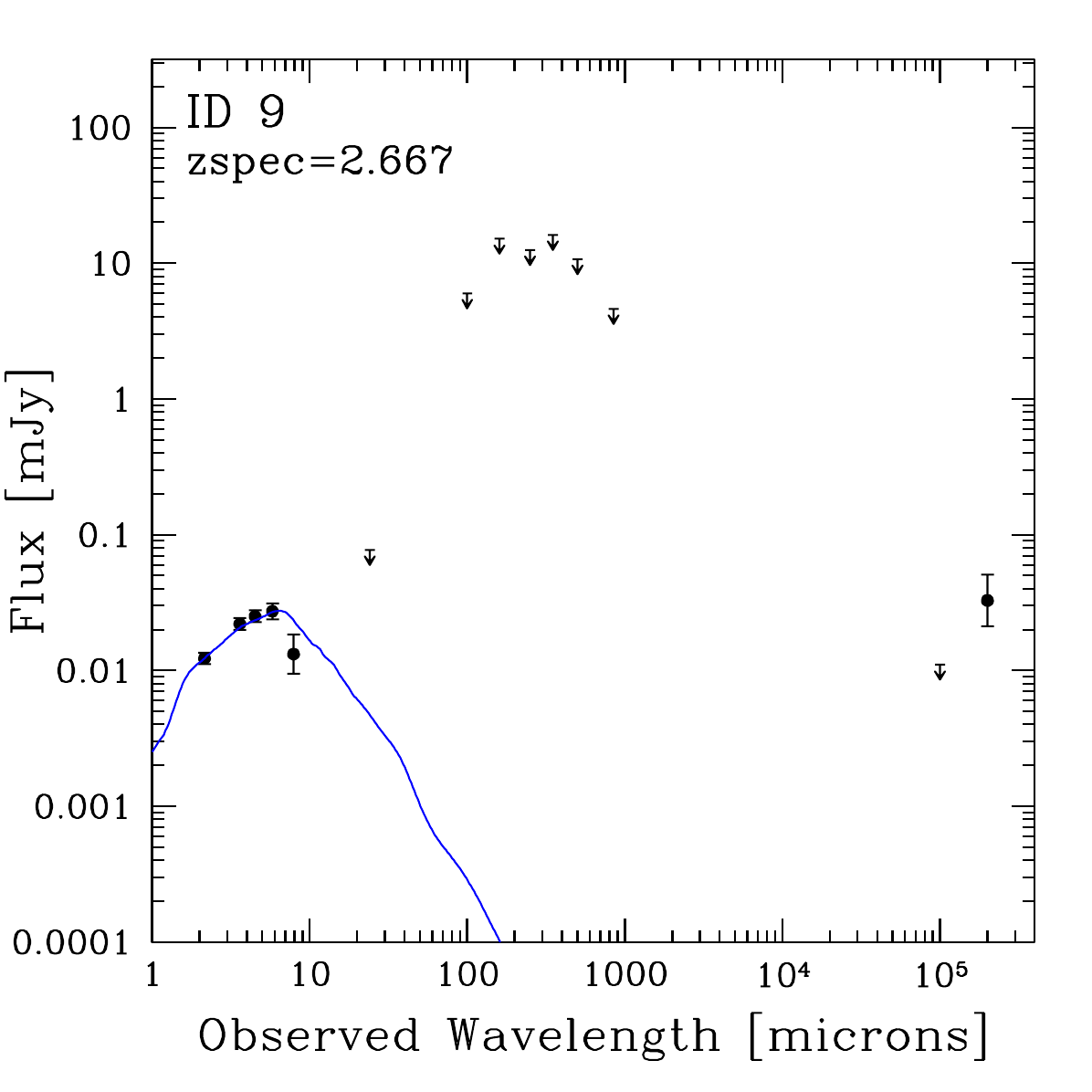}
\end{subfigure}%

\caption{Continued.}
\label{fig:cl12301a}
\end{figure*}

\begin{figure*}
\centering
\begin{subfigure}{.4\textwidth}
  \centering
  \includegraphics[width=0.85\linewidth]{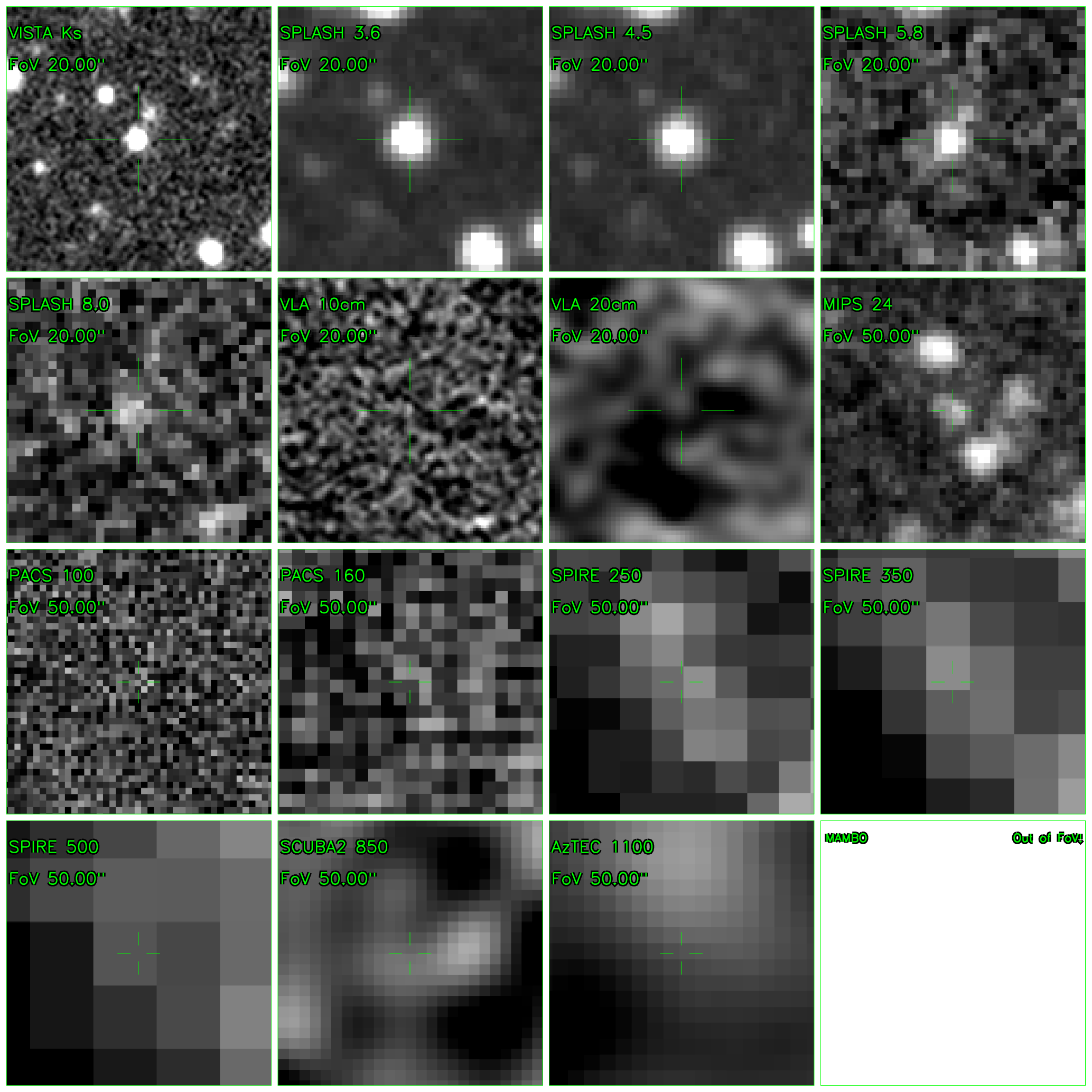}
\end{subfigure}%
\begin{subfigure}{.4\textwidth}
  \centering
  \includegraphics[width=0.85\linewidth]{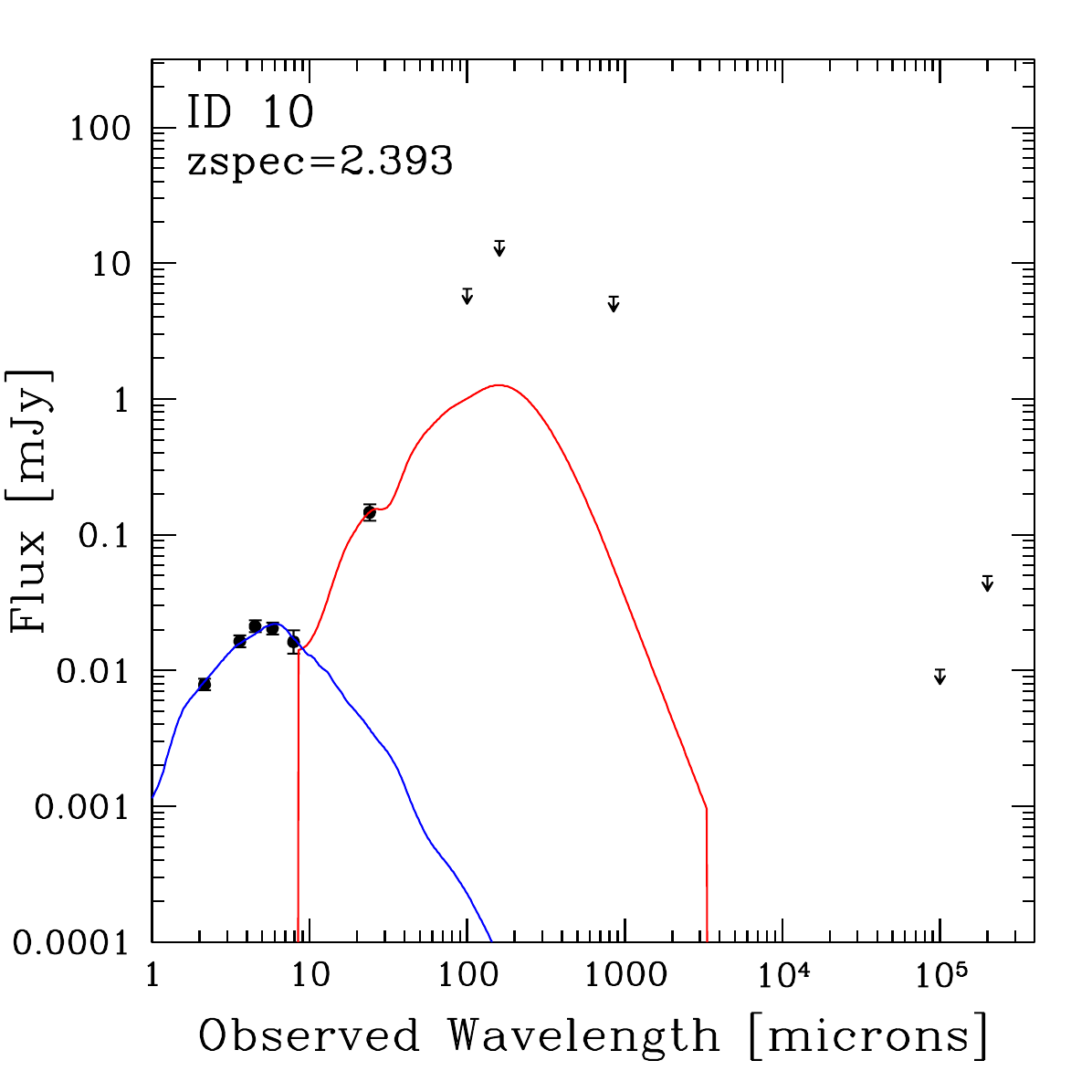}
\end{subfigure}%
\caption{Continued.}
\end{figure*}

\end{appendix}

\end{document}